\documentclass[twocolumn]{aastex63}
\usepackage{natbib}
\usepackage{color}
\usepackage{mathtools}
\usepackage{hyperref} 
\usepackage{multirow}

\def    \aa  		{\rm {A\&A}}
\def    \apjl  		{\rm {ApJL}}
\def    \apj  		{\rm {ApJ}}
\def    \mnras  	{\rm {MNRAS}}
\def    \araa  		{\rm {ARA\& A}}
\def    \apjl  		{\rm {ApJL}}

\def	\cm		{\,{\rm {cm}}}
\def	\K		{\,{\rm {K}}}
\def	\g		{\,{\rm {g}}}
\def	\mum	{\,{\mu \rm{m}}}

\def \bea {\begin{eqnarray}}
\def \ena {\end{eqnarray}}                  

\def    \ba     {\bf  a}

\def	\ba	{{\bf a}}

\def	\B	{{\rm B}}

\def	\C	{{\rm C}}

\def	\cm	{\,{\rm cm}}
\def	\d	{{\rm d}}

\def	\D	{{\rm D}}

\def	\erg	{\,{\rm erg}}
\def	\eV	{\,{\rm eV}\,}
\def	\g	{\,{\rm g}}
\def	\gas	{\,{\rm gas}}

\def	\G	{{\rm G}}

\def	\H	{{\rm H}}

\def	\N	{{\rm N}}

\def	\s	{\,{\rm s}}

\def	\AU	{\,{\rm au}}

\def	\V	{{\rm V}}

\def	\rad	{{\rm rad}}
\def	\yr	    {\,{\rm yr}}

\def	\xhat		{\hat{\bf x}}
\def	\yhat		{\hat{\bf y}}

\def	\ba			{{\bf a}}

\def    \gas     	{{\rm gas}}

\graphicspath{{./Figures/}}  

\begin{document}
\shorttitle{Dust and Ice Disruption in Protoplanetary Disks}
\shortauthors{Tung and Hoang}
\title{Rotational Disruption of Dust and Ice by Radiative Torques in Protoplanetary Disks and Implications for Observations}
\author{Ngo-Duy Tung}
\affil{University of Science and Technology of Hanoi, VAST, 18 Hoang Quoc Viet, Hanoi, Vietnam}
\author{Thiem Hoang$^{\star}$}
\affil{Korea Astronomy and Space Science Institute, Daejeon 34055, Republic of Korea; \href{mailto:thiemhoang@kasi.re.kr}{thiemhoang@kasi.re.kr}}
\affil{University of Science and Technology, Korea, (UST), 217 Gajeong-ro Yuseong-gu, Daejeon 34113, Republic of Korea}

\correspondingauthor{Thiem Hoang$^{\star}$}
\email{thiemhoang@kasi.re.kr}

\begin{abstract}
Dust and ice mantles on dust grains play an important role in various processes in protoplanetary disks (PPDs) around a young star, including planetesimal formation, surface chemistry, and being the reservoir of water in habitable zones. In this paper, we will perform two-dimensional modeling of rotational disruption of dust grains and ice mantles due to centrifugal force within suprathermally rotating grains spun-up by radiative torques for disks around T-Tauri and Herbig Ae/Be stars.
We first study rotational disruption of large composite grains and find that large aggregates could be disrupted into individual nanoparticles via the RAdiative Torque Disruption (RATD) mechanism. We then study rotational desorption of ice mantles and ro-thermal desorption of molecules from the ice mantle. 
We will show that ice mantles in the warm surface layer and above of the disk can be disrupted into small icy fragments, followed by rapid evaporation of molecules. We suggest that the rotational disruption mechanism can replenish the ubiquitous presence of polycyclic aromatic hydrogen (PAHs)/ nanoparticles in the hot surface layers of PPDs as observed in mid-IR emission, which are presumably destroyed by extreme ultraviolet (UV) stellar photons. We show that the water snowline is more extended in the presence of rotational desorption, which would decrease the number of comets but increase the number of asteroids formed in the solar nebula. Finally, we suggest that the more efficient breakup of carbonaceous grains than silicates by RATD might resolve the carbon deficit problem measured on the Earth and rocky bodies.

\end{abstract}

\section{Introduction}\label{sec:intro}
Dust and ice mantles on dust grains play an important role in various processes in protoplanetary disks (PPDs) around young stars, including the formation of planetesimals (\citealt{2008ARA&A..46...21B}), surface chemistry (\citealt{Henning2013}), and being the reservoir of water in habitable zones. Ice mantles also affect the formation of super-Earth (\citealt{2015ApJ...808..150H}) and giant-planet cores \citep{2008ApJ...673..502K}. In particular, the desorption of water and volatiles from ice mantles affects the chemical composition of giant planet atmospheres (\citealt{2011ApJ...743L..16O}; \citealt{2014ApJ...794L..12M}) and the delivery of water to the surfaces of terrestrial planets (\citealt{2007ApJ...669..606R}). 

PPDs exhibit a strong gradient of local density and temperatures in the vertical and radial directions due to the effect of stellar radiation and gravity. Near the star, water ice sublimates within a region of temperatures of $T_{\rm sub}\gtrsim 150\K$, and the boundary defines the water snowline about 2.7 au from a sun-like star \citep{1981PThPS..70...35H}. The physical structure of a PPD can be characterized by three distinct regions, including the hot upper layer directly irradiated by stellar and interstellar photons, partly shielded warm intermediate layer, and the disk interior near the midplane which is completely shielded from stellar photons (see e.g., \citealt{Henning2013}). 

Very large molecules, such as polycyclic aromatic hydrocarbon (PAH) molecules, are usually detected from the surface layers of disks around Herbig Ae/Be stars and some T-Tauri stars (\citealt{2004A&A...427..179H}; \citealt{2017ApJ...835..291S}) via mid-infrared emission features at 3.3, 6.2, 7.7, 8.6, 11.3, and 17 $\mu$m (\citealt{1984A&A...137L...5L}; \citealt{1985ApJ...290L..25A}; \citealt{2007ApJ...656..770S}; \citealt{2007ApJ...657..810D}). Furthermore, near-infrared (near-IR) observations from the Very Large Telescope (VLT) reveal the presence of carbonaceous nanoparticles throughout the surface of PPDs, even in the central cavity where large grains are depleted due to grain growth \citep{2019A&A...623A.135B}. Moreover, microwave observations reveal the existence of nanoparticles, including PAHs, nanosilicates, nanodiamonds, which are explained by spinning dust emission (\citealt{Hoang:2018hc}).

The widespread presence of PAHs/nanoparticles in the surface layer is unexpected because those nanoparticles are thought to be efficiently destroyed by extreme UV and X-ray radiation from the central star (\citealt{2010A&A...511A...6S}). Thus, the origin of such PAHs/nanoparticles in the disk surface layer remains unclear. Previously, PAHs are suggested to form in the high temperature and density regions behind the rim of PPDs (see \citealt{2011EAS....46..271K}). An alternative explanation is that nanoparticles may follow a different evolution from classical grains (size of $0.1\mu$m). As a result, while the classical grains are depleted in the disk due to coagulation and settling, PAHs/nanoparticles that are well mixed to the gas can exempt from grain settling and coagulation, and turbulence mixing can frequently transport nanoparticles from the disk interior to the surface (see \citealt{2007A&A...473..457D}). In this paper, we will show that PAHs/nanoparticles can be reproduced by rotational disruption of dust grains ($a\gtrsim 0.1\mum$) via the RAdiative Torque Disruption (RATD) mechanism as discovered by \cite{Hoang:2019da}. 

Destruction of ice mantles essentially occurs in the hot surface layer via thermal sublimation where ice mantles are heated to above $150\K$. In the warm molecular layer, with grain temperatures of $T_{d}\sim 30-100\K$, ice mantles can survive against thermal sublimation and are usually thought to play a central role for formation and desorption of complex molecules (see e.g., \citealt{Henning2013}). Indeed, ice mantles are considered the main route to form complex organic molecules (COMs), organic molecules that contain more than 6 atoms, such as $\rm CH_{3}OH$, $\rm HCOOH$, $\rm CH_{3}CHO$, $\rm C_{2}H_{5}OH$ (see \citealt{Herbst:2009go} for a review). To date, several complex molecules, including CH$_3$CN (\citealt{Oberg2015}), CH$_3$OH (\citealt{2016ApJ...823L..10W}), and formic acid (HCOOH) \citep{Favre:2018kf} are detected from PPDs. Interestingly, these molecules are observed at distant locations of $R>10 $au where grain temperatures are lower than the sublimation threshold of observed molecules.\footnote{\cite{Lee:2019} detected five COMs from V883 Ori from the stellar outburst during which an abrupt increase in the stellar luminosity expands the sublimation front.} The question is how such complex molecules can be released to the gas at large distances?

Recently, \cite{Hoang:2019td} showed that the entire icy grain mantle could be disrupted due to centrifugal stress induced by suprathermal rotation of grains spun-up by radiative torques (RATs, see \citealt{Hoangreview} for a review). Depending on the local gas density, rotational desorption of ice can occur at temperatures much lower than the water sublimation limit at $\sim 150\K$. Subsequently, water and complex molecules rapidly desorb from the resulting icy fragments due to thermal spikes. This rotational disruption mechanism is found to release water and complex molecules at much lower temperatures than previously predicted by thermal sublimation. Additionally, \cite{HoangTung} discovered that centrifugal potential induced by grain suprathermal rotation acts to reduce the potential barrier of adsorbed molecules onto the ice mantle and significantly enhances the rate of desorption. This mechanism is termed ro-thermal desorption.\footnote{In this paper, rotational disruption is used to refer to the split of a grain, while rotational desorption is referred to the separation of the ice mantle from the grain core.} As a result, we expect that these mechanisms would be efficient in the disk conditions even in the warm layer, which would dramatically affect the physical and chemical properties of PPDs. In this paper, we will apply these mechanisms to perform two-dimensional modeling of ice evolution and molecule desorption for PPDs around T-Tauri and Herbig Ae/Be stars. Moreover, PAHs/nanoparticles locked-up in the ice mantle could also be released by rotational and ro-thermal desorption of ice mantles (\citealt{HoangTung}). 
 
The structure of this paper is as follows. In Section \ref{sec:model}, we will describe a physical model of PPDs. In Section \ref{sec:RATD}, we will briefly review the physical mechanisms of RATD, rotational desorption of ice mantles, and ro-thermal desorption, respectively. Then, we will apply these mechanisms for the PPDs conditions and calculate disruption and desorption sizes of grains in Section \ref{sec:result}. We study the effect of rotational disruption on absorption and scattering opacity in Section \ref{sec:kappa}. Section \ref{sec:discussion} is devoted to discussing the implications of the applied mechanism on the detection of COMs and carbonaceous nanoparticles in the disks. A summary of our main findings is given in Section \ref{sec:summary}.

\section{Physical model of a protoplanetary disk}\label{sec:model}

\subsection{A passive irradiated disk model}
\begin{figure}
\includegraphics[width=0.48\textwidth]{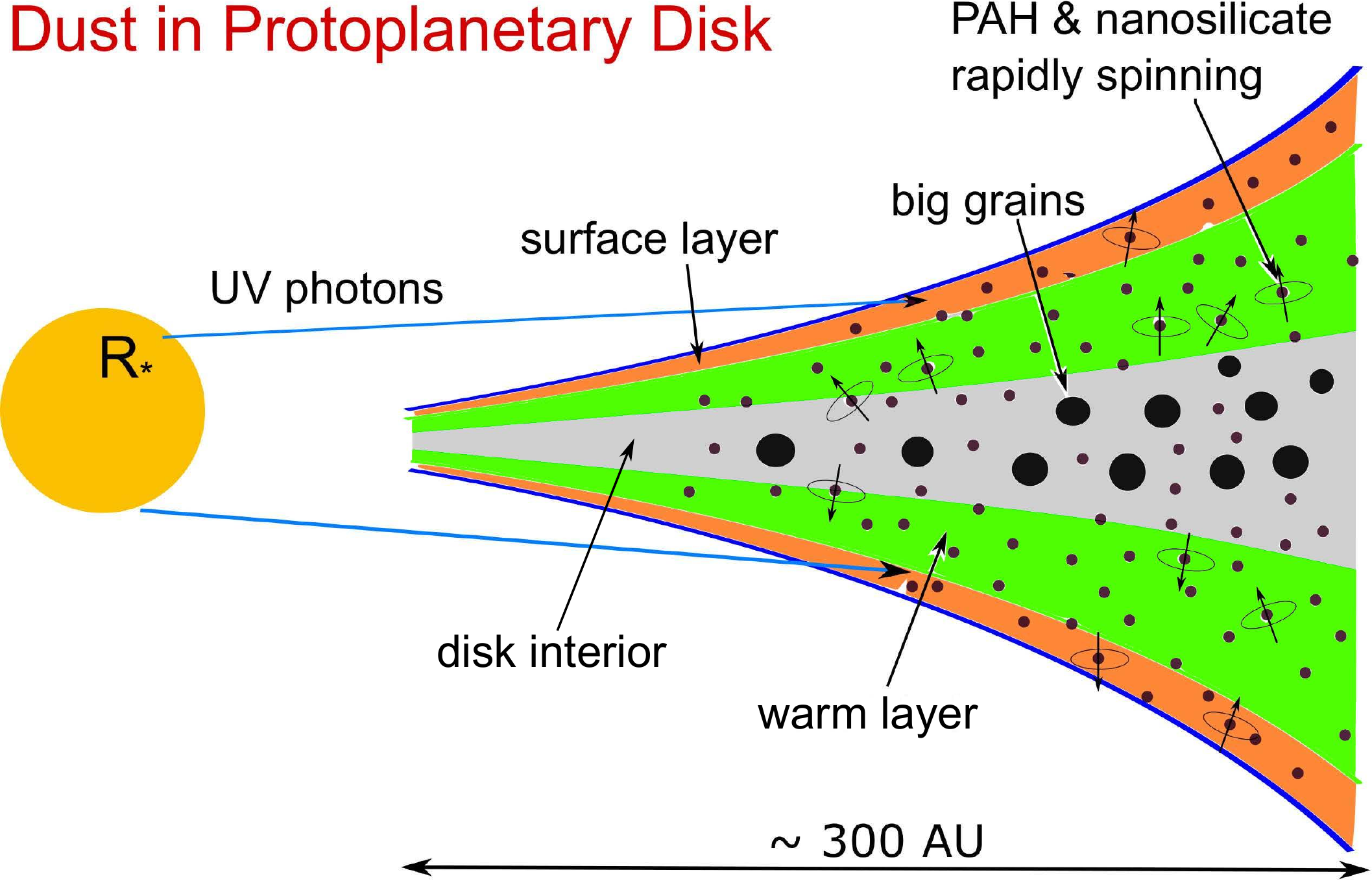}
\caption{A schematic illustration of a flared disk around a young star, consisting of three vertical layers: hot surface layer, warm intermediate layer, and disk interior.}
\label{fig:disk}
\end{figure}

For our calculations, we adopt a two-dimensional (2D) flared, radiative and hydrostatic equilibrium disk model (\citealt{1997ApJ...490..368C}) as illustrated in Figure \ref{fig:disk}. The disk surface is defined by a slant path of unity optical depth of $\tau_{V}\sim 1$. This layer is directly heated by stellar radiation, which is considered a photodissociated region. Dust and gas below this layer are heated by attenuated stellar radiation and define a warm layer ($T\sim 30-100\K)$. The final region is the disk interior, which is completely shielded from stellar radiation and only heated by infrared emission from hot dust in the surface layer.

\subsection{Gas Density Profile}

The gas number density at disk radius $R$ and height $z$ for the hydrostatic disk model, assuming a Gaussian vertical profile \citep{1974MNRAS.168..603L}, is given by
\bea
n_{\H}(R,z)&\approx&\frac{1}{2m_{\H}}\frac{\Sigma(R)}{H_{p}\sqrt{2\pi}}\exp\left(-\frac{z^{2}}{2H_{p}^{2}}\right), \label{eq:ngas0}
\ena
where $\Sigma(R)$ is the total surface mass density at radius $R$ given by
\bea
\Sigma(R)=\Sigma_{1}\left(\frac{R}{1~\AU}\right)^{-\alpha},\label{eq:sigmar}
\ena
where $\alpha$ is the model constant, and $\Sigma_{1}$ is the surface mass density at $R=1\AU$.

The pressure height scale $H_{p}$ is described by
\bea
\frac{H_{p}}{R}=\frac{H_{0}}{R_{0}}\left(\frac{R}{R_{0}}\right)^{1/7},\label{eq:Hp}
\ena
where $H_{0}$ is the aspect ratio at the reference radius $R_{0}$.

For $R_{0} =100$ au, $H_{0}/R_{0}$ is taken to be $0.1$ as a fiducial model, which corresponds to $H_{p}/R=0.1\times 3^{1/7}$ at $R_{\rm out}=300$ au. The chosen aspect ratio is much lower than predicted by \cite{1997ApJ...490..368C} but is comparable to observations (\citealt{2018ApJ...863...44A}). Here, we assume $\alpha = 1$ and $\Sigma_{1}$ varied to cover a wide range of disk mass. Other physical parameters are listed in Table \ref{tab:diskmod}, including the stellar temperature $T_{\star}$, stellar mass $M_{\star}$, stellar radius $R_{\star}$, and the inner and outer disk radii $R_{\rm in}$ and $R_{\rm out}$.

\begin{table}
\begin{center}
\caption{Physical parameters of a PPD}\label{tab:diskmod}
\begin{tabular}{llllll} 
\hline\hline
{\it Objects} & $T_{\star}$ & $M_{\star}$ & $R_{\star}$ & $R_{\rm in}$ & $R_{\rm out}$ \\
 & $(\K)$ & $(M_{\odot})$ & $(R_{\odot})$ & $(\AU)$ & $(\AU)$ \\[1mm]

\hline\\
Herbig Ae/Be & 10000	& 2 & 2  &  1 & 300\\[1mm]
T-Tauri	 & 4000	&  0.5 & 2 &  0.1 & 300 \\[1mm]
\hline\hline\\
\end{tabular}
\end{center}
\end{table}

\begin{figure*}
\includegraphics[width=0.5\textwidth]{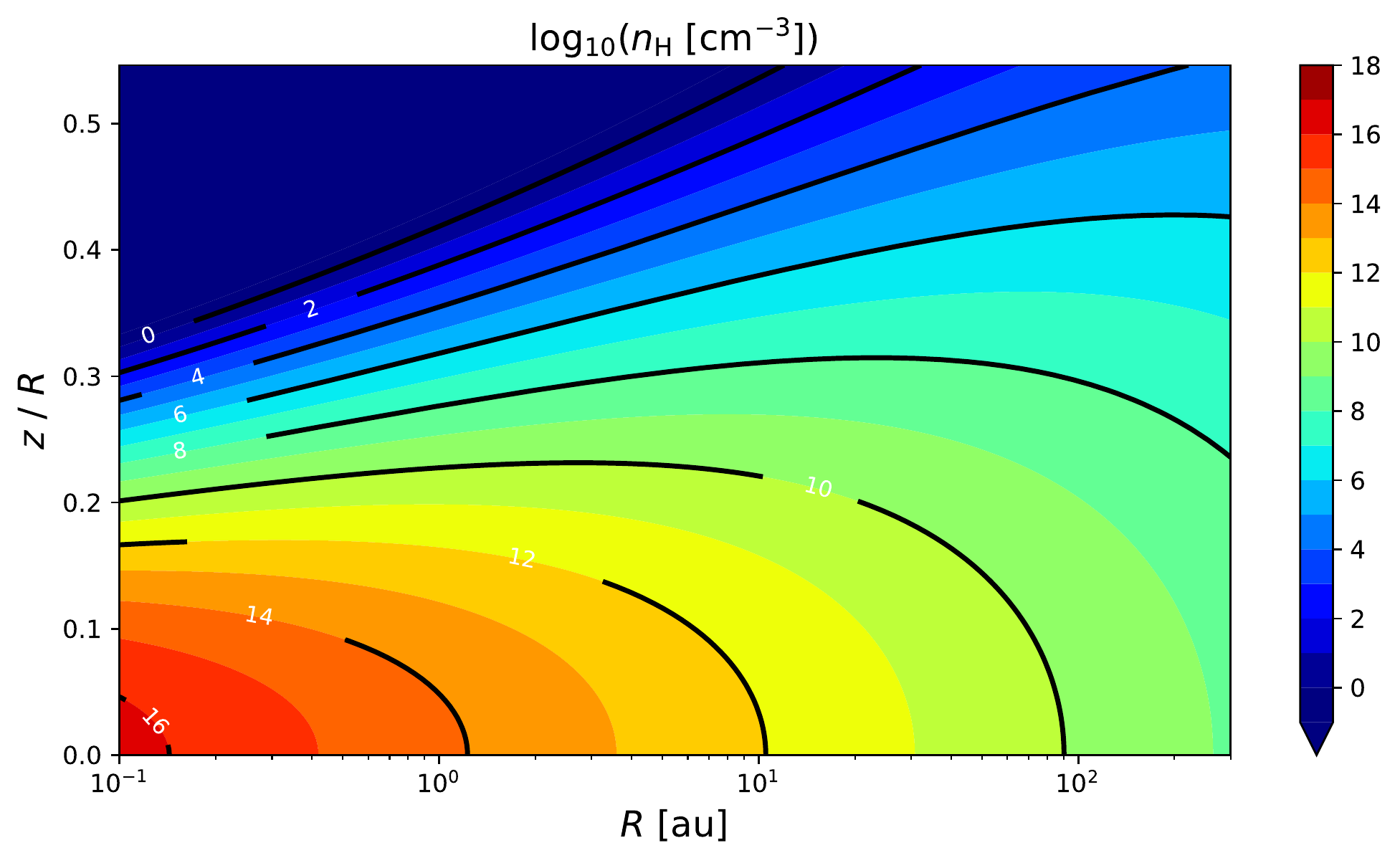}
\includegraphics[width=0.5\textwidth]{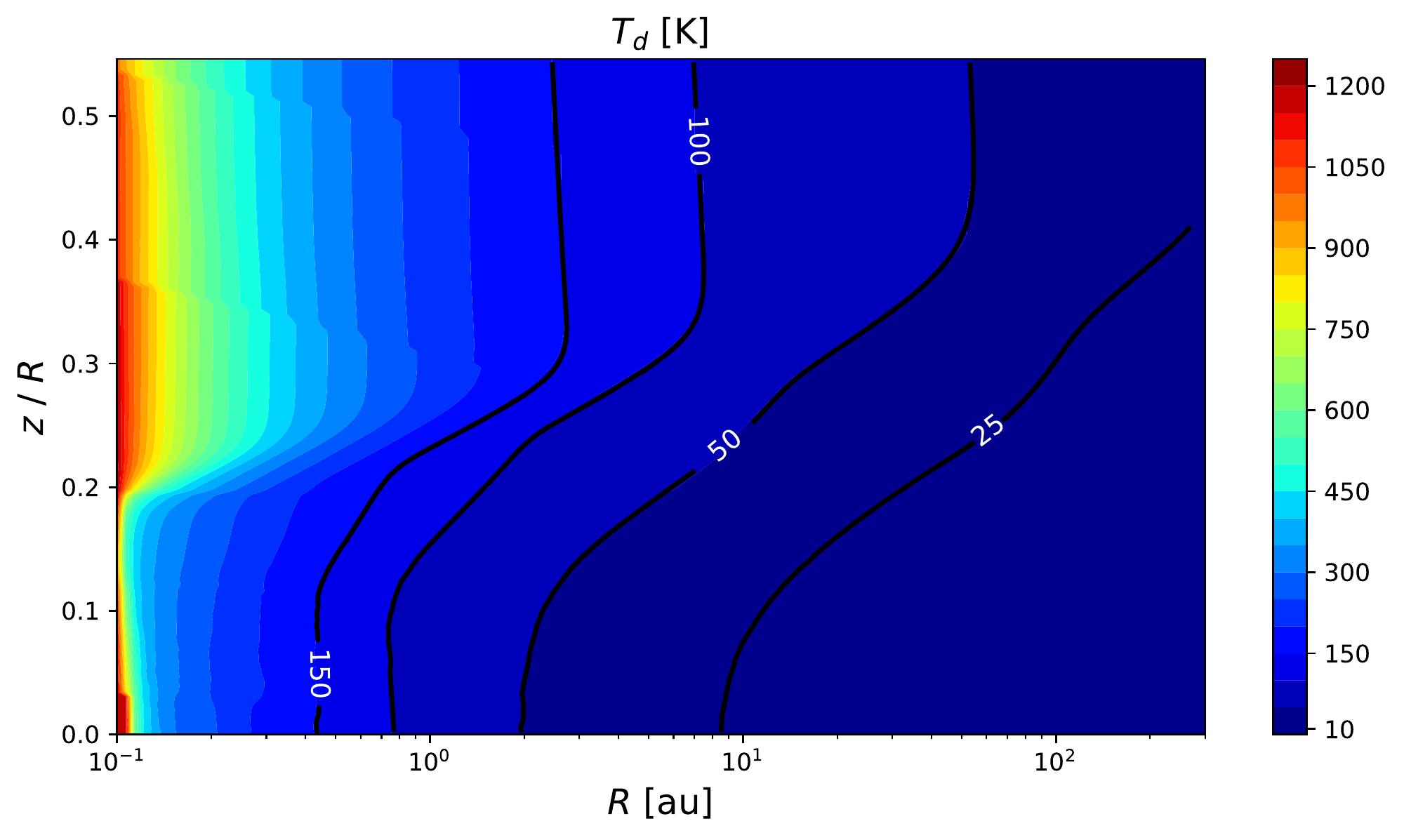}
\caption{Gas density (left panel) and grain temperature obtained from \textsc{radmc-3d} (right panel) for a PPD around T-Tauri stars as a function of the disk radius $R$ and disk height scaled by radius $z/R$ with $\Sigma_1 = 1000\g\cm^{-2}$. The black lines show the grain temperatures of $T_{d}=150, 100, 50, 25 \K$ where the line $T_{d}=150\K$ represents the water snowline. Water ice sublimates in the region of $T_{d}>150\K$.}
\label{fig:nH_T_Tauri}
\end{figure*}

\begin{figure*}
\includegraphics[width=0.5\textwidth]{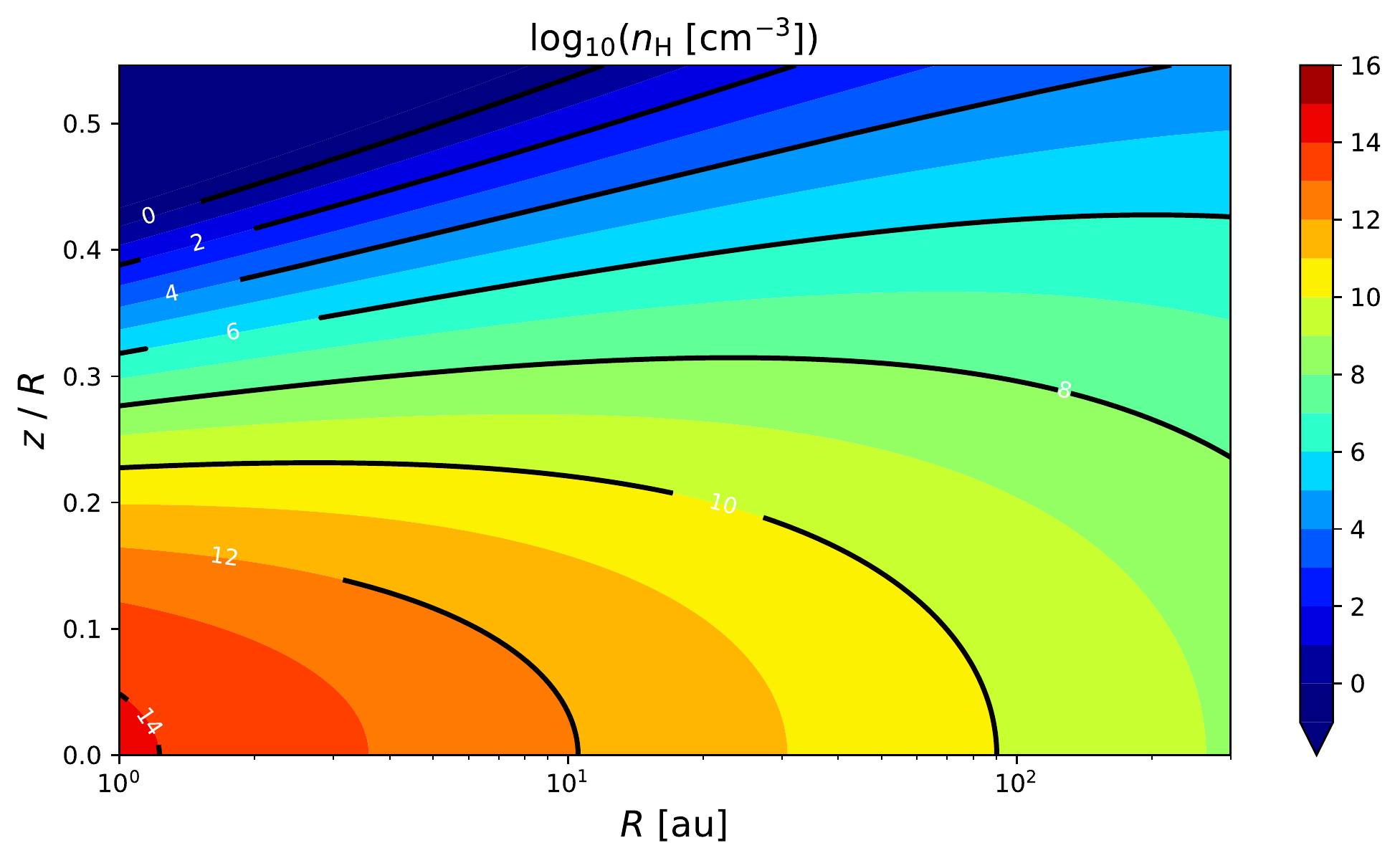}
\includegraphics[width=0.5\textwidth]{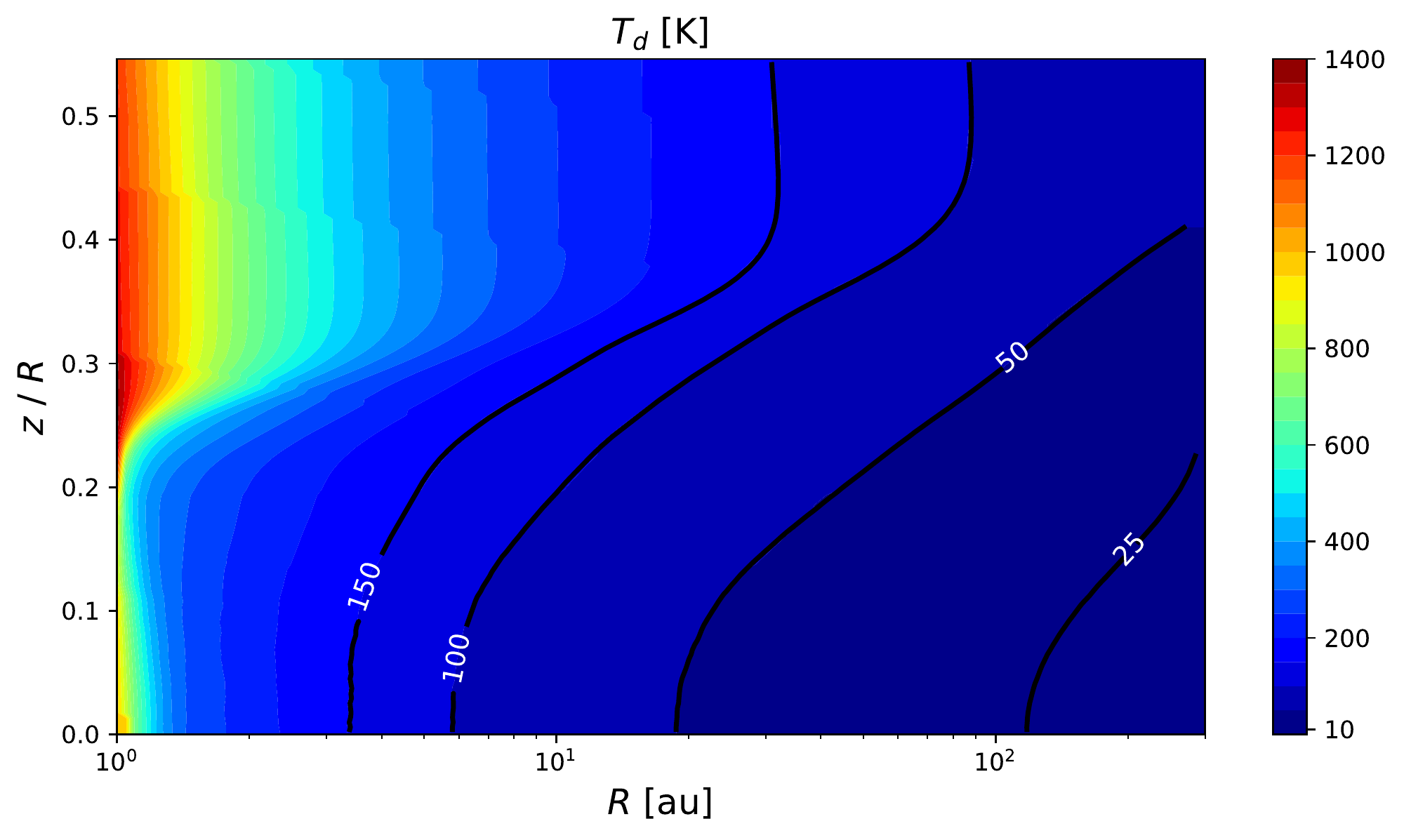}
\caption{Same as Figure \ref{fig:nH_T_Tauri} but for a disk around Herbig Ae/Be stars. Grain temperatures are higher due to the higher central star temperature, and the water snowline is further.}
\label{fig:nH_T_Herbig}
\end{figure*}

\subsection{Grain Temperature and Radiation Field}\label{sec:Td-U}
With the disk model defined, we compute the grain temperature throughout the disk by performing a radiative transfer calculation using the publicly available 3D Monte Carlo radiative transfer code (\textsc{radmc-3d}; \citealt{2012ascl.soft02015D}).\footnote{The code and user guide are available at \url{http://www.ita.uni-heidelberg.de/~dullemond/software/radmc-3d/}.} The grid resolutions are $N_r = 128$ for $R_{\rm in} < R < R_{\rm out}$ and $N_{\theta} = 128$ for $0 < \theta < 2\pi$. The number of photons is $N_{\rm phot} \sim 10^9$. The dust opacity is calculated assuming a power-law size distribution,
$n(a) \propto a^{-q}$ ($a_{\rm min} < a < a_{\rm max}$)
with the absorption cross-section for spherical grains computed using the Mie theory coded from \citet{1983asls.book.....B}, assuming the optical constant of amorphous silicate (Mg$_{0.7}$Fe$_{0.3}$SiO$_{3}$)\footnote{\url{http://www.astro.uni-jena.de/Laboratory/OCDB/data/silicate/amorph/pyrmg70.lnk}}. We adopt the lower and upper cutoffs of the size distribution of big grains $a_{\rm min} = 0.1\mum$ and $a_{\rm max} = 10\mum$, and the standard MRN distribution $q = 3.5$ \citep{1983A&A...128..212M}. The dust has a total mass ranging from $2\times10^{-6}$ to $2\times10^{-3} M_{\odot}$ for $\Sigma_1=1-1000\g\cm^{-2}$ and the same vertical structure as the gas.

Figures \ref{fig:nH_T_Tauri} and \ref{fig:nH_T_Herbig} show the gas density and grain temperature from \textsc{radmc-3d} for the disk around T-Tauri and Herbig Ae/Be stars, respectively, with $\Sigma_1 = 1000\g\cm^{-2}$.

The mean wavelength of the radiation field of a young star can be calculated by
\bea
\bar{\lambda} = \frac{\int \lambda u_{\lambda}(T_{\star}) d\lambda}{\int u_{\lambda}(T_{\star}) d\lambda}, \label{eq:lamb}
\ena
where the radiation intensity $u_{\lambda} (T_{\star})$ is described by the Planck function. This yields $\overline{\lambda} = 1.33\mum$ for the T-Tauri disks and $\overline{\lambda} = 0.53\mum$ for the Herbig Ae/Be disks. The radiation energy density is $u_{\rm rad}=\int u_{\lambda} d\lambda$.

The strength of the radiation field in the disk is characterized by a dimensionless parameter $U=u_{\rm rad}/u_{\rm ISRF}$ where $u_{\rm ISRF}=8.64\times 10^{-13}\erg\cm^{-3}$ is the energy density of the average interstellar radiation field (ISRF) in the solar neighborhood from \citet{1983A&A...128..212M}. The local value of $U$ in the disk can be approximately calculated using the grain temperature obtained from \textsc{radmc-3d} as follows:
\bea
U \simeq \left(\frac{a}{0.1\mum}\right)^{6/15} \left(\frac{T_{d}}{16.4\K}\right)^6, \label{eq:Urad}
\ena
for silicate grains (see \citealt{2011piim.book.....D}).

\section{Rotational disruption of dust and ice by radiative torques}\label{sec:RATD}
\subsection{Grain suprathermal rotation by RATs}
When exposed to an anisotropic radiation field, a dust grain of irregular shape experiences RATs (\citealt{1976Ap&SS..43..291D}) that can spin-up it to suprathermal rotation (\citealt{1996ApJ...470..551D}; \citealt{2009ApJ...695.1457H}). Numerical calculations reveal that RATs are weakly dependent on the dust composition (\citealt{2007MNRAS.378..910L}; \citealt{2019ApJ...878...96H}), thus, the following discussion is applied to both carbonaceous, silicate grains or carbon-silicate mixture.

Let $a$ be the effective grain size, which is defined as the radius of the equivalent sphere of the same volume. Following \citet{2019ApJ...876...13H}, the maximum rotation rate of dust grains spun-up by a radiation field of anisotropy $\gamma$, mean wavelength $\bar{\lambda}$, and radiation strength $U$ is given by
\bea
\omega_{\rm RAT}&\simeq&9.6\times 10^{8}\gamma a_{-5}^{0.7}\bar{\lambda}_{0.5}^{-1.7}\nonumber\\
&&\times\left(\frac{U}{n_{1}T_{2}^{1/2}}\right)\left(\frac{1}{1+F_{\rm IR}}\right)\rad\s^{-1},~~~\label{eq:omega_RAT1}
\ena
for grains with $a\lesssim \bar{\lambda}/1.8$, and
\bea
\omega_{\rm RAT}&\simeq &1.78\times 10^{10}\gamma a_{-5}^{-2}\bar{\lambda}_{0.5}\nonumber\\
&&\times \left(\frac{U}{n_{1}T_{2}^{1/2}}\right)\left(\frac{1}{1+F_{\rm IR}}\right)\rad\s^{-1},~~~\label{eq:omega_RAT2}
\ena
for grains with $a> \overline{\lambda}/1.8$. Here, $n_{1}=n_{\H}/(10\cm^{-3})$, $T_2=T_{\rm gas}/(100\K)$, $\bar{\lambda}_{0.5}=\bar{\lambda}/(0.5\mum)$, $a_{-5} = a/(10^{-5}\cm)$ and $F_{\rm IR}$ is the dimensionless parameter describing the grain rotational damping by infrared emission that depends on $n_{\rm H}$, $T_{\rm gas}$, and $U$ (\citealt{1998ApJ...508..157D}; \citealt{2010ApJ...715.1462H}). The rotation rate depends on the parameter $U/n_{\H}T_{\gas}^{1/2}$ and the damping by far-infrared emission $F_{\rm IR}$. 

For convenience, let $a_{\rm trans}=\bar{\lambda}/1.8$ which denotes the grain size at which the RAT efficiency changes between the power law and flat stages (see e.g., \citealt{2007MNRAS.378..910L}; \citealt{{Hoang:2019da}}), and $\omega_{\rm RAT}$ changes from Equation (\ref{eq:omega_RAT1}) to (\ref{eq:omega_RAT2}).

\subsection{Rotational disruption of composite grains}
The centrifugal force applied to a spinning grain induces centrifugal stress $S=\rho \omega^{2}a^{2}/4$ with $\rho$ being the mass density of the grain, which tends to tear the grain apart. When the centrifugal stress exceeds the maximum tensile strength of the grain material, the grain is instantaneously disrupted into small fragments. This mechanism of dust destruction, developed by \citet{{Hoang:2019da}}, is termed Radiative Torque Disruption (RATD).

The critical rotation rate required to disrupt the composite grain is described by
\bea
\omega_{\rm disr}&=&\frac{2}{a}\left(\frac{S_{\max}}{\rho} \right)^{1/2}\nonumber\\
&\simeq& \frac{3.6\times 10^{8}}{a_{-5}}S_{\max,7}^{1/2}\hat{\rho}^{-1/2}~\rad\s^{-1},\label{eq:omega_disr}
\ena
where $S_{\max,7}=S_{\max}/(10^{7} \erg \cm^{-3}$) with $S_{\max}$ being the maximum tensile strength of the grain and $\hat{\rho}=\rho/(3\g\cm^{-3})$ \citep{Hoang:2019da}.

The value of $S_{\max}$ depends on the composition, internal structures, and grain size. In PPDs, dust particles can have a large range of sizes, from nanoparticles to mm/cm sized pebbles owing to the effect of grain coagulation and growth. To model RATD in the surface and intermediate layers of PPDs, we consider the size dependence of the tensile strength. Grains of size $a>0.1\mum$ can be considered as porous/composite. For these grains, we adopt a composite grain model as proposed by \citet{1989ApJ...341..808M}, in which individual particles of silicate or carbonaceous materials are assumed to be compact and spherical of radius $a_{p}$.

Following \citet{1995A&A...295L..35G}, the maximum tensile strength of the grain is given by
\bea
S_{\max}=3\beta(1-P)\frac{\bar{E}}{2ha_{p}^{2}},\label{eq:Smax_comp1}
\ena
where $\bar{E}=\alpha10^{-3} \eV$ is the mean intermolecular interaction energy at the contact surface between two particles, $h$ is the mean intermolecular distance, $P$ is the porosity of the grain, and $\beta$ is the mean number of contact points per particle between 1-10. For our estimates, we fix the porosity $P=0.2$, as previously assumed for Planck data modeling \citep{2018A&A...610A..16G}, and adopt the typical values $\alpha=1$, $\beta=5$, $h=0.3$nm. Equation (\ref{eq:Smax_comp1}) yields $S_{\max} = 3.2\times10^5 \erg \cm^{-3}$ for $a_{p}=10$ nm.

Grains of $a \leq 0.05\mum$ are likely compact and have higher $S_{\rm max}$ than that of composite grains. We take the typical value $S_{\max} = 10^9 \erg \cm^{-3}$ for these grains. For grain sizes in between ($0.1\mum \geq a > 0.05\mum$), we assume an intermediate value of $S_{\max} = 10^7 \erg \cm^{-3}$.

Grain disruption sizes can be computed by solving the equation $\omega_{\rm RAT}=\omega_{\rm disr}$. From Equation (\ref{eq:omega_disr}), one can see that $\omega_{\rm disr}$ decreases as $\propto a^{-1}$, whereas Equations (\ref{eq:omega_RAT1}) and (\ref{eq:omega_RAT2}) imply that $\omega_{\rm RAT}$ first increases with $a$ until $a = a_{\rm trans}$ and then decreases more rapidly than $\omega_{\rm disr}$ as $a$ goes beyond $a_{\rm trans}$. Hence, one can expect there are two intersections of $\omega_{\rm RAT}$ and $\omega_{\rm disr}$, one for $a<a_{\rm trans}$ and one for $a>a_{\rm trans}$. Upon reaching the critical size determined by the first intersection of $\omega_{\rm RAT}$ and $\omega_{\rm disr}$, grains are disrupted. This value can be obtained from Equations (\ref{eq:omega_RAT1}) and (\ref{eq:omega_disr}), as given by 
\bea
a_{\rm disr}&\simeq&0.06\gamma^{-1/1.7}\bar{\lambda}_
{0.5}(S_{\max,7}/\hat{\rho})^{1/3.4}\nonumber\\
&&\times (1+F_{\rm IR})^{1/1.7}\left(\frac{n_{1}T_{2}^{1/2}}{U}\right)^{1/1.7}\mum,~~~\label{eq:adisr_low}
\ena
for $a \lesssim a_{\rm trans}$. The maximum size of grains that can still be disrupted is determined by the second intersection and can be estimated from Equations (\ref{eq:omega_RAT2}) and (\ref{eq:omega_disr}), as given by
\bea
a_{\rm disr,max}&\simeq&4.9\gamma\bar{\lambda}_{0.5}\hat{\rho}^{1/2}S_{\max,7}^{-1/2}\nonumber\\
&&\times \left(\frac{1}{1+F_{\rm IR}}\right)\left(\frac{U}{n_{1}{T_{2}}^{1/2}}\right)\mum.~~~\label{eq:adisr_up}
\ena
which depends on the local gas density and temperature due to gas damping. Under the effect of rotational disruption, all grains within the size range $a_{\rm disr} \le a \le a_{\rm disr,max}$ would be disrupted.

In the absence of rotational damping, the time required to spin-up the grains of size $a_{\rm disr}$ to $\omega_{\rm disr}$ defines the disruption time:
\bea
t_{\rm disr}&=&\frac{I\omega_{\rm disr}}{dJ/dt}=\frac{I\omega_{\rm disr}}{\Gamma_{\rm RAT}}\nonumber\\
&\simeq& \left(\gamma U_{5}\right)^{-1}\bar{\lambda}_
{0.5}^{1.7} \hat{\rho}^{1/2} S_{\rm max,7}^{1/2}\left(\frac{a_{\rm disr}}{0.1~\mu\rm m}\right)^{-0.7} {\rm yr},\label{eq:tdisr}
\ena
where $U_{5}=U/10^5$.

When rotational damping is present, solving the equation $\omega(t)=\omega_{\rm disr}$ yields the disruption timescale
\bea
t_{\rm disr}&=&-\tau_{\rm damp}\ln \left(1-\frac{\omega_{\rm disr}}{\omega_{\rm RAT}}\right)\nonumber\\
&=&-\tau_{\rm damp}\ln \left(1-\frac{t_{\rm disr,0}}{\tau_{\rm damp}}\right),\label{eq:tdisr_exact}
\ena
which is applicable for $a_{\rm disr,max}>a>a_{\rm disr}$. Note that, for $a=[a_{\rm disr}, a_{\rm disr,max}]$, $t_{\rm disr}\rightarrow \infty$ since it takes $t\gg t_{\rm damp}$ to reach $\omega=\omega_{\rm RAT}$. In strong radiation fields, when $t_{\rm disr,0}\ll \tau_{\rm damp}$, $t_{\rm disr}$ returns to $t_{\rm disr,0}$.

\subsection{Rotational desorption of ice mantles}\label{sec:ro-desp}

\begin{figure}
\includegraphics[width=0.5\textwidth]{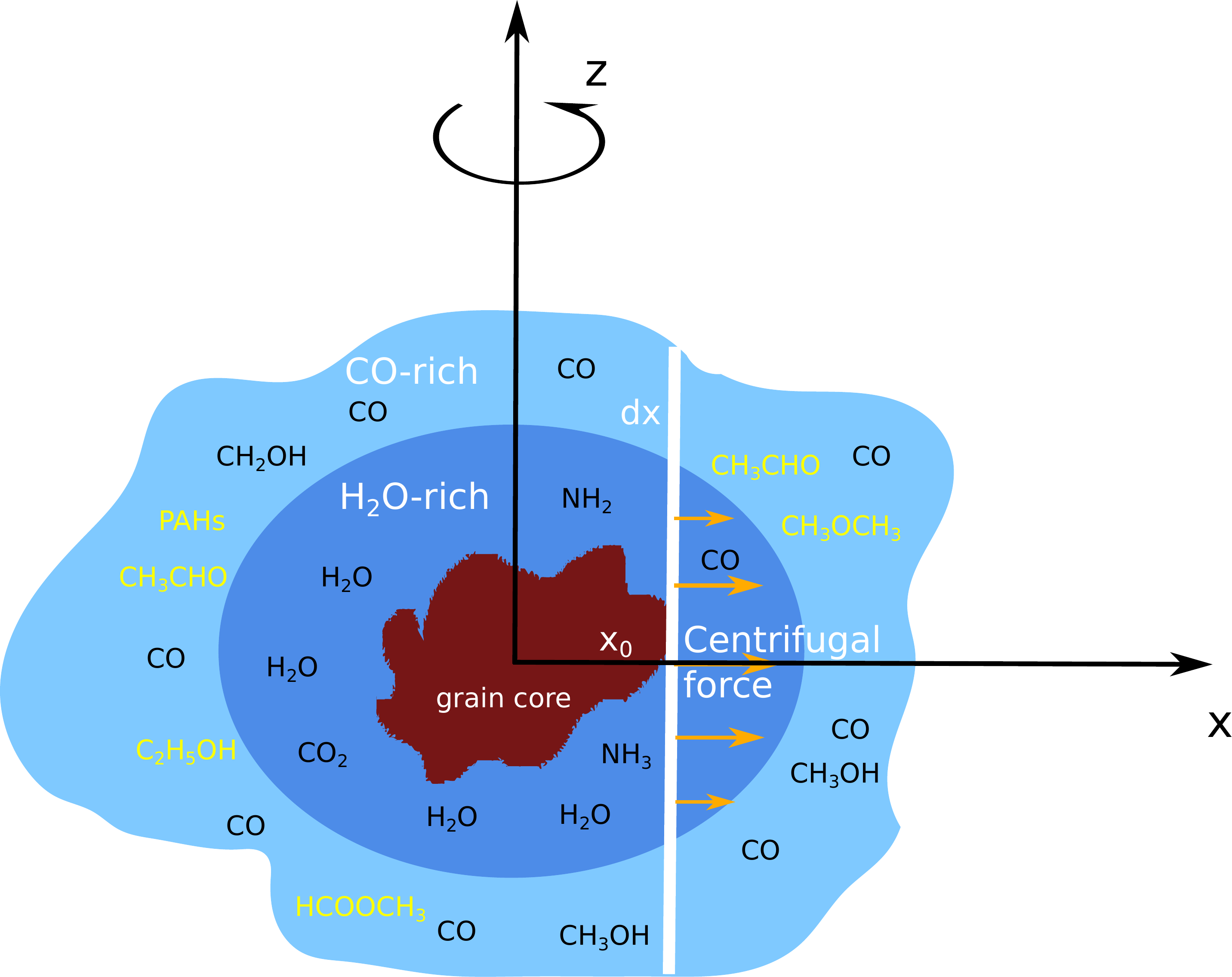}
\caption{Schematic illustration of a rapidly spinning core-ice mantle grain of irregular shape, comprising a water rich (dark blue) and a CO-rich (light blue) mantle layer. Complex molecules and PAHs are also present in the ice mantle.}
\label{fig:core-icemantle}
\end{figure}

The rotational desorption mechanism of icy grain mantles is first studied by \cite{Hoang:2019td}, who found that the centrifugal stress induced by grain suprathermal rotation can desorb the entire ice mantle into tiny icy fragments, provided that the grain core is compact (see Figure \ref{fig:core-icemantle}).

The grain model considered in \cite{Hoang:2019td} is the one made of an amorphous silicate/carbonaceous core covered by a thick double-layer ice mantle \citep{2010ApJ...716..825O}, where complex organic molecules are formed, as illustrated in Figure \ref{fig:core-icemantle}. It is worth noting that the formation of ice mantles takes place on the grain surface due to accretion of gas molecules in cold and dense regions of hydrogen density $n_{\H}=n(\H)+2n(\H_{2}) \sim 10^{3}-10^{5}\cm^{-3}$ or the visual extinction $A_{V}> 3$ \citep{1983Natur.303..218W}. Although icy grain mantles have non-spherical shape as demonstrated by dust polarization, let us assume that the grain shape can be described by an equivalent sphere of the same volume with effective radius $a$.

As discovered in \citet{Hoang:2019td}, the centrifugal stress applied on a spinning core-ice mantle grain tends to pull off the ice mantle from the grain core at sufficiently fast rotation. Upon reaching the critical disruption limit $\omega_{\rm disr}$, the centrifugal stress is equal to the tensile strength of the mantle onto the grain core, which causes the ice mantle near the equator to detach. When the rotation rate increases beyond $\omega_{\rm disr}$ such that the centrifugal stress exceeds the ice tensile strength that holds different parts of the mantle together, the mantle is disrupted into small fragments.

Let $x_{0}$ be the distance from the interface between core-mantle to the spinning axis. From Equation (7) in \citet{Hoang:2019td}, one obtains the centrifugal stress on the ice mantle:
\bea
S_{x}\simeq 2.5\times 10^{9}\hat{\rho}_{\rm ice}\omega_{10}^{2}a_{-5}^{2} \left[1-\left(\frac{x_{0}}{a}\right)^{2} \right] \erg \cm^{-3},\label{eq:Sx}
\ena
where $\hat{\rho}_{\rm ice}=\rho_{\rm ice}/(1\g\cm^{-3})$ with $\rho_{\rm ice}$ being the mass density of the ice mantle, which is $\rho_{\rm ice}\sim 1\g\cm^{-3}$ for pure ice \citep{Hoang:2019td}, and $\omega_{10}=\omega/(10^{10}\rm rad\s^{-1})$.

The critical rotational velocity for disruption is given by $S_{x}=S_{\rm max}$:
\bea
\omega_{\rm disr}&=&\frac{2}{a(1-x_{0}^{2}/a^{2})^{1/2}}\left(\frac{S_{\max}}{\rho_{\rm ice}} \right)^{1/2}\nonumber\\
&\simeq& \frac{6.3\times 10^{8}}{a_{-5}(1-x_{0}^{2})^{1/2}}\hat{\rho}_{\rm ice}^{-1/2}S_{\max,7}^{1/2}~\rad\s^{-1}.\label{eq:omega_disr1}
\ena

Similar to rotational disruption, one can obtain the desorption sizes of icy grains, as given by
\bea
a_{\rm desp}&\simeq&0.13\gamma^{-1/1.7}\bar{\lambda}_
{0.5}(S_{\max,7}/\hat{\rho}_{\rm ice})^{1/3.4}\nonumber\\
&&\times (1+F_{\rm IR})^{1/1.7}\left(\frac{n_{1}T_{2}^{1/2}}{U}\right)^{1/1.7}\mum,~~~\label{eq:adesp_low}
\ena
for $a\le \overline{\lambda}/1.8$, and the maximum size of icy grains that can still be disrupted by centrifugal stress caused by RATs is described by
\bea
a_{\rm desp,max}&\simeq&2.9\gamma \bar{\lambda}_{0.5}\hat{\rho}_{\rm ice}^{1/2}S_{\max,7}^{-1/2}\nonumber\\
&&\times \left(\frac{1}{1+F_{\rm IR}}\right) \left(\frac{U}{n_{1}T_{2}^{1/2}}\right)~\mum.~~~\label{eq:adesp_up}
\ena

In the absence of rotational damping, the characteristic timescale for rotational desorption can be estimated as:
\bea
t_{\rm desp} \simeq 0.6 \left(\gamma U_{5}\right)^{-1}\bar{\lambda}_
{0.5}^{1.7} \hat{\rho}_{\rm ice}^{1/2} S_{\rm max,7}^{1/2}\left(\frac{a_{\rm desp}}{0.1~\mu\rm m}\right)^{-0.7} {\rm yr}.~~~\label{eq:tdesp}
\ena

\subsection{Ro-thermal desorption of molecules from ice mantles}\label{sec:ro-thermal}

In addition to rotational desorption, water and complex molecules can also be desorbed from icy grain mantles by the so-called \textit{rotational-thermal} or {\it ro-thermal} desorption mechanism. Note that, in star-forming regions, molecules are assumed to be physically adsorbed to the ice mantle via van der Waals force, so-called physisorption. In this mechanism, instead of evaporating from the disrupted icy fragments of the original mantle as studied in the previous subsection, molecules sublimate directly from the intact icy grain mantle with the support of centrifugal force. A detailed description and formulation of the ro-thermal desorption mechanism is presented in \citet{HoangTung}. Here we briefly describe the mechanism to apply it for the disks.

Let $\tau_{\rm sub,0}$ be the desorption rate of absorbed molecules with binding energy $E_{b}$ from a grain at rest ($\omega=0$) which is heated to temperatures $T_{d}$. Following \cite{1972ApJ...174..321W}, one has
\bea
\tau_{\rm sub,0}^{-1}=\nu\exp\left(-\frac{E_{b}}{kT_{d}}\right),
\ena
where $\nu$ is the characteristic vibration frequency of adsorbed molecules in the perpendicular direction to the lattice surface and given by
\bea
\nu =\left(\frac{2N_{s}E_{b}}{\pi^{2}m}\right)^{1/2}
\ena
with $N_{s}$ being the surface density of binding sites \citep{1987ppic.proc..333T}. Typically, $N_{s}\sim 2\times 10^{15}$ site$\cm^{-2}$.

Following \cite{HoangTung}, in the presence of grain suprathermal rotation, the centrifugal force acting on the absorbed molecule of mass $m$ at distance $r\sin\theta$ from the spinning axis is
\bea
{\bf F}_{\rm cen}=m{\bf a}_{\rm cen}=m\omega^{2}(x\xhat+y\yhat),\label{eq:Fcen}
\ena
where the unit vectors ($\xhat,\yhat$) describe the plane perpendicular to the spinning axis, and $\ba_{\rm cen}=-\nabla \phi_{cen}$ is the centrifugal acceleration. Here the centrifugal potential 
\bea
\phi_{cen}=\frac{1}{2}\omega^{2}r^{2}\sin\theta^{2}=\frac{\omega^{2}a^{2}}{3},
\ena
which gives the average centrifugal potential
\bea
\langle \phi_{\rm cen}\rangle=\frac{\omega^{2}a^{2}}{3},\label{eq:phi_cen}
\ena
where $r\equiv a$ and $\langle \sin^{2}\theta\rangle=2/3$ is taken.

As a result, the effective binding energy of the absorbed molecule becomes
\bea
E_{b,rot}=E_{b}-m\langle \phi_{\rm cen}\rangle,\label{eq:Ebind_rot}
\ena

The rate of ro-thermal desorption is then given by
\bea
\tau_{\rm sub,rot}^{-1}=\nu\exp\left(-\frac{E_{b}-m\langle \phi_{\rm cen}\rangle}{kT_{d}}\right),\label{eq:tsub_rot}
\ena
where the second exponential term describes the probability of desorption induced by centrifugal potential.

\section{Numerical results}\label{sec:result}
\subsection{Rotational disruption of dust grains}
We first show the results for disruption of dust grains in the surface and intermediate layers, which is a top-down mechanism to reproduce PAHs/nanoparticles as observed via mid-IR emission.

\subsubsection{Disruption sizes}\label{subsec:adisr}

To compute the disruption sizes of grains by RATD for a disk, we solve numerically the equation $\omega_{\rm RAT}=\omega_{\rm disr}$. We consider the maximum grain size $a_{\rm max} = 10\mum$. We adopt $\gamma=0.7$ for the anisotropy degree of the radiation field, which is appropriate for the surface and intermediate layers that is illuminated mostly by the stellar and interstellar radiation fields. 

\begin{figure*}
\includegraphics[width=0.5\textwidth]{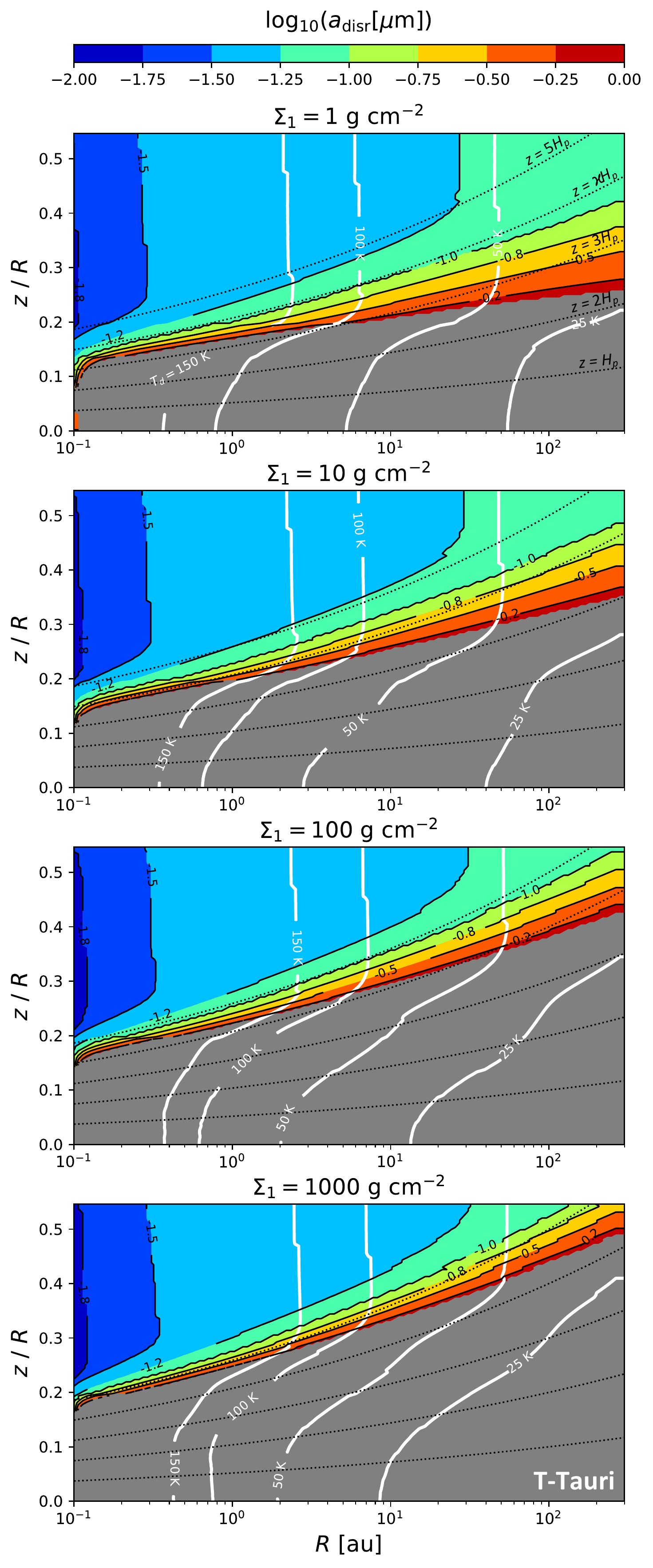}
\includegraphics[width=0.5\textwidth]{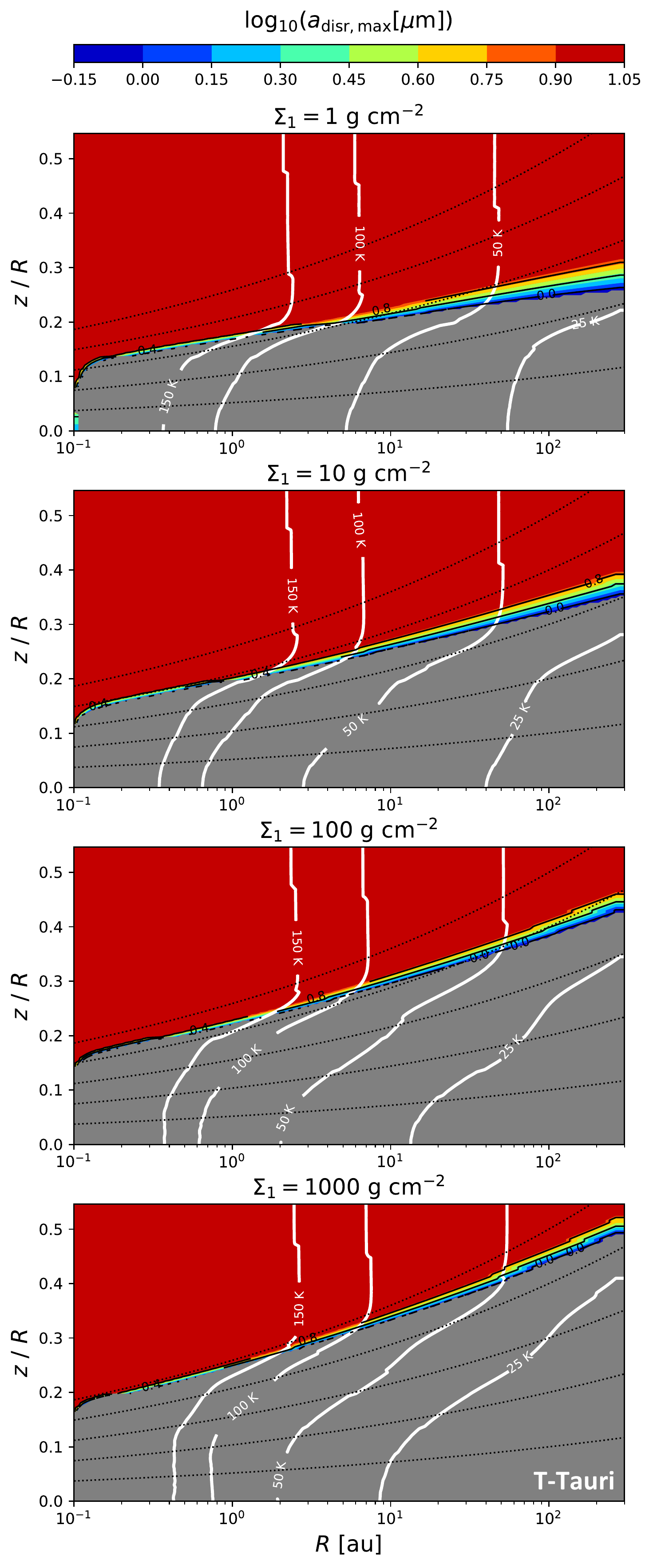}
\caption{Disruption sizes of composite grains for a PPD around T-Tauri stars as functions of $R$ and $z/R$, assuming different values of the surface mass density $\Sigma_{1}=1-10^{3}\g\cm^{-2}$. White lines show the locations of the disk that have $T_d = 150,100,50,25 \K$. Black dotted lines marks the heights $z = 1,2,3,4,5H_{p}$.}
\label{fig:a_disr_TTauri}
\end{figure*}

\begin{figure*}
\includegraphics[width=0.5\textwidth]{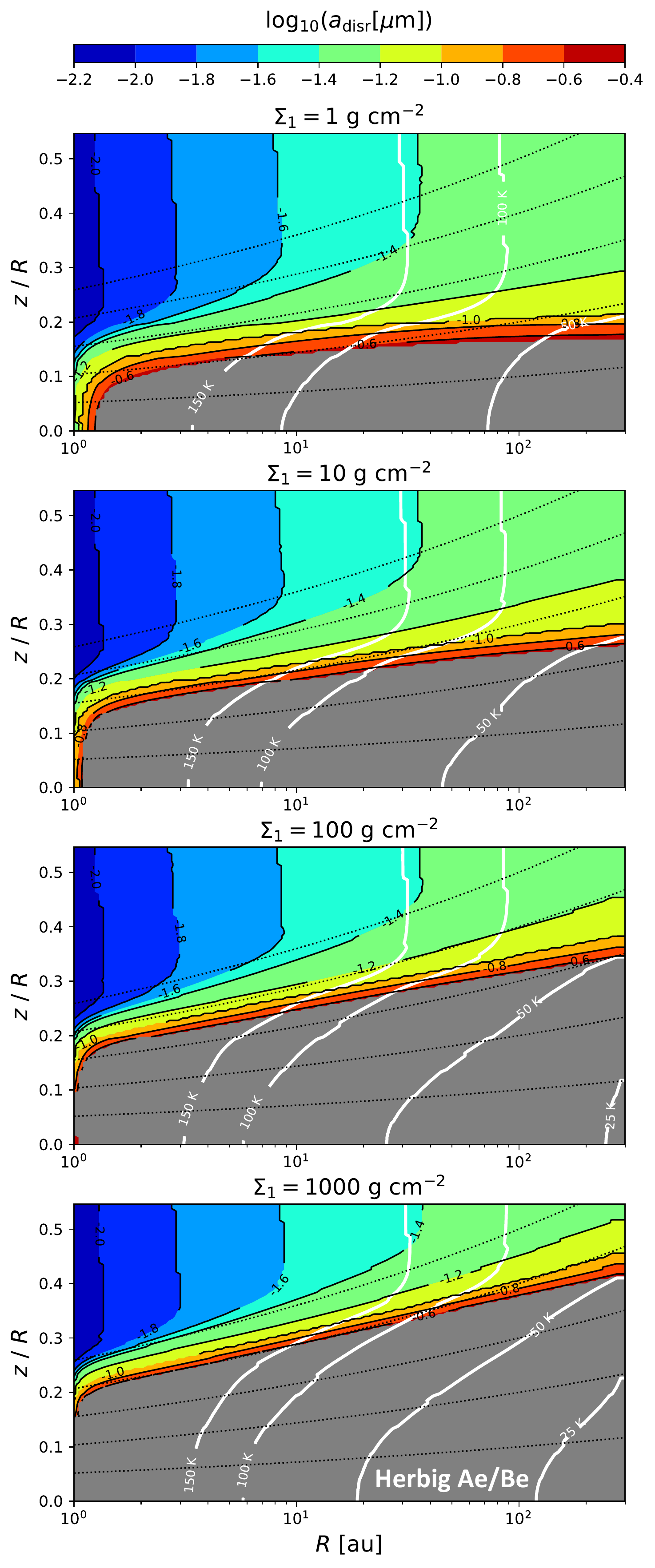}
\includegraphics[width=0.5\textwidth]{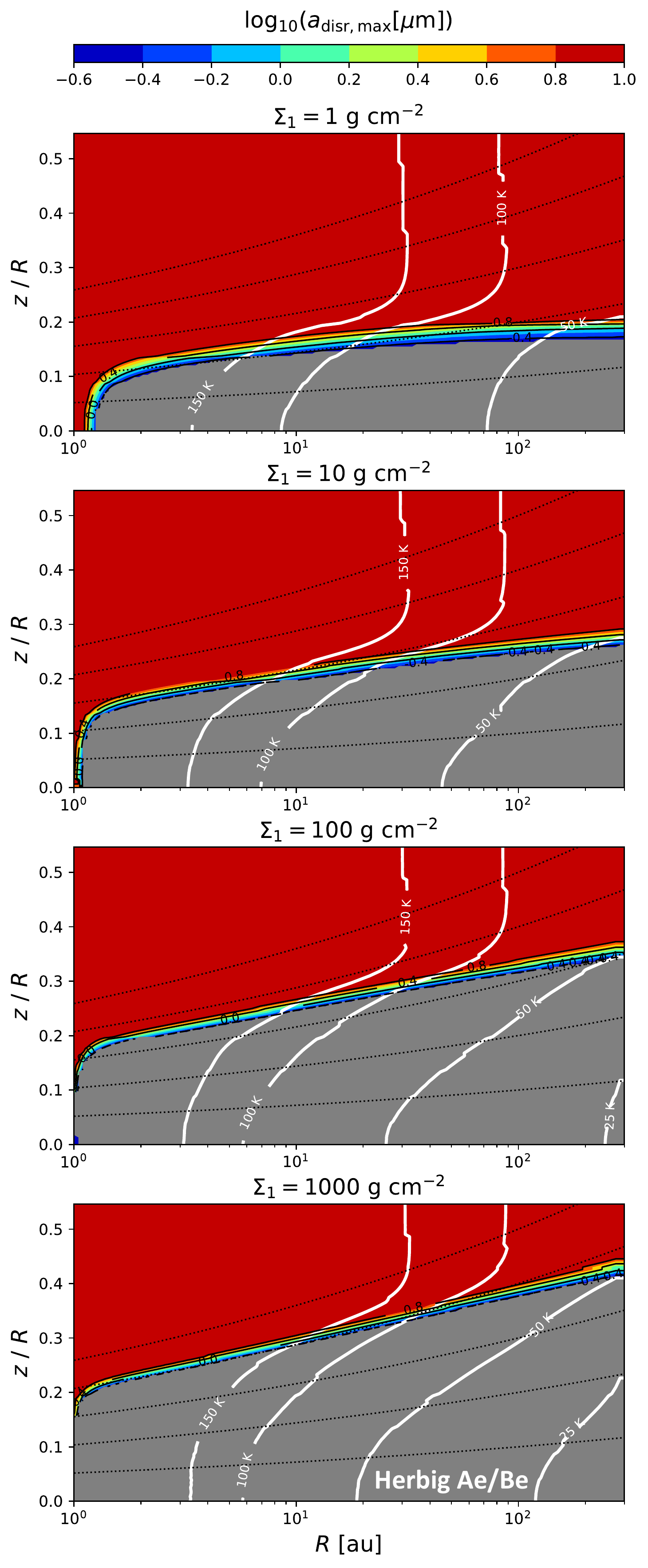}
\caption{Same as Figure \ref{fig:a_disr_TTauri} but for a disk around Herbig Ae/Be stars. Rotational disruption is more efficient because of stronger radiation fields from the central star.}
\label{fig:a_disr_Herbig}
\end{figure*}

Figure \ref{fig:a_disr_TTauri} shows the obtained grain disruption sizes for a disk around T-Tauri stars with different values of $\Sigma_1$ ranging from $1$ to $1000\cm^{-2}$. The gray shaded region indicates that no disruption occurs due to the low radiation intensity and/or high gas density. Depending on the surface mass density of the disks, rotational disruption occurs in both the warm intermediate and surface layers with $T_{d}>30\K$. Lower value of $\Sigma_1$ results in an optically thin disk in which the disruption of grains is more prominent. In the disk surface, where the gas density $n_{\rm H} \lesssim 10^{9} \cm^{-3}$ and grain temperature $T_{d}>150\K$ (see Figure \ref{fig:nH_T_Tauri}) for which grains could be spun-up to extremely fast rotation, rotational disruption is very efficient. The grain disruption size $a_{\rm disr}$ decreases with the disk height $z$ from $a_{\rm trans} \sim 0.74\mum$ to a minimum value $\sim 0.01 \mum$ in the central region ($R \lesssim 0.2\AU$), indicating that all grains with sizes larger than $\sim 10$ nm would be disrupted in this region to form nanoparticles. In the warm layer, rotational disruption is still pronounced in the region where $T_d = 50-150\K$ and $n_{\rm H} \lesssim 10^{6} \cm^{-3}$, and near the upper boundary of the disk height, small grains of $a \gtrsim 0.05\mum$ are already disrupted. For lower grain temperatures and higher gas density (i.e, $T_d = 30-50\K$ and $n_{\rm H} \lesssim 10^{2} \cm^{-3}$ for the case of $\Sigma_1=1000\g\cm^{-2}$), rotational disruption become less efficient and only occurs at the surface layer with $z/R = 0.4-0.5$. Nevertheless, grains of $a \gtrsim 0.1\mum$ are still destroyed under the effect of RATD at the height boundary. The maximum disruption size $a_{\rm disr,max}$ increases rapidly with $z$ and exceeds the threshold $a_{\rm max} = 10 \mum$ in the surface layer, which means only very large grains could survive. Overall, the closer to the central star, the smaller $a_{\rm disr}$ and the greater $a_{\rm disr,max}$, which means that a broader range of grain sizes is disrupted by RATD.

Same as Figure \ref{fig:a_disr_TTauri}, but Figure \ref{fig:a_disr_Herbig} shows the results for the disk around Herbig Ae/Be stars. The active region of rotational disruption is larger due to higher temperatures in the surface layer. At $R \sim 30\AU$, small grains of $a \sim 0.03\mum$ can still be disrupted compared to $R \sim 0.3\AU$ in the case of T-Tauri disks. Beyond that radius, disruption size is $\sim  0.05\mum$ at the top of the scale height and remains relatively small up to the outer radius ($0.01-0.05\mum$). 

\begin{figure*}
\centering
\includegraphics[width=0.45\textwidth]{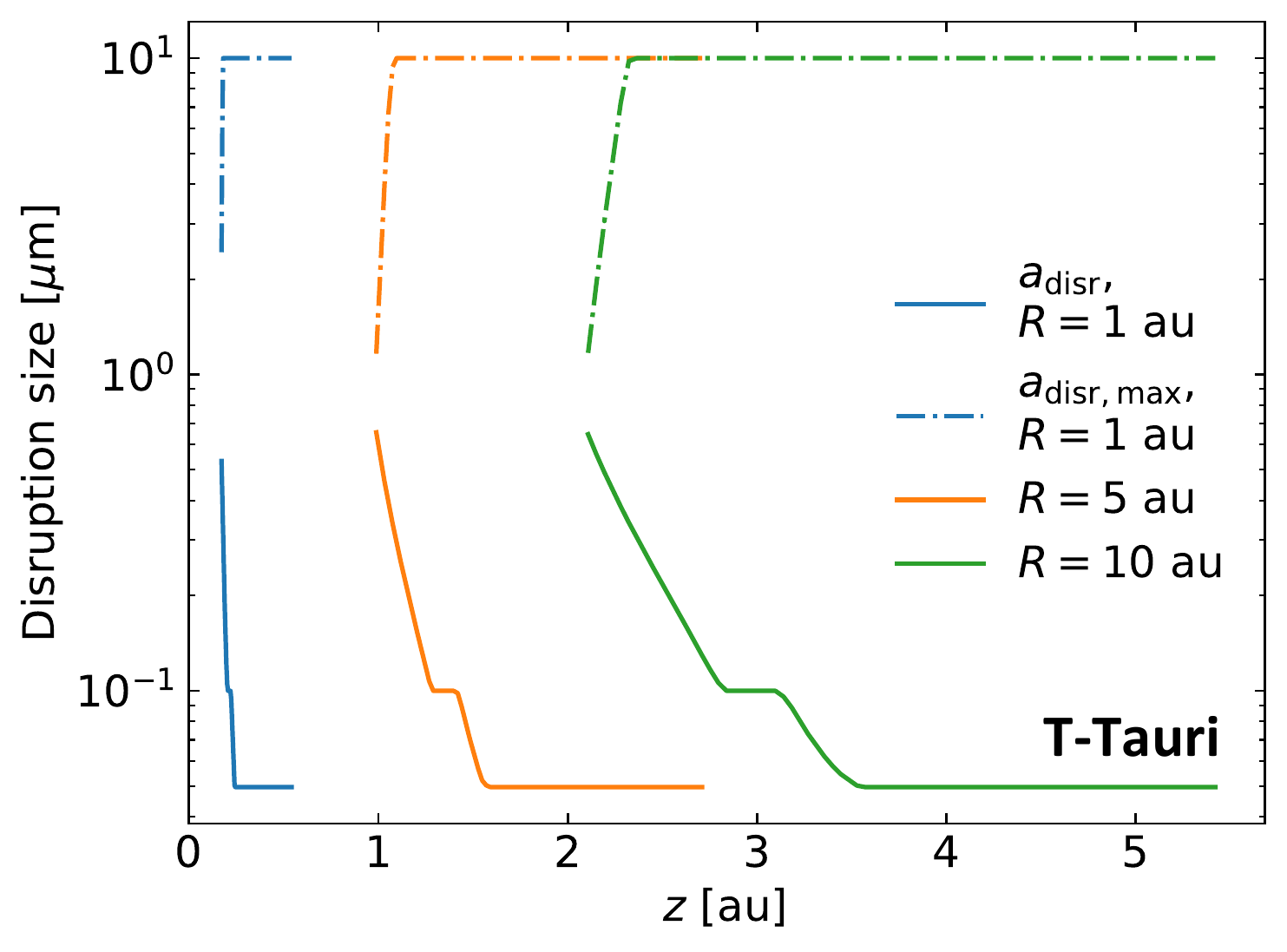}
\includegraphics[width=0.45\textwidth]{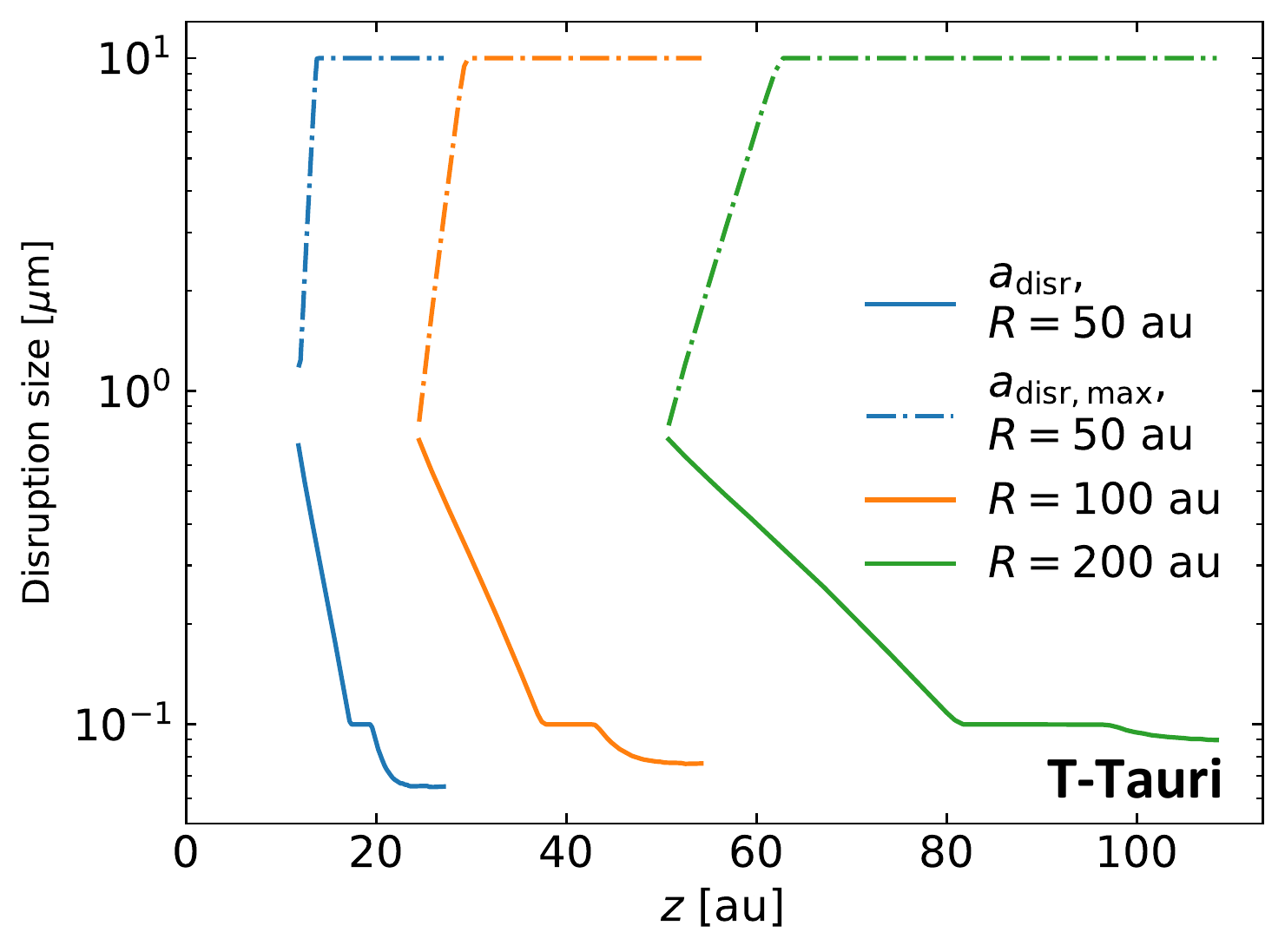}
\includegraphics[width=0.45\textwidth]{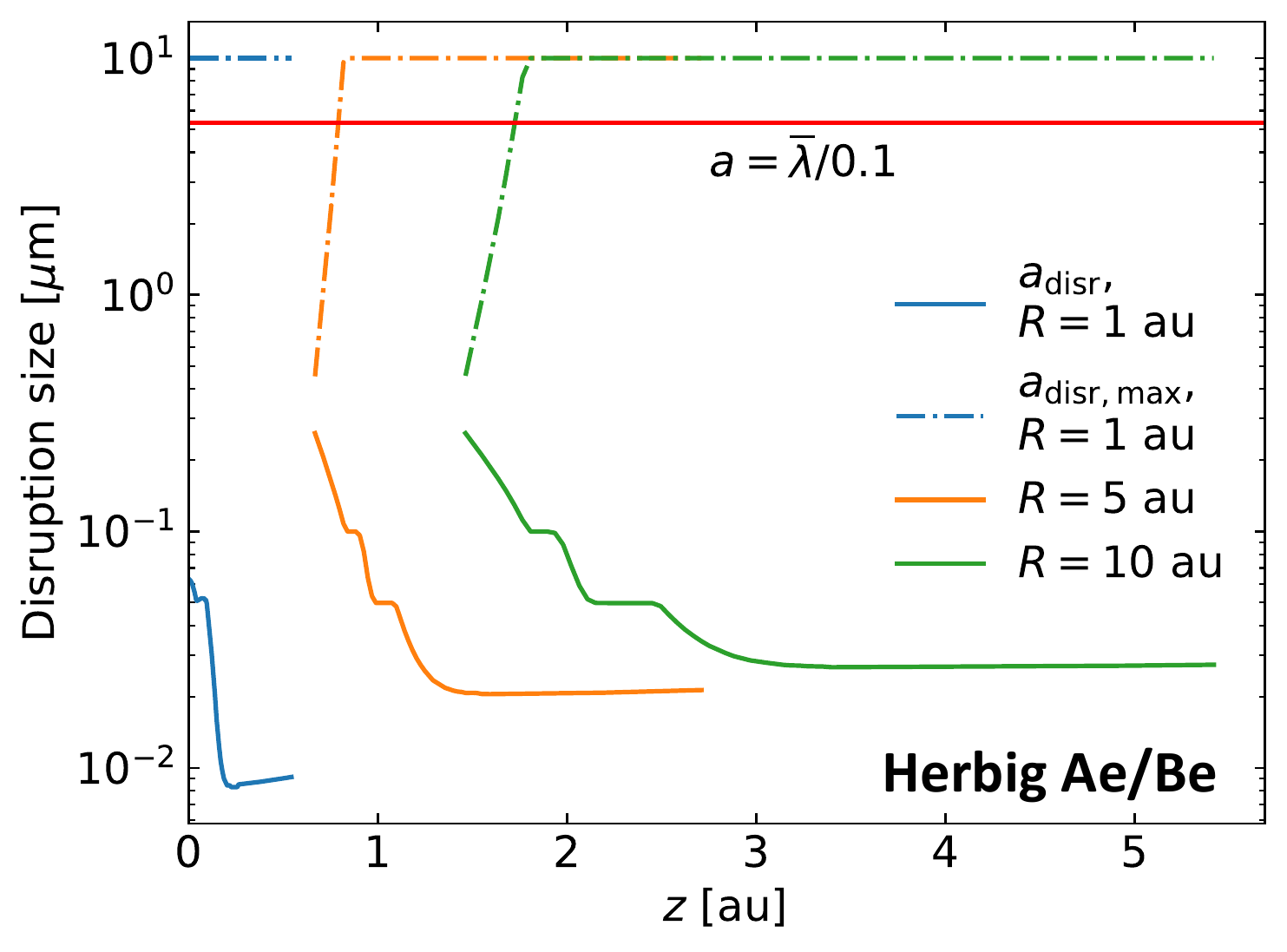}
\includegraphics[width=0.45\textwidth]{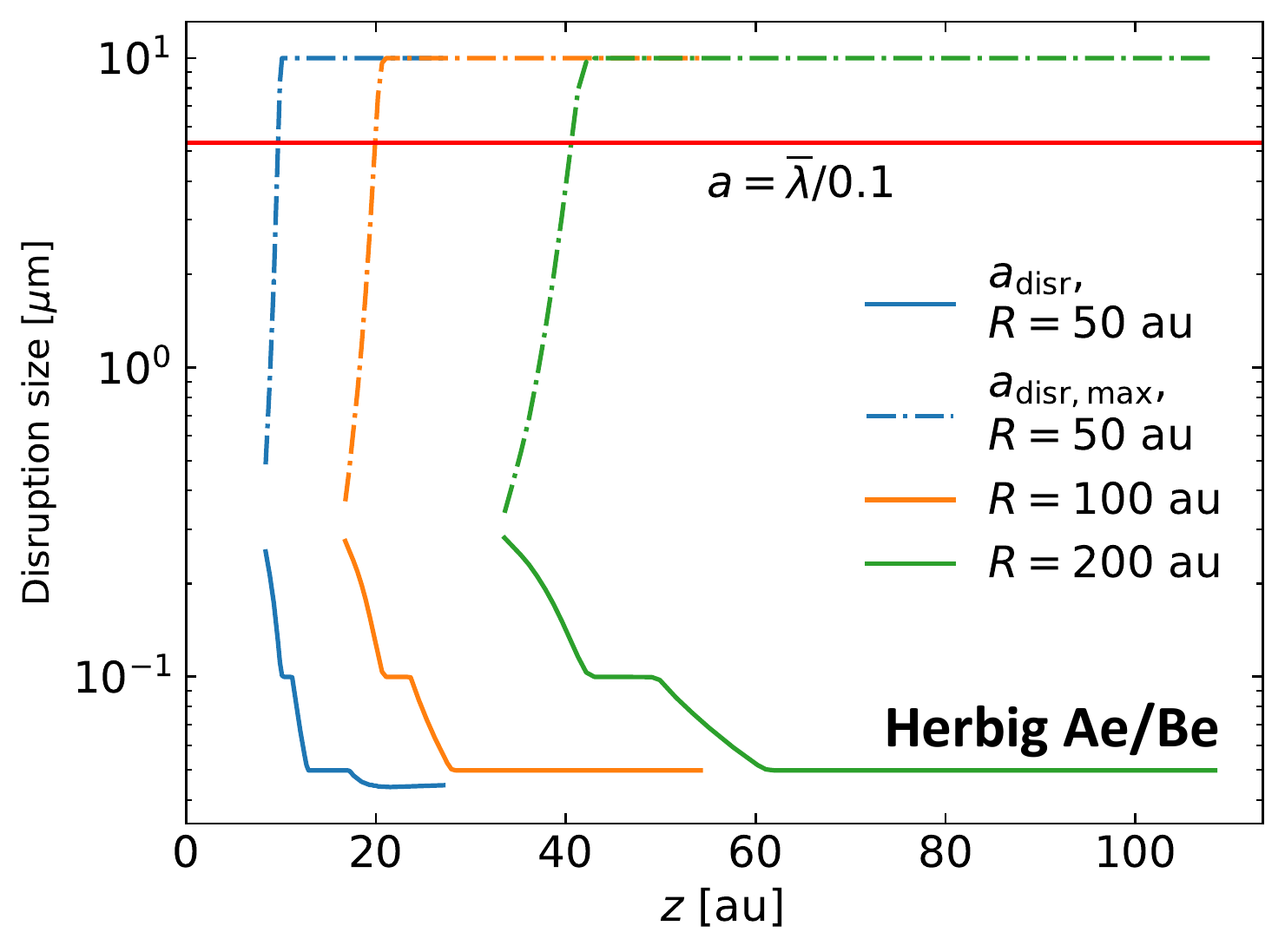}
\caption{Disruption sizes $a_{\rm disr}$ and $a_{\rm disr,max}$ as a function of the disk height $z$ computed for different disk radii $R$, with $\Sigma_1 = 1 \g\cm^{-2}$. Upper and lower panels show the results for T-Tauri disks and Herbig Ae/Be disks, respectively. $a_{\rm disr}$ decreases and $a_{\rm disr,max}$ increases with increasing $z$. The red horizontal line shows $a_{\rm disr,max} = \bar{\lambda}/0.1$.}
\label{fig:adisrvsz}
\end{figure*}

To better understand rotational disruption in the different layers of the disk, in Figure \ref{fig:adisrvsz}, we show the disruption sizes as a function of the disk height $z$ for the different disk radii, with $\Sigma_1 = 1 \g\cm^{-2}$. From Equations (\ref{eq:adisr_low}) and (\ref{eq:adisr_up}), one can see that $a_{\rm disr}$ decreases when $n_{\rm H}$ decreases as well as $T_d$ and $U \sim T_{d}^6$ increases, whereas the opposite tendency is expected for $a_{\rm disr,max}$. As $n_{\rm H}$ exponentially decreases and $T_d$ is significantly greater as one goes higher to the exposed surface layer (see Figures \ref{fig:nH_T_Tauri} and \ref{fig:nH_T_Herbig}), $a_{\rm disr}$ declines while $a_{\rm disr,max}$ increases rapidly with $z$, making the range [$a_{\rm disr}$,$a_{\rm disr,max}$] larger. The discontinuity of the plots for $a_{\rm disr}$ at $a=0.1\mum$ and $a=0.05\mum$ originates from the 3 separate values assumed for $S_{\rm max}$ for 3 ranges of grain sizes. Generally speaking, $a_{\rm disr}$ mostly ceases at $0.05 \mum$ due to the large tensile strength $S_{\rm max} \sim 10^9 \erg\cm^{-3}$ of compact grain cores of $a \leq 0.05\mum$. Only near the central region (e.g., $R=1\AU$) can disruption sizes be smaller than $0.05\mum$. Note that the maximum grain size is set to be $a_{\rm max} = 10\mum$, which accounts for the stop at $10 \mu$m of $a_{\rm disr, max}$. However, such large grains may not experience lower RATs due to incoherent contributions of RATs from the different grain facets (see, e.g., \citealt{2007ApJ...669.1085C}). Since the calculations of RATs for very large grains of size $a > \bar{\lambda}/0.1$ are not yet available due to computing limitations (see e.g., \citealt{2019ApJ...878...96H}), in the case of Herbig disks where $\bar{\lambda}/0.1 = 5.33\mum$, we plot a horizontal line where $a = \bar{\lambda}/0.1$ below which RATs are previously calculated for irregular grains (\citealt{2007MNRAS.378..910L}).

\subsubsection{Rotational disruption time vs. grain-grain collision destruction time}

We now calculate the time it takes for rotational disruption to disrupt the grains $t_{\rm disr}$ in the disks from Equation (\ref{eq:tdisr}), using the numerical results for $a_{\rm disr}$ and $U$ from Section \ref{subsec:adisr}. 

Other possible dust destruction mechanism is grain shattering caused by grain-grain collisions induced by radiation pressure \citep{2017ApJ...847...77H}. The mean time between two collisions defines the destruction time by grain-grain collisions
\bea
\tau_{\rm gg} &=& \frac{1}{\pi a^2 n_{\rm gr} v_{\rm gg}} = \frac{4\rho a M_{\rm g/d}}{3n_{\rm H} m_{\rm H} v_{\rm gg}}\nonumber\\
&\simeq& 2.5\times 10^7\hat{\rho} a_{-5} \left(\frac{30 \cm^{-3}}{n_{\rm H}}\right) \left(\frac{1~ {\rm km} \s^{-1}}{v_{\rm gg}}\right)~\rm yr,~~~
\label{eq:tgg}
\ena
where a single size $a$ distribution with the gas-to-dust mass ratio $M_{\rm g/d}=100$ and the number density of dust grains $n_{\rm gr}$ is assumed, and $v_{\rm gg}$ is the relative velocity of grains. For turbulent effect, one has $v_{\rm gg} =\alpha^{p} c_{\rm s}$ with the turbulence parameters $\alpha=0.01$ and $p=1/2$ (see, e.g. \citealt{2004A&A...421.1075D}), and $c_{\rm s}$ being the gas sound speed. Comparing Equation (\ref{eq:tgg}) with (\ref{eq:tdisr}), one can see that grain-grain collisions require a much longer timescale to produce small grains compared to RATD.

\begin{figure*}
\includegraphics[width=0.5\textwidth]{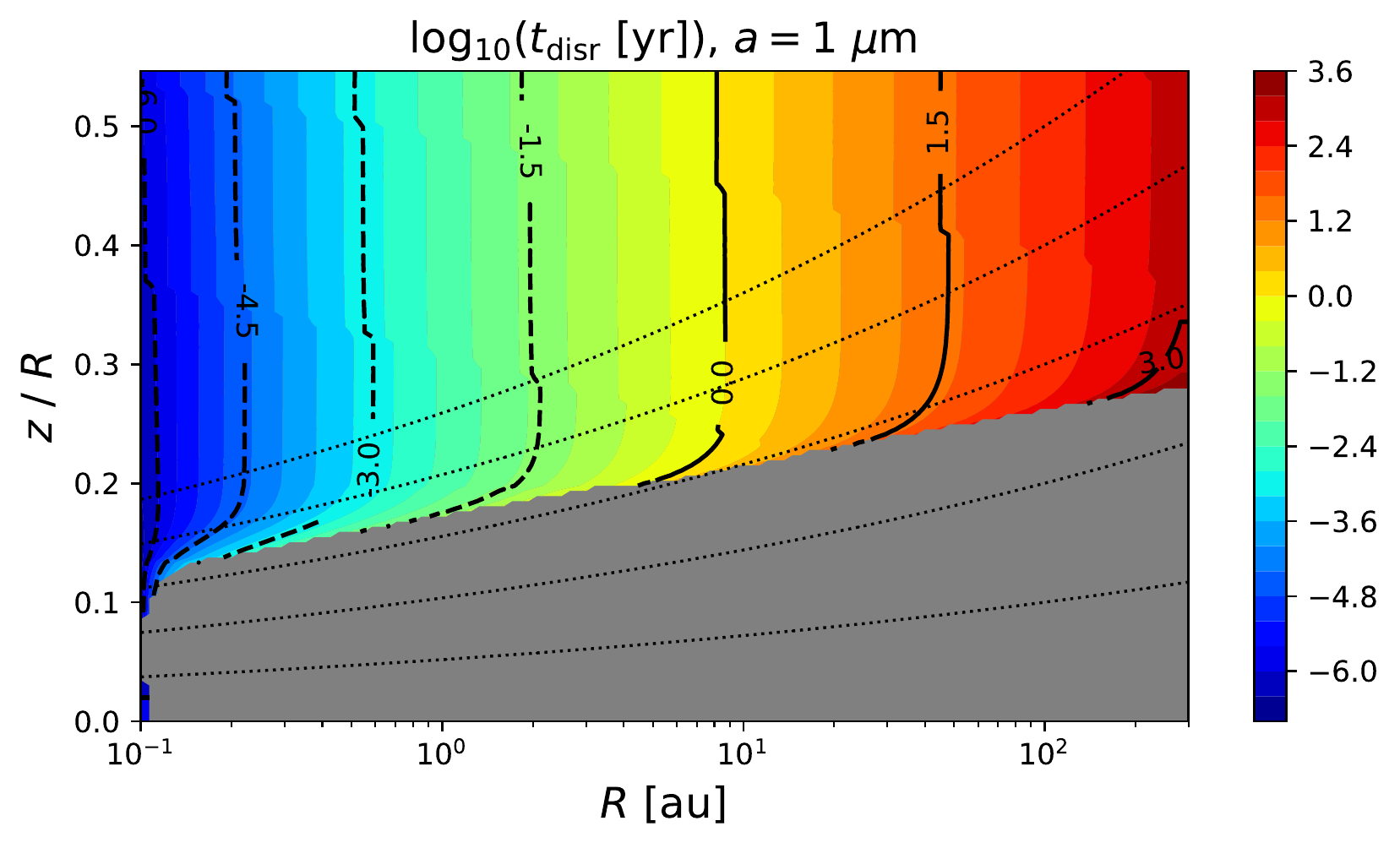}
\includegraphics[width=0.5\textwidth]{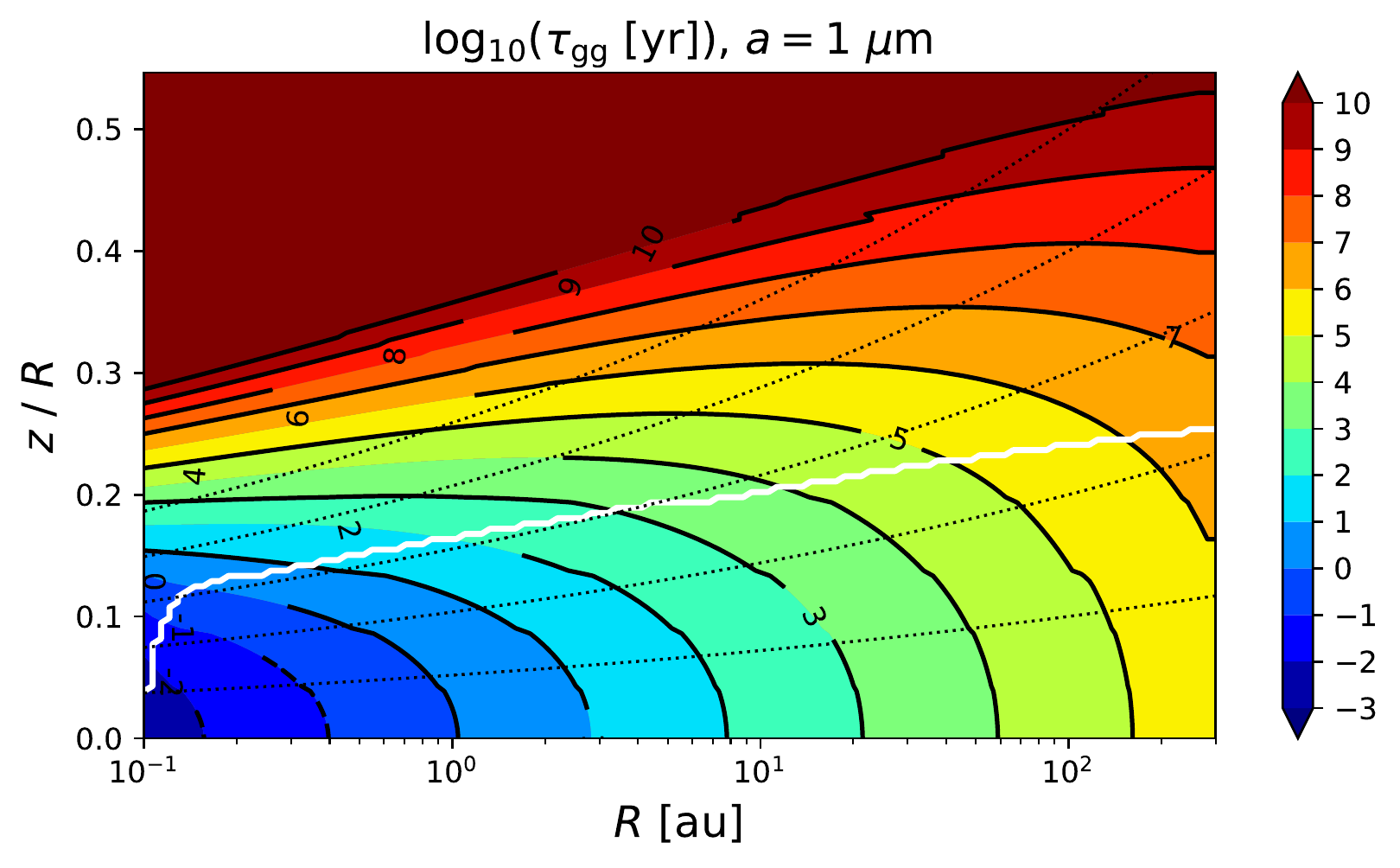}
\caption{Comparison of rotational disruption time $t_{\rm disr}$ (left panel) with shattering time by grain−grain collisions $\tau_{\rm gg}$ (right panel) for grain size $a=1\mum$ and $\Sigma_1 = 1\g\cm^{-2}$. The T-Tauri disk is considered. White solid line marks the boundary above which rotational disruption is effective.}
\label{fig:tdisr}
\end{figure*}

In Figure \ref{fig:tdisr}, we compare the disruption time by RATD (left panel) with the shattering time by grain-grain collisions (right panel) for different grain sizes $a$ for the T-Tauri disks, with $\Sigma_1 = 1 \g\cm^{-2}$. In the surface layer, where the gas and dust density are relatively low, the shattering time is extremely high as $\tau_{\rm gg}$ is most governed by $n_{\rm H}$. In contrast, the disruption time, which depends on disruption sizes, is much lower due to high radiation intensity from the central star that causes grains to disrupt easier and hence, $a_{\rm disr}$ becomes smaller.

It is worth noting that, the higher $\Sigma_1$, the more massive the disk would be. A massive PPD, particularly around a T-Tauri star of $0.5 M_{\odot}$, is likely to be gravitationally unstable. In such case, internal heating from gravitational instability-driven spirals and/or turbulence could dominate stellar radiation, giving rise to a disk model different from the passive irradiate disk adopted in Section \ref{sec:model}. Additionally, the Gaussian vertical profile of the gas described by Equation (\ref{eq:ngas0}) assumes a vertically isothermal temperature structure, whereas the temperature profile from \textsc{radmc-3d} has a vertically stratified structure due to stellar radiation. As a result, the disk is not in hydrostatic equilibrium. However, the disruption time given by Equation (\ref{eq:tdisr}) is smaller than the dynamical timescale, which is defined as shortest timescale on which the disk structure can vary,
\bea
\tau_{D} &=& \frac{1}{\Omega} = \left( \frac{G M_{\star} }{R^3} \right)^{-1/2} \nonumber\\
&\simeq& 159.2 \left(\frac{R}{100\AU}\right)^{3/2}\left(\frac{M_{\star}}{M_{\odot}}\right)^{-1/2} {\rm yr},
\ena
where $\Omega$ is the Keplerian orbital period. The problem of disk instability is well studied in the literature (see, e.g, \citealt{1994ApJ...421..640N}, \citealt{2005A&A...432L..31U}, \citealt{2019ApJ...871..150P}) and it is well established that the thermal and viscous timescales are much larger than $\tau_{D}$. Therefore, we expect the disruption of grains would not be significantly affected in such events.

\subsection{Rotational desorption of ice mantles and the new location of the snowline}\label{subsec:adesp}
Now we move on to show our results for rotational disruption and desorption of ice mantles which is relevant to the warm intermediate layer.

\subsubsection{Desorption sizes of ice mantles}\label{sec:adesp}

We first consider a core-ice mantle grain model with a fixed core radius $a_{\rm c} = 0.05\mum$ and the mantle thickness $\Delta a_{m}$ can vary so that the grain size is $a=a_{c}+\Delta a_{m}$. Similar to disruption sizes, numerical calculations for desorption sizes are performed with conservative values for the tensile strength of the grain core $S_{\rm max,core} = 10^9\erg\cm^{-3}$ and of the ice mantle $S_{\rm max,mantle} = 10^{7}\erg\cm^{-3}$ which is comparable to the tensile strength of bulk ice (see e.g., \citealt{2012JGRE..117.8013L}).

\begin{figure*}
\includegraphics[width=0.5\textwidth]{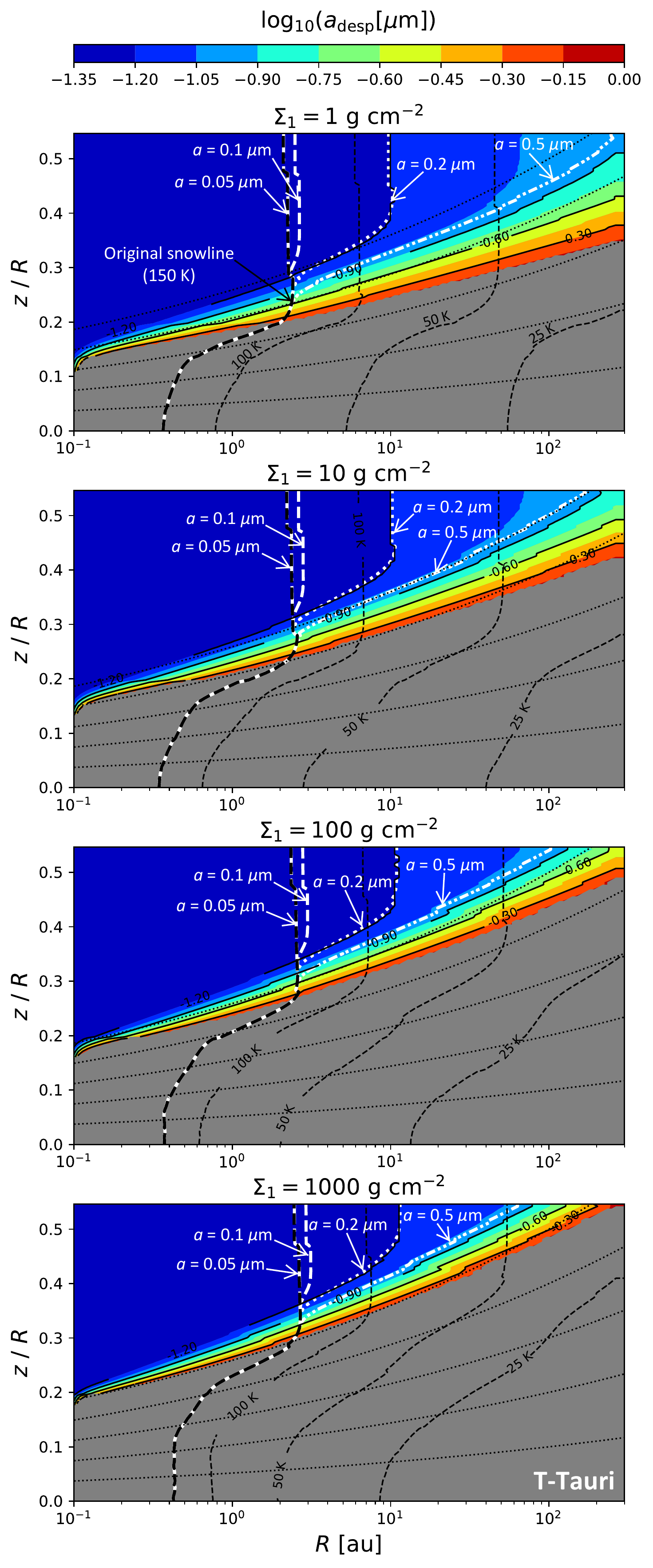}
\includegraphics[width=0.5\textwidth]{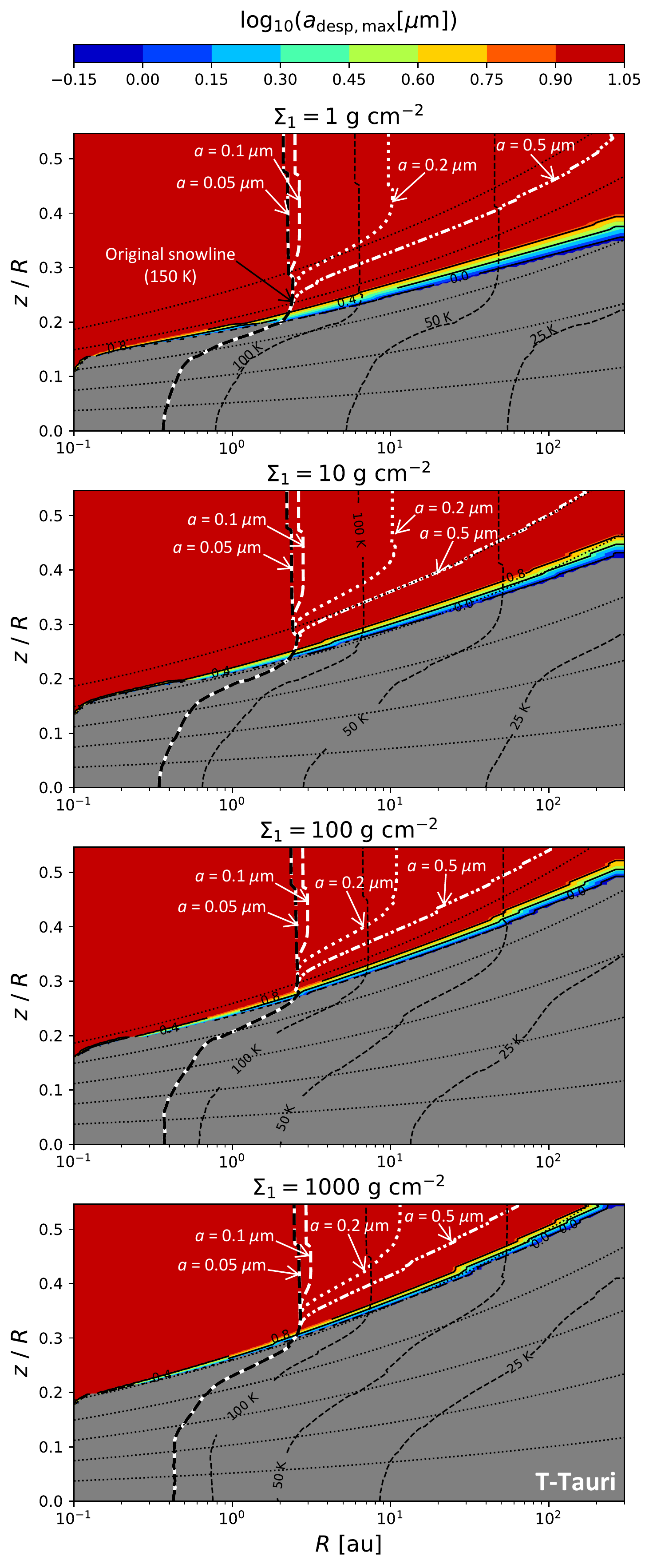}
\caption{Desorption sizes of ice mantles for a PPD around T-Tauri stars as a function of $R$ and $z/R$, assuming a fixed core radius $a_{\rm c} = 0.05\mum$ and the varying mantle thickness. White dashed and dotted lines illustrate the region where $\tau_{\rm sub,rot}^{-1}(T_d) = \tau_{\rm sub,0}^{-1}(T_{\rm sub})$ for different grain sizes $a$, assuming water ice of $T_{\rm sub}=150\K$. Different values of the surface mass density $\Sigma_{1}$ are considered.}
\label{fig:a_desp_TTauri}
\end{figure*}

The results for the T-Tauri disk are plotted in Figure \ref{fig:a_desp_TTauri}. The snowline ($T_{d} = 150\K$) obtained from \textsc{radmc-3d} is shown for comparison. The similar variation features as rotational disruption can be observed here. Rotational desorption of ice mantles also occurs in the warm intermediate and surface layers with $T_d > 30\K$, where grains are spun-up to suprathermal rotation and the ice mantles are disrupted due to centrifugal stress. The desorption size $a_{\rm desp}$ is smaller and $a_{\rm desp,max}$ is greater near the central star because of the higher radiation intensity and lower gas number density. In the surface layer where $T_d > 150\K$, icy grains of sizes larger than the grain core $a_{c}=0.05\mum$ are mostly desorbed. The intermediate layer where $T_d = 30-150\K$ also witnesses great efficiency of rotational desorption, with $a_{\rm desp}$ drastically increasing from $a_{\rm trans} = 0.74\mum$ to $\sim a_{c}=0.05\mum$ as $R$ declines and $z$ increases. Interestingly, rotational desorption takes place in the region beyond the water snowline (around $3 \AU$ from the star) as well. Consequently, the ice mantles of dust grains in this region are already disrupted into smaller fragments, giving rise to the subsequent evaporation into water vapor. As a result, rotational desorption can destroy the ice mantles of grains beyond the snowline and therefore push down the snowline in the vertical direction to the boundary beyond which the ice mantles start to be disrupted under the RATD effect.

Furthermore, to demonstrate the effect of ro-thermal desorption on the release of COMs beyond the water snowline, we plot also in Figure \ref{fig:a_desp_TTauri} the region where $\tau_{\rm sub,rot}^{-1}(T_d) = \tau_{\rm sub,0}^{-1}(T_{\rm sub})$ for different grain sizes $a$, where $T_{\rm sub} = 150 \K$ is taken to be the sublimation temperature and $E_b = 5700 \K$ the binding energy of water ice \citep{2013ApJ...765...60G}. In the surface layer, the snowlines in the cases with and without the rotation of grains become separated due to high $T_d$ and low $n_{\rm H}$ that increases the rotation rate and thus, decreases the binding energy of molecules. As a result, the region that has the desorption rate equal to that of water ice by thermal desorption is extended further from the central star. The effect is more significant for larger grain sizes. Our model of ro-thermal desorption is not limited only to the water snowline. Plugging the analogous parameters for CO molecule, i.e. $E_b = 1150\K$ \citep{2013ApJ...765...60G} and $T_{\rm sub} = 25\K$ \citep{1993prpl.conf.1177M}, one can see that the rotation of grains induced by RAT also increases the desorption rate of CO ice. Thus, the CO iceline under the framework of ro-thermal desorption would also be extended compared to the classical one. It is worth noting that the effect of rotational desorption is more efficient than that of ro-thermal desorption, i.e., rotational desorption can occur at more distant regions.

\begin{figure*}
\includegraphics[width=0.5\textwidth]{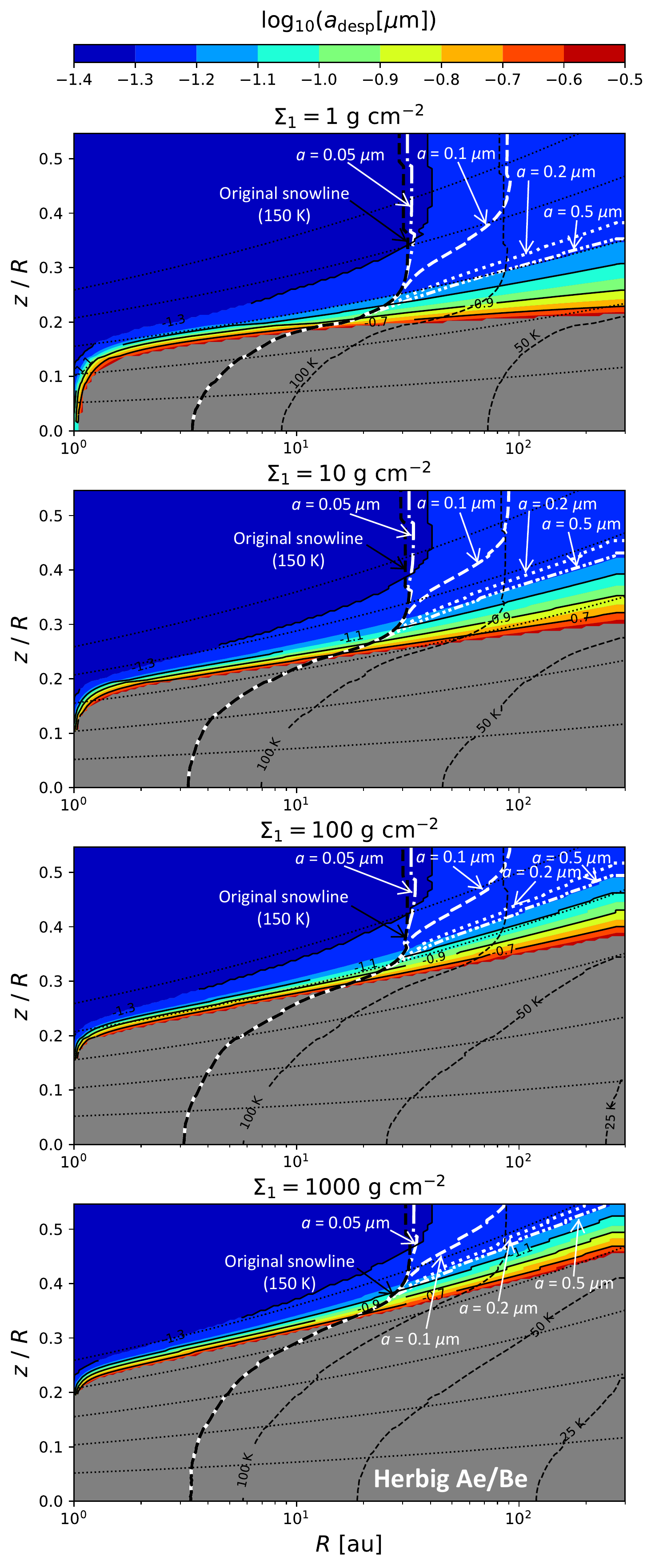}
\includegraphics[width=0.5\textwidth]{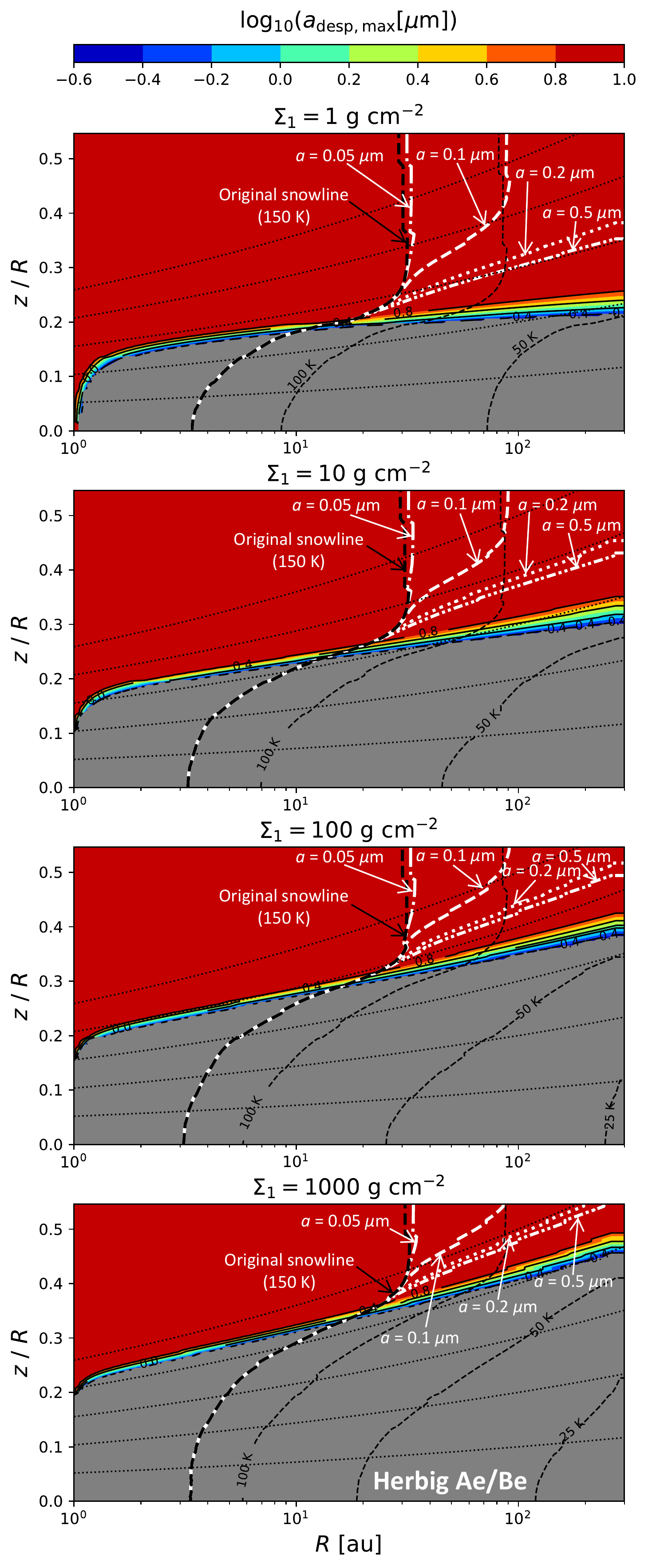}
\caption{Same as Figure \ref{fig:a_desp_TTauri} but for a disk around Herbig Ae/Be stars. Due to higher dust temperatures induced by a stronger radiation field, rotational desorption takes place in a broader region.}
\label{fig:a_desp_Herbig}
\end{figure*}

Figure \ref{fig:a_desp_Herbig} shows similar results as Figure \ref{fig:a_desp_TTauri} but for the disk around Herbig Ae/Be stars. The active region of rotational desorption is larger due to higher temperatures in the surface layer. Rotational desorption is also very efficient in the region beyond the snowline, which is much further from the star in this case (around $40 \AU$ from the central star). Up to $\sim 200\AU$, desorption sizes still can reach $~0.05\mum$ in the upper layer. At the outer radius $R=300\AU$, the desorption of icy grains has not eased for high values of $\Sigma_1$ (e.g., $100$ and $1000 \g\cm^{-2}$) as in the case of T-Tauri disks. Thus, the snowline is pushed down further by rotational desorption in this case. 

\begin{figure*}
\includegraphics[width=0.5\textwidth]{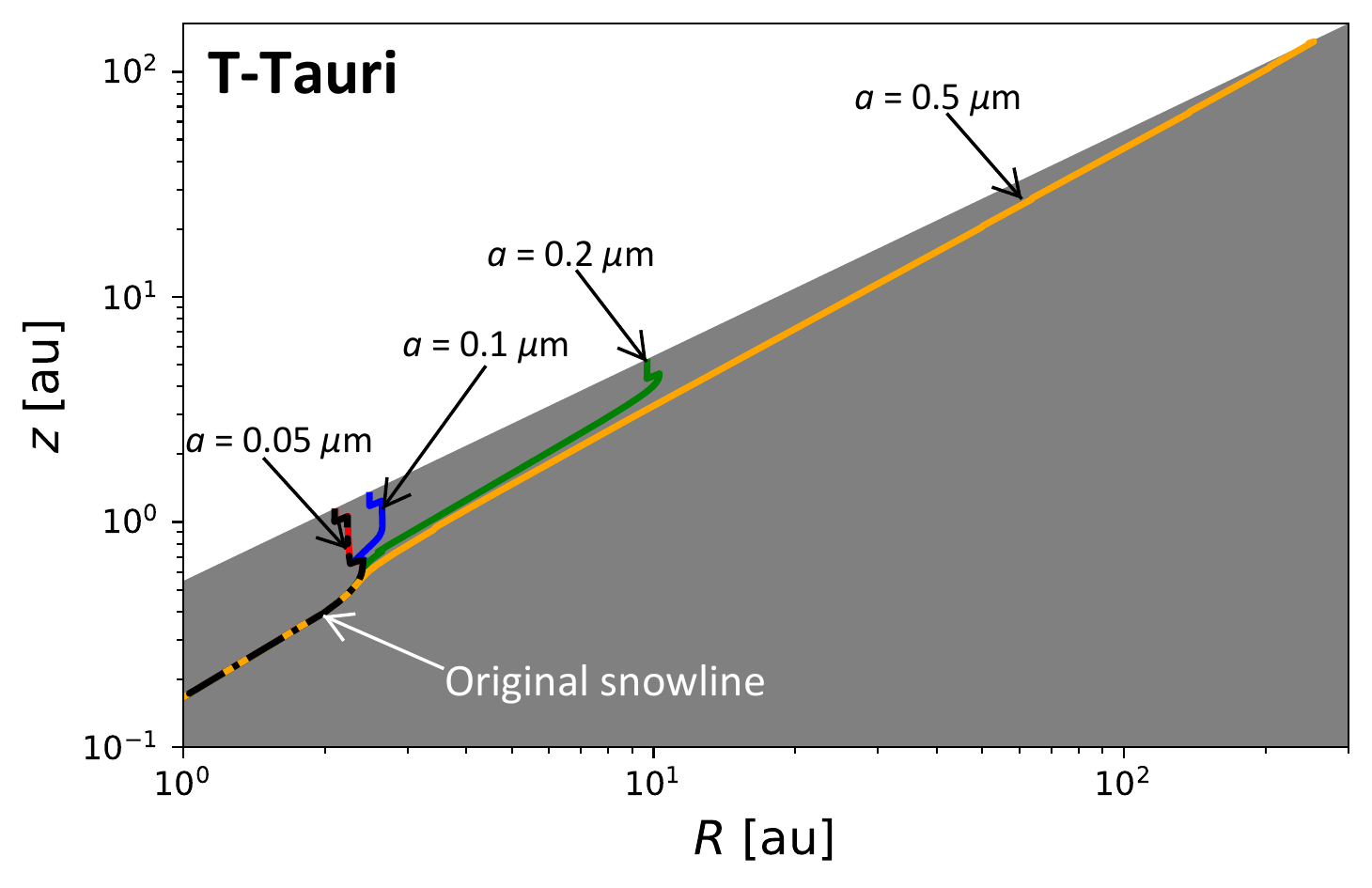}
\includegraphics[width=0.5\textwidth]{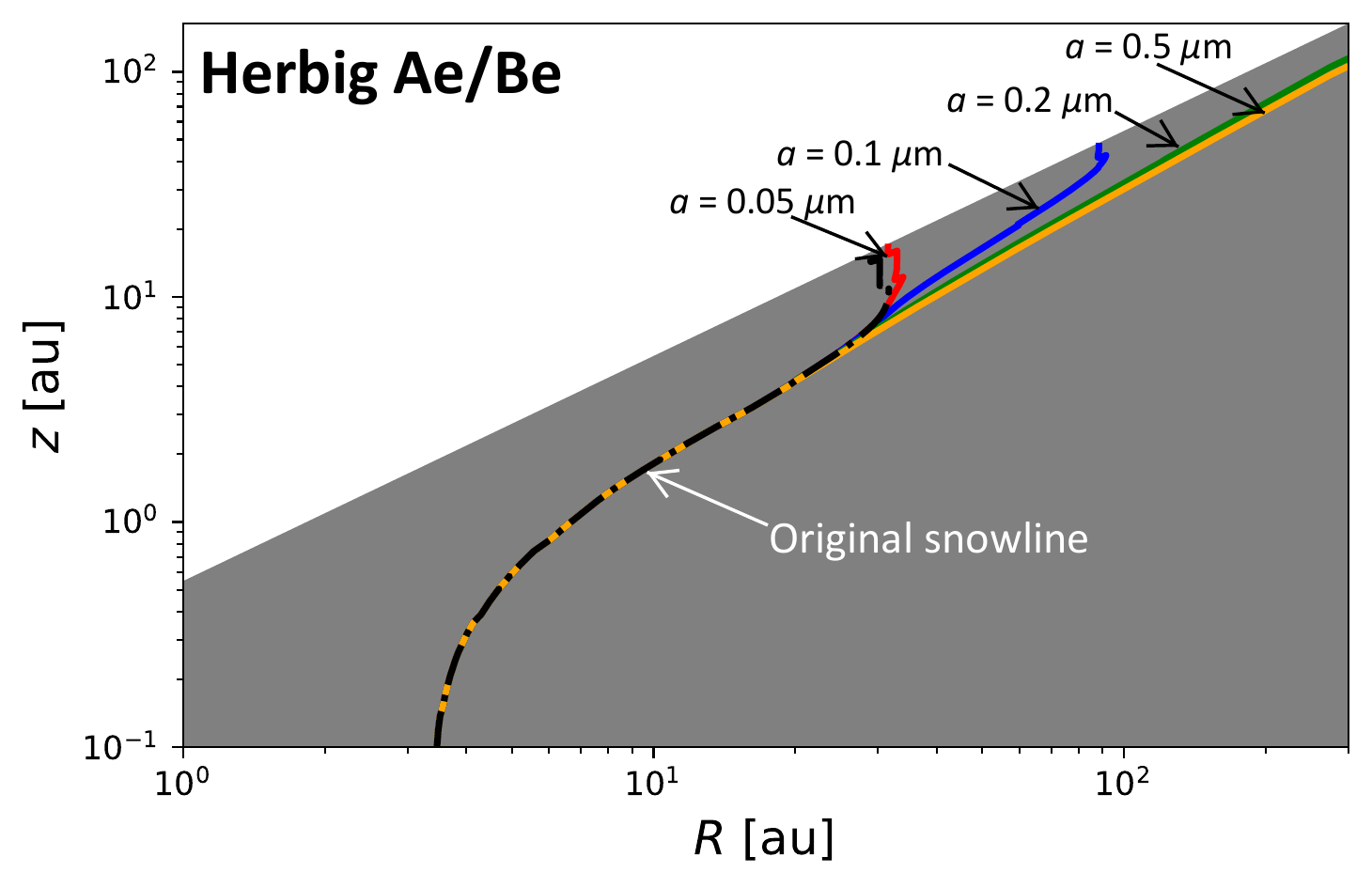}
\caption{Illustration of the water snowline for disks around T-Tauri (left panel) and Herbig stars (right panel) due to ro-thermal desorption. Ro-thermal desorption of ice mantles extends the original snowline, depending on the radius of icy grains.}
\label{fig:T_subrot}
\end{figure*}

In Figure \ref{fig:T_subrot}, we plot the results for ro-thermal desorption with $\Sigma_1 = 1 \g\cm^{-2}$ in regular $R-z$ coordinates for a better illustration of this effect on the snowline location.

\begin{figure*}
\centering
\includegraphics[width=0.45\textwidth]{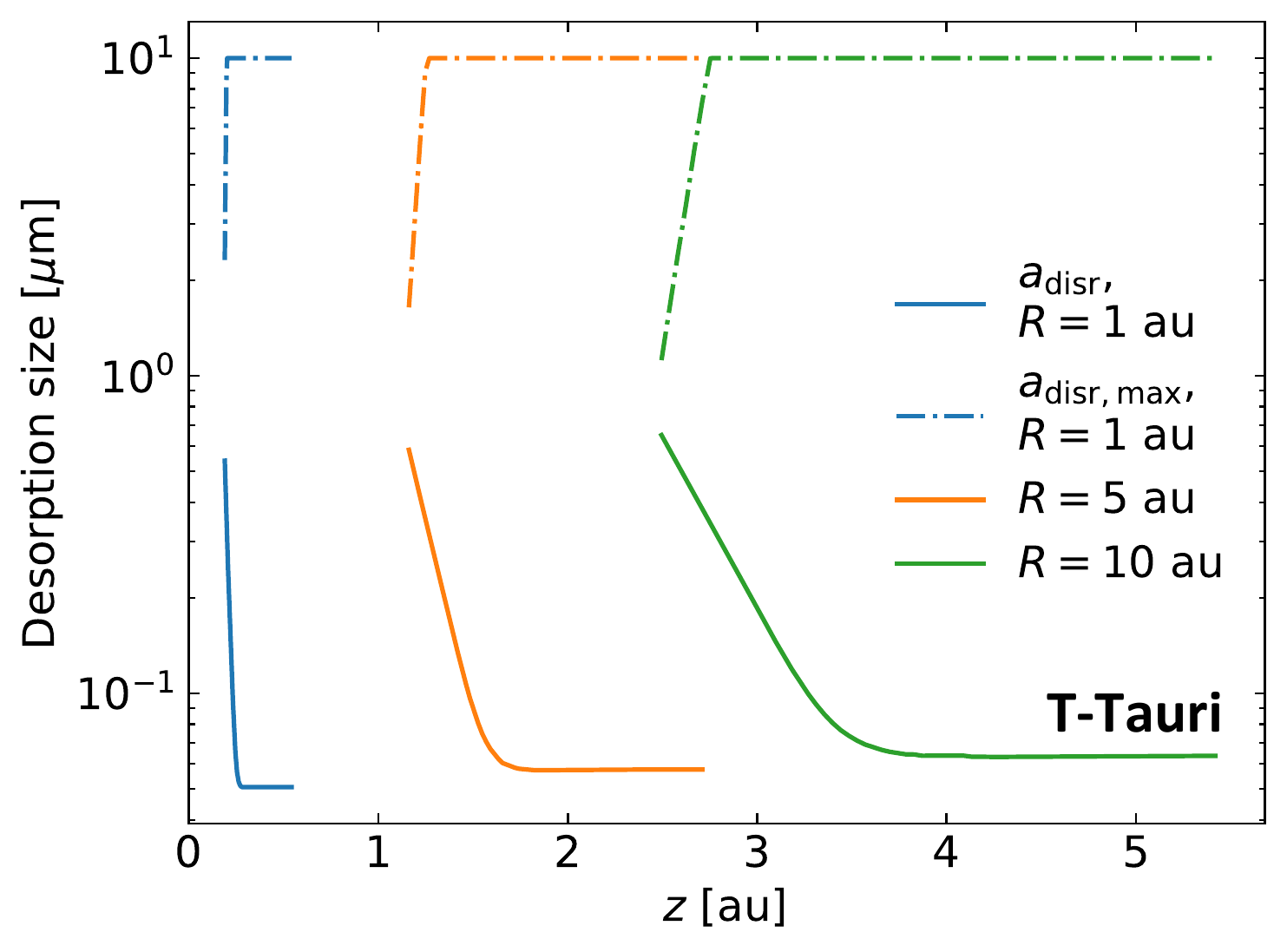}
\includegraphics[width=0.45\textwidth]{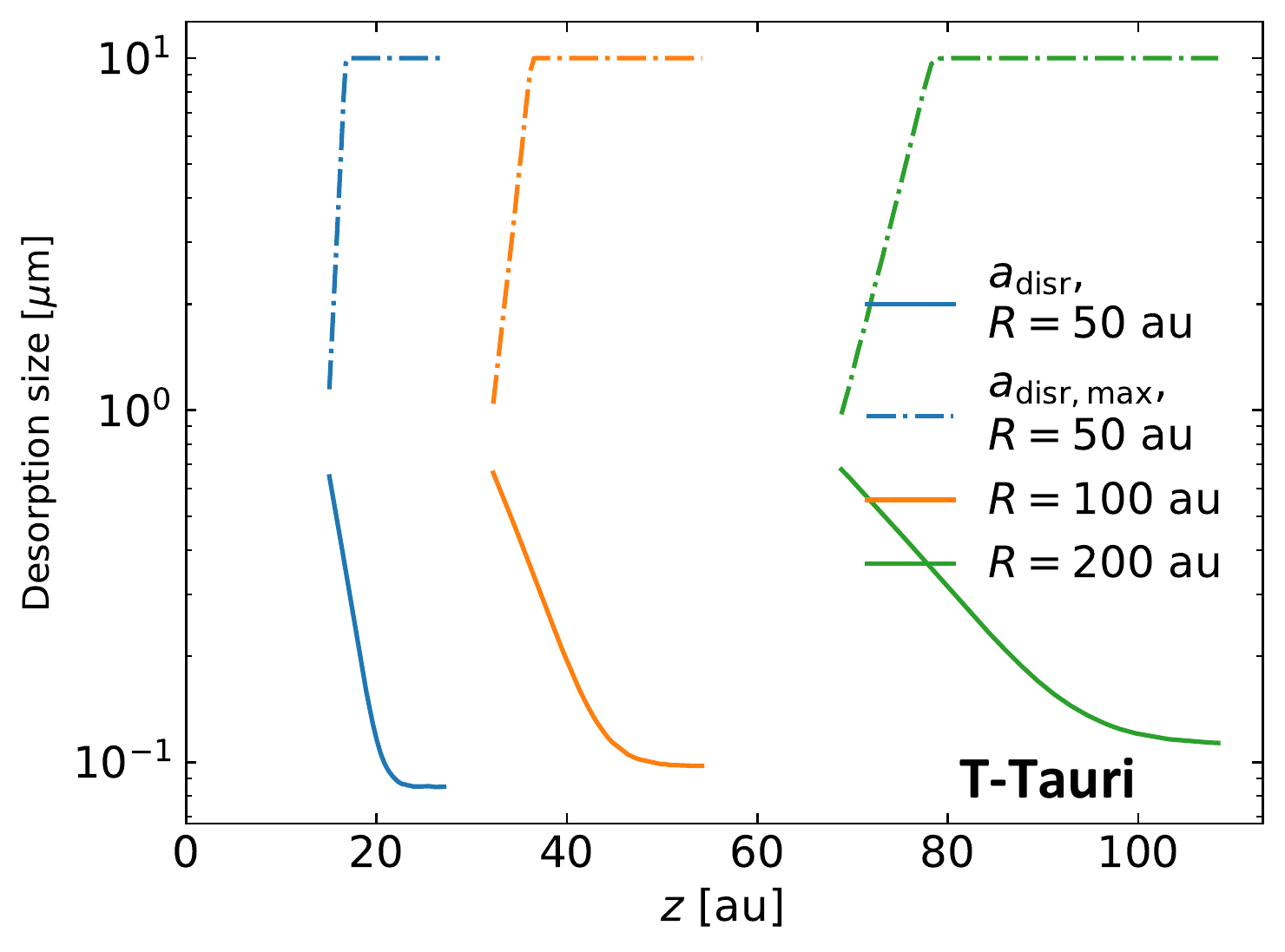}
\includegraphics[width=0.45\textwidth]{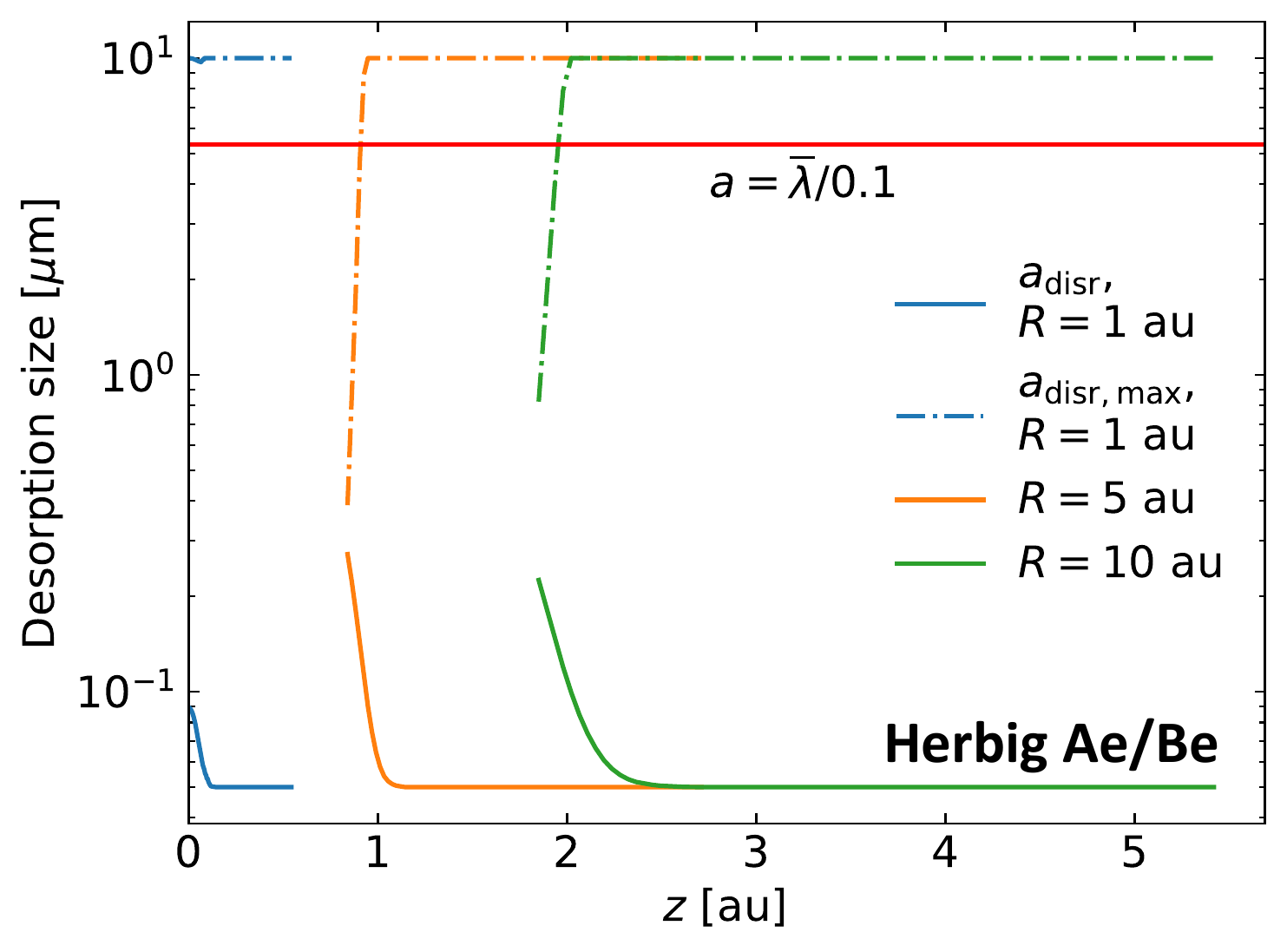}
\includegraphics[width=0.45\textwidth]{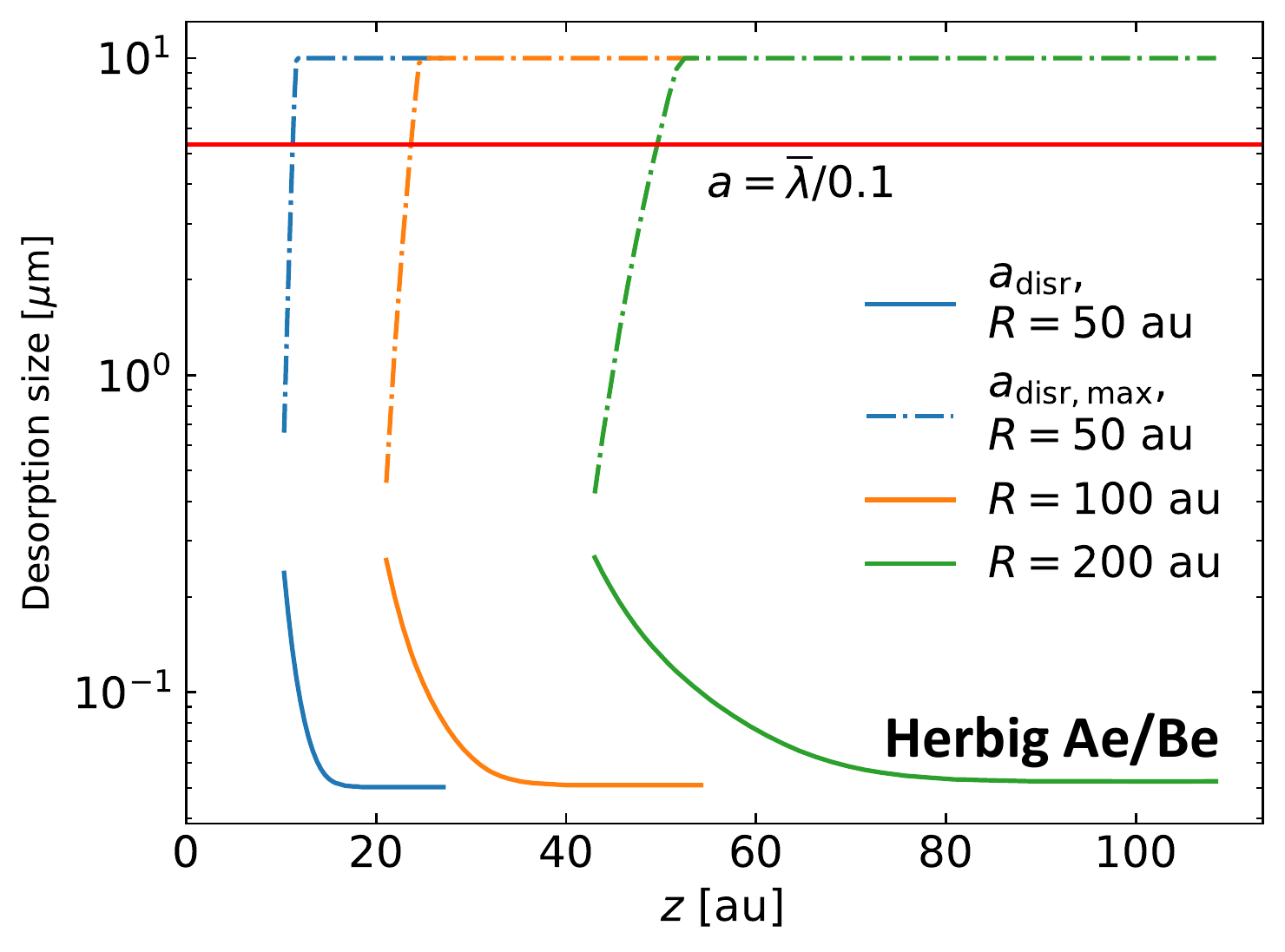}
\caption{Desorption sizes $a_{\rm desp}$ and $a_{\rm desp,max}$ as a function of the disk height $z$ computed for different radii $R$. Upper and lower panels show the results for T-Tauri disks and Herbig Ae/Be disks, respectively. $a_{\rm desp}$ decreases and $a_{\rm desp,max}$ increases with increasing $z$. The red horizontal line shows $a_{\rm disr,max} = \bar{\lambda}/0.1$.}
\label{fig:adespvsz}
\end{figure*}

Figure \ref{fig:adespvsz} shows the desorption sizes of icy grains as a function of the disk height $z$ at the different positions along the disk radius, with $\Sigma_1 = 1 \g\cm^{-2}$. One can see that $a_{\rm desp}$ decreases exponentially with $z$ to the lower boundary $~0.05\mum$, while $a_{\rm desp,max}$ increases exponentially until reaching the maximum value of $\sim10 \mum$ due to the decrease of $n_{\rm H}$ and the increase of $T_d$, similar to rotational disruption.


\begin{figure*}
\centering
\includegraphics[width=0.45\textwidth]{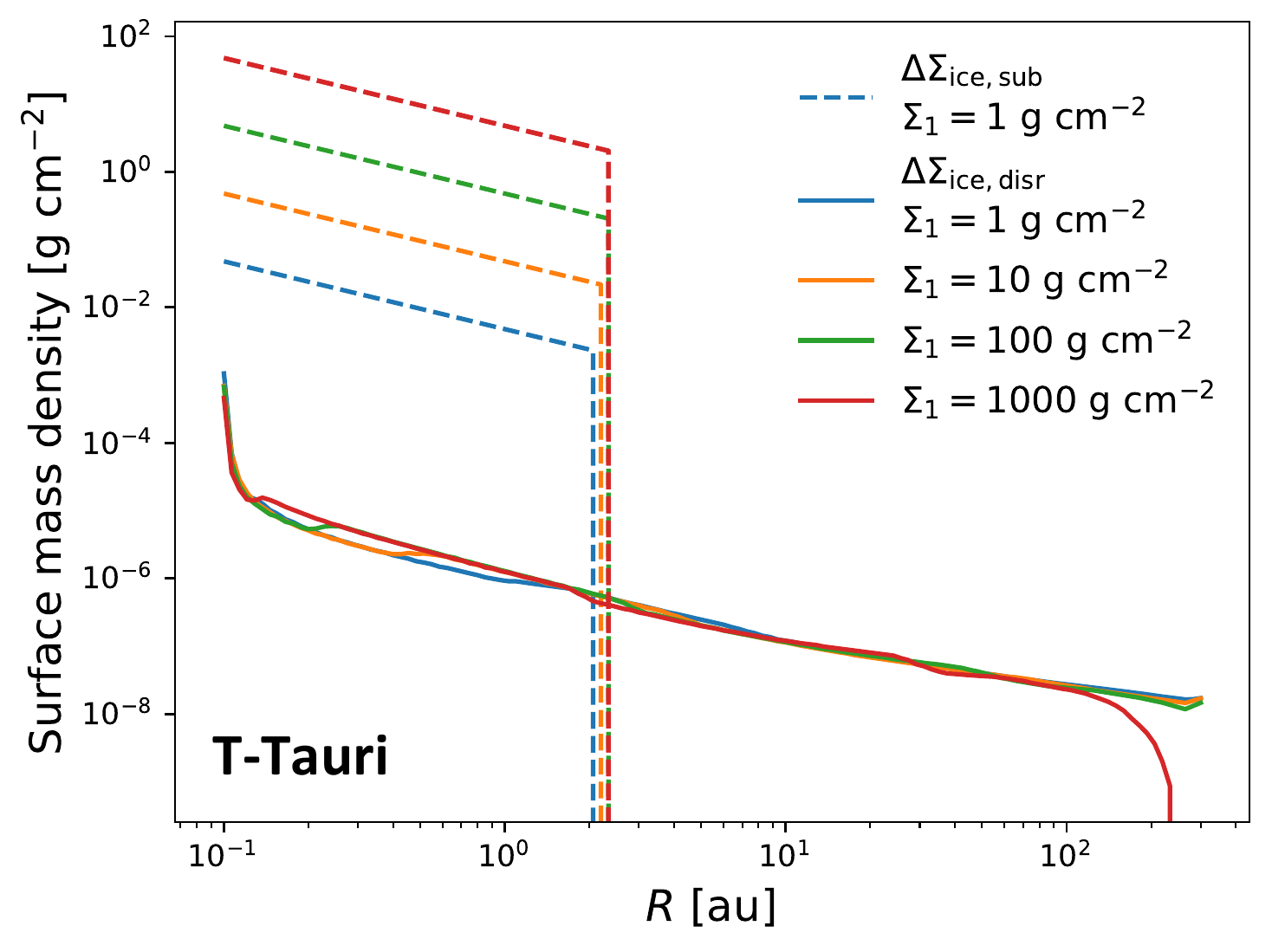}
\includegraphics[width=0.45\textwidth]{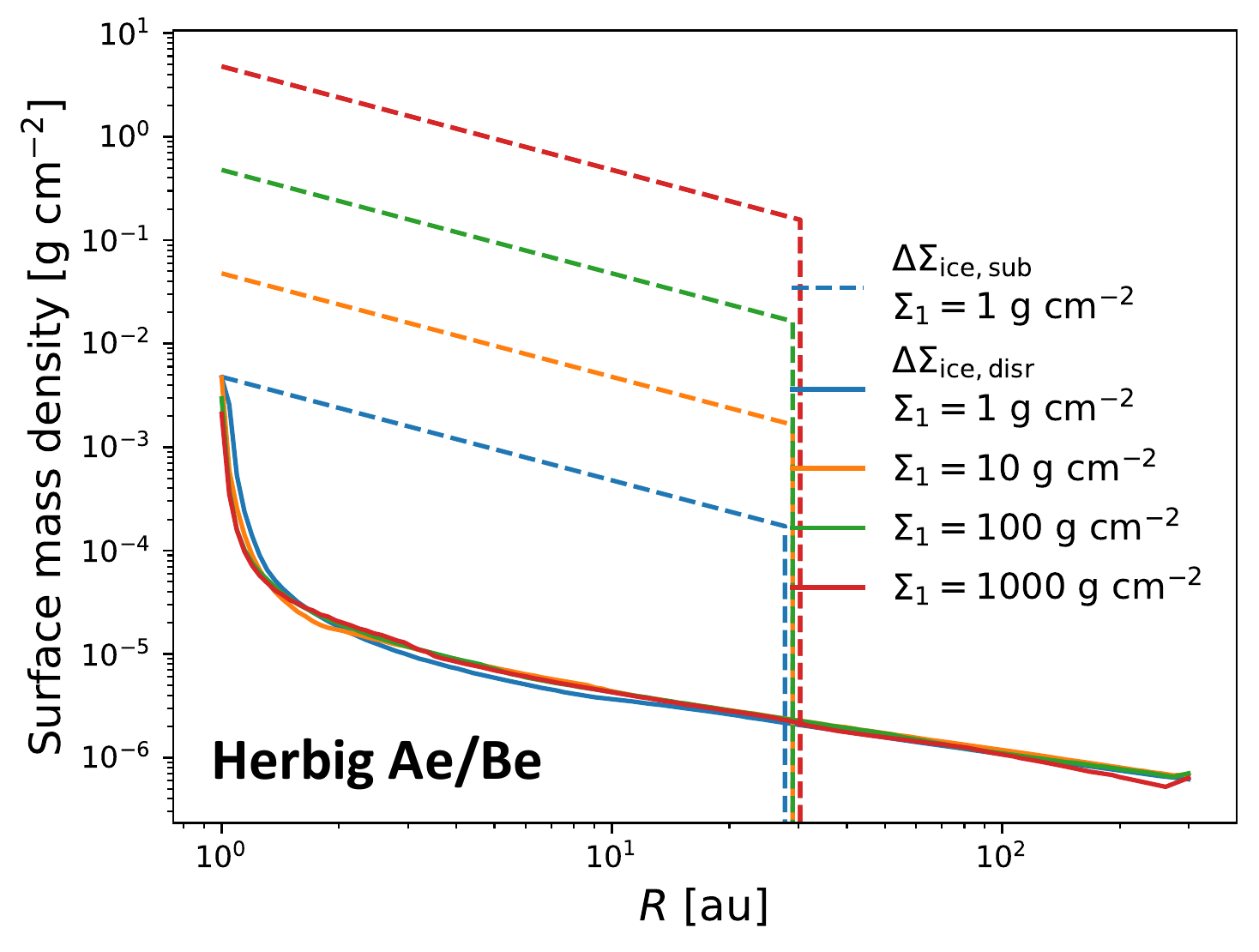}
\caption{Surface mass density $\Delta \Sigma_{\rm ice}$ of disrupted water ice by rotational desorption (solid lines) and of sublimated water ice by thermal sublimation (dashed lines). Thermal sublimation is the main mechanism to desorb water prior to the snowline, but rotational desorption takes over beyond the snowline. Left and right panels show the results for T-Tauri disks and Herbig Ae/Be disks, respectively.}
\label{fig:Delta_Sigma}
\end{figure*}

To quantify the effect of rotational desorption on the presence of ice in the disks, we calculate the total surface mass density of water ice disrupted by rotational desorption
\bea
\Delta \Sigma_{\rm ice,disr}(R) =  \int_{-\infty}^{+\infty} dz \int_{a_{\rm disr}}^{a_{\rm disr,max}} \rho_{ice} V_{\rm ice}(a) \frac{dn_{\rm gr}(a)}{da} da,
\ena
where $V_{\rm ice} = 4\pi(a^3-a_{\rm c}^3)/3$ is the volume of the ice mantle of grain of size $a$ with the core radius $a_{\rm c}$, and the power-law distribution $dn_{\rm gr}(a)/da = Ca^{-q}$ is again assumed for icy grains. We consider the minimum grain size $a_{\rm min} = 0.05\mum$ and the maximum grain size $a_{\rm max} = 0.10\mum$. The normalization coefficient $C$ is calculated assuming the ratio of gas-to-dust mass of 100. For thermal sublimation, we assume that the ice mantles of all grains with $T_d > 150\K$ are removed and compute the amount of sublimated ice $\Delta \Sigma_{\rm ice,sub}$. Results for $\Delta \Sigma_{\rm ice,disr}$ and $\Delta \Sigma_{\rm ice,sub}$ are shown in Figure \ref{fig:Delta_Sigma}. Prior to the snowline (at $R \sim 3\AU$ for the T-Tauri disks and $R \sim 40\AU$ for the Herbig Ae/Be disks), the amount of water ice sublimates through thermal sublimation is higher than that of ice disrupted by rotational desorption. This is expected given that ice is believed to be present only in the region beyond the snowline where grain temperature is lower than the sublimation threshold $T_d \sim 150\K$ of H$_2$O. Therefore, according to our assumption, the mantles of all grains located before the snowline from the disk midplane to the surface are destroyed by thermal sublimation, whereas rotational does not occur in the disk interior and thus, can affect a smaller surface mass density of water ice. However, beyond the snowline, rotational desorption continues to happen in the upper layer, and ice can still be removed in this region. The amount of disrupted ice is higher for smaller maximum grain size, for which more dust grains in the range size $[a_{\min},a_{\max}]$ are under the effect of RATD.

\subsubsection{Rotational desorption time vs. sublimation time}

Now let us compare the time it takes for rotational desorption to disrupt the ice mantles with the thermal sublimation time.

Following \citet{Hoang:2019td}, the sublimation time of the ice mantle of thickness $\Delta a_{m}$ is given by
\bea
t_{\rm sub}(T_d)&=&-\frac{\Delta a_{m}}{da/dt}
= \frac{\Delta a_{m}}{l\nu}\exp\left(\frac{E_{b}}{T_d}\right)\nonumber\\
&\simeq& 1.5\times 10^{3}\left(\frac{\Delta a_{m}}{500\AA}\right)\exp\left(\frac{E_{b}}{5700\K}\frac{100\K}{T_{d}} \right) \yr.,\label{eq:tsub}
\ena
where $da/dt\sim l/\tau_{\rm sub}$ with $l$ being the thickness of ice monolayer, is the rate of decrease in the mantle thickness due to thermal sublimation.

We compute $t_{\rm sub}$ for desorption of water molecules with the binding energy $E_b=5700 \K$ using the dust temperature from \textsc{radmc-3d} and the desorption time for the disks in the region where rotational desorption of ice mantles takes place from Equation (\ref{eq:tdesp}), using numerical results for $a_{\rm desp}$ from Section \ref{subsec:adesp}. A comparison is provided in Figure \ref{fig:tdesp} for grain size $a = 1\mum$ and $\Sigma_1 = 1 \g\cm^{-2}$. Rotational desorption of ice mantles is much faster than thermal sublimation in the surface and intermediate layers. The difference could be of order $10$ in the inner central region.

\subsection{Ro-thermal desorption of molecules from ice mantles}\label{sec:rothervsROTD}

\subsubsection{Ro-thermal desorption of molecules}

We have seen that rotational desorption and ro-thermal desorption of ice mantles are both very efficient mechanisms to desorb COMs. As shown in Figures \ref{fig:a_desp_TTauri} and \ref{fig:a_desp_Herbig}, for the grain model with a fixed core radius $a_{\rm c} = 0.05\mum$ and the varying mantle thickness, the region where rotational desorption is important appears to be larger than that of ro-thermal desorption. However, for a grain model of thin ice mantle, the resulting tensile stress by suprathermal rotation may be insufficient to overcome the adhesive energy between the mantle and the grain core surface. Therefore, it is hard to disrupt the entire ice mantle by RATD. In this case, the tensile strength is replaced by the adhesive strength, which depends on the mechanical property of the surface and grain temperature. The adhesive strength is low for a clean surface, but it can reach $\sim 10^{9}\erg\cm^{-3}$ for some rough surface \citep{Work:2018bu}. As shown in \citet{HoangTung}, ro-thermal desorption of individual molecules can occur before the disruption of ice mantle if the ice mantle thickness is below 100 monolayers of water ice (i.e., $\Delta a_{m} < 200$ {\AA}).

\begin{figure*}
\includegraphics[width=0.5\textwidth]{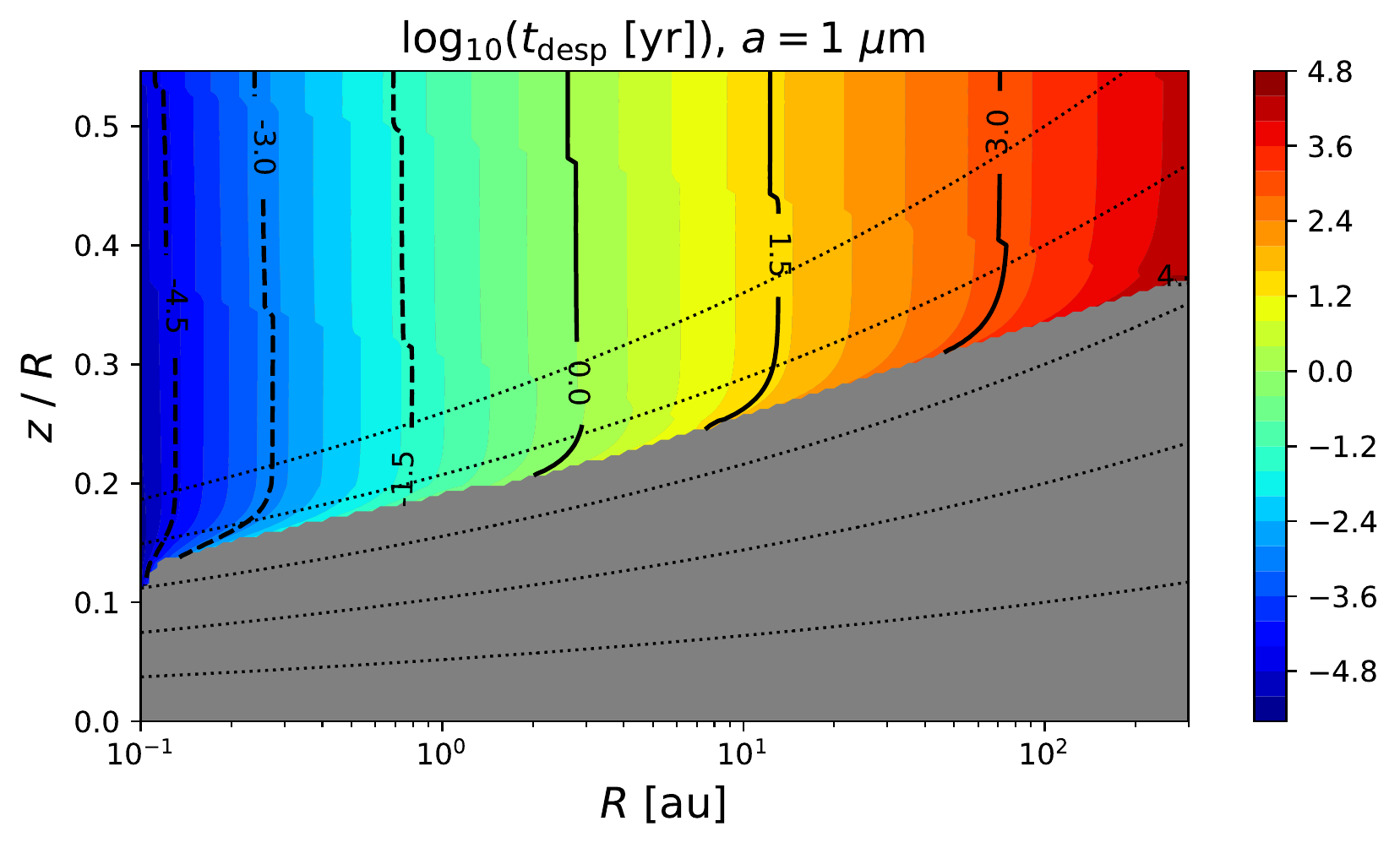}
\includegraphics[width=0.5\textwidth]{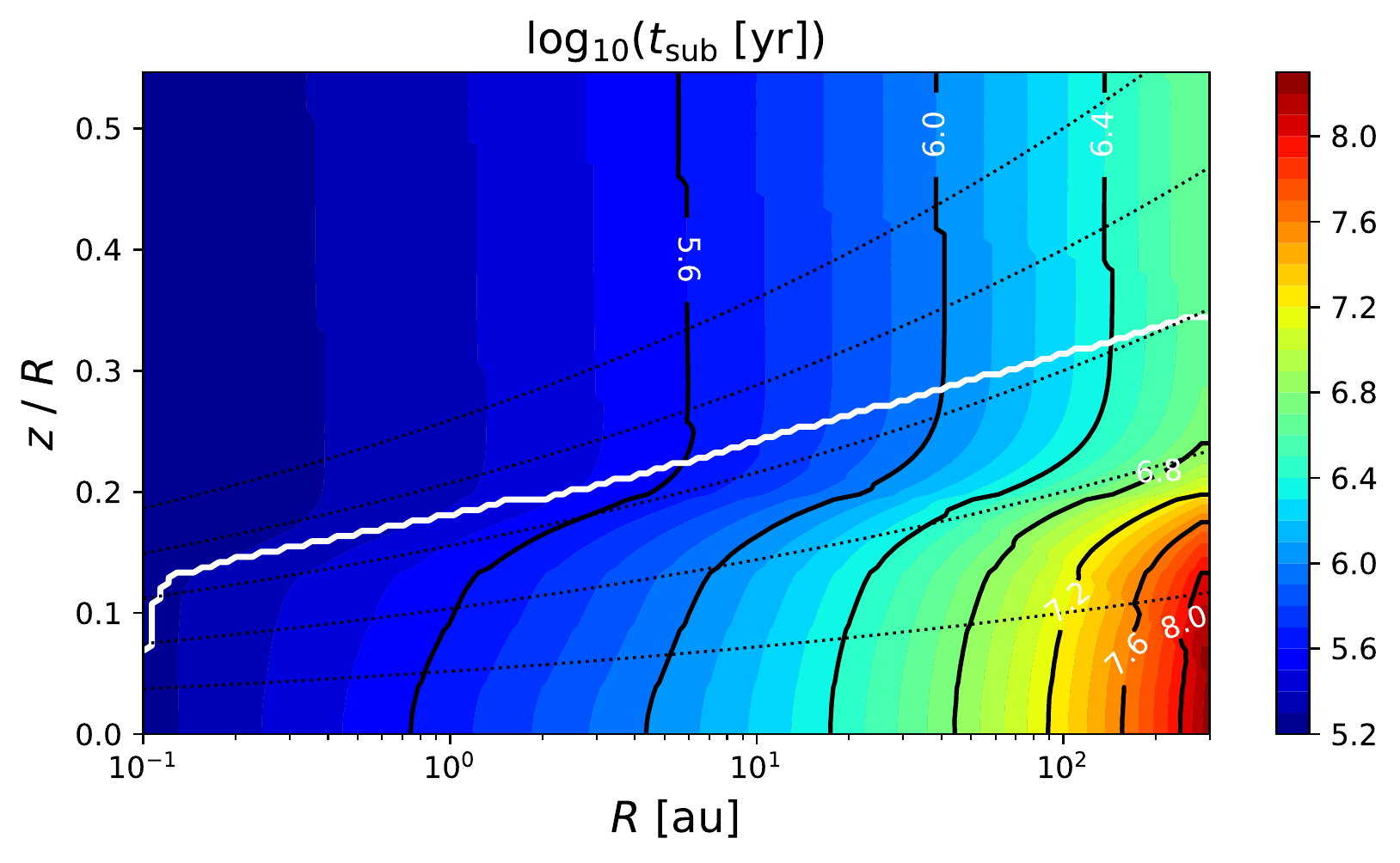}
\caption{Comparison of desorption time of water ice $t_{\rm desp}$ (left panel) with the classical sublimation time (right panel) for a disk around T-Tauri star. Grain size $a = 1\mum$ and $\Sigma_1 = 1 \g\cm^{-2}$ are considered. Rotational desorption is much faster than sublimation in the active region (above the white line, right panel).}
\label{fig:tdesp}
\end{figure*}

\begin{figure*}
\includegraphics[width=0.5\textwidth]{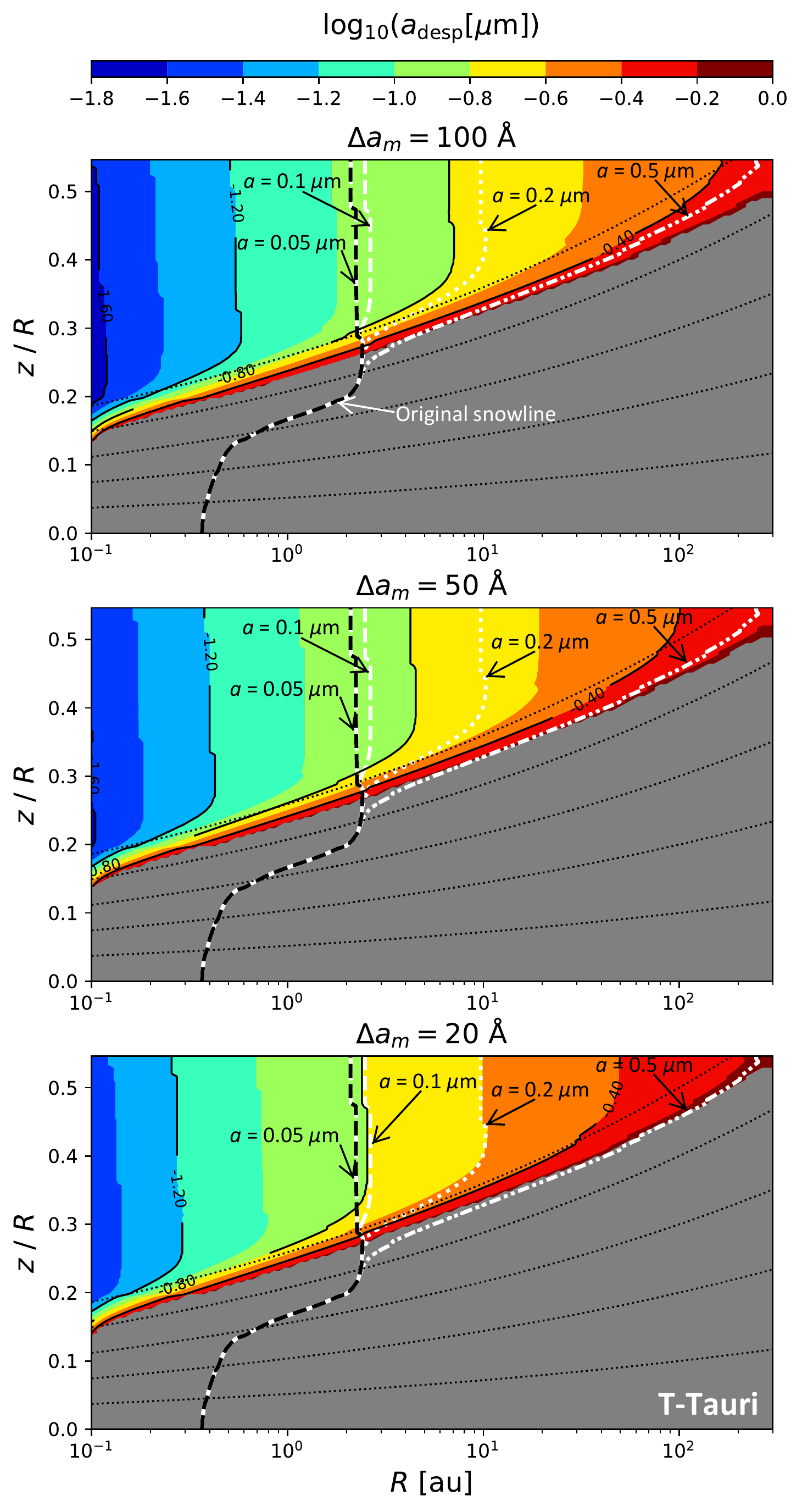}
\includegraphics[width=0.5\textwidth]{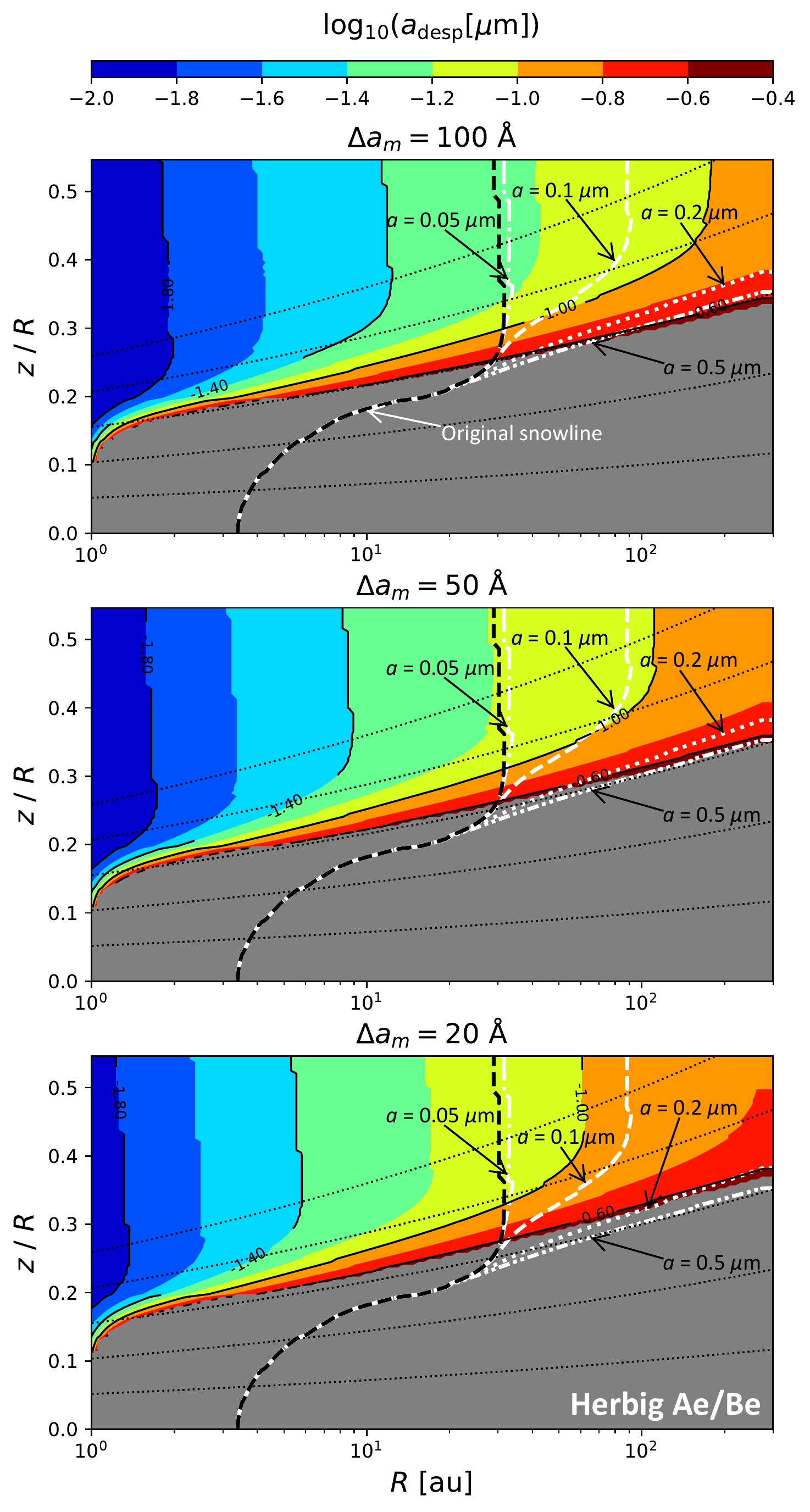}
\caption{Desorption sizes of ice mantles $a_{\rm desp}$ as a function of $R$ and $z/R$ for T-Tauri disks (left panels) and Herbig Ae/Be disks (right panels) for the model with a fixed mantle thickness $\Delta a_{m}$ and the varying core radius. The adhesive strength $S_{\rm max,mantle} = 10^9\erg\cm^{-3}$ is assumed for the ice mantle. Different values of $\Delta a_{m}$ are considered. Rotational desorption is less efficient for thinner ice mantle.}
\label{fig:adesp_1e9}
\end{figure*}

We thus consider a core-ice mantle model in which the mantle thickness $\Delta a_{m}$ is fixed to a certain value while the core radius is varied. Results with $\Sigma_1 = 1\g\cm^{-2}$ are shown in \ref{fig:adesp_1e9}. For $\Delta a_{m} = 100${\AA}, the effect of rotational desorption of ice mantles becomes less significant than ro-thermal desorption for $a=0.5\mum$ for the case of T-Tauri disks and $a=0.2\mum$ for the case of Herbig Ae/Be disks. For thinner mantles, the region where ice mantles are disrupted is narrower, and the active region of ro-thermal desorption is sufficiently broader for larger grain sizes. In that case, ro-thermal desorption takes over rotational desorption of ice mantles to release water and COMs into the gas phase. Furthermore, for this case of thin mantle, the effect of rotational desorption is comparable to ro-thermal desorption, but less efficient than for the case of thick ice mantle (see Figure \ref{fig:tdesp}).

\subsubsection{Ro-thermal desorption of PAHs/nanoparticles from the ice mantle}
Ro-thermal desorption is found to be an efficient mechanism to desorb PAHs that are weakly bound to the ice mantle via van der Waals force \citep{HoangTung}. Since ro-thermal desorption requires lower radiation strength to desorb than rotational desorption, one can describe the efficiency of ro-thermal desorption by considering the ejection threshold. Following \cite{HoangTung}, one obtains the ejection threshold of PAHs:
\bea
\omega_{\rm ej}=\left(\frac{3E_{b}}{ma^{2}}\right)^{1/2}\simeq \frac{10^{10}}{a_{-5}}\left(\frac{(E_{b}/k)}{4000\K}\frac{m_{\rm C6H6}}{m}\right)^{1/2}\rm rad\s^{-1},\label{eq:omega_ej_PAH}~~~~
\ena
where the binding energy of benzene C$_{6}$H$_{6}$ and naphthalene (C$_{10}$H$_{8}$) to ice is $E_{b}/k\sim 4000\K$ and $6000\K$ (see Table 4 and 5 in \citealt{Michoulier:2018cx}).
 
The ejection radiation strength is then
\bea
U_{\rm ej}\simeq 35n_{1}T_{2}^{1/2}(1+F_{\rm IR})\frac{\lambda_{0.5}^{1.7}}{\gamma a_{-5}^{1.7}}\left(\frac{(E_{b}/k)}{4000\K}\frac{m_{\rm C6H6}}{m}\right)^{1/2}~\label{eq:Uej_PAH}
\ena
for $a\lesssim a_{\rm trans}$, and 
\bea
U_{\rm ej}\simeq 1.8n_{1}T_{2}^{1/2}(1+F_{\rm IR})\frac{\lambda_{0.5}^{1.7} a_{-5}}{\gamma}\left(\frac{(E_{b}/k)}{4000\K}\frac{m_{\rm C6H6}}{m}\right)^{1/2}
\ena
for $a> a_{\rm trans}$. 
Clearly, the ejection threshold is much lower than that of water and COMs. Therefore, the ro-thermal desorption is efficient for desorption of PAHs in star-forming regions.

\section{Effect of Rotational Disruption on Dust Opacity}\label{sec:kappa}

Assuming no disruption, the dust opacity, defined as the total absorption cross-section per unit of dust mass, is given by
\bea
\kappa_{\rm abs,sca}(\lambda) &=& \frac{\int_{a_{\rm min}}^{a_{\max}} \pi a^{2}Q_{\rm abs,sca}(a,\lambda)(dn_{\rm gr}/da) da}{\int_{a_{\rm min}}^{a_{\max}} (4\pi \rho a^{3}/3)(dn_{gr}/da) da}\nonumber\\
&=&\frac{\int_{a_{\rm min}}^{a_{\max}} X(a)a^{-q}da}{\int_{a_{\rm min}}^{a_{\max}} Y(a)a^{-q}da}, \label{eq:kappa0}
\ena
where $Q_{\rm abs,sca}(a,\lambda)$ is the absorption/scattering efficiency for a grain of radius $a$ at wavelength $\lambda$, $dn_{\rm gr}/da$ is the grain size distribution, $X(a)=\pi a^{2}Q_{\rm abs,sca}(a,\lambda)(dn_{\rm gr}/da)$ and $Y(a)=(4\pi \rho a^{3}/3)(dn_{gr}/da)$. Here we also assume the power-law distribution grain size $dn_{\rm gr}/da=Cn_{\rm H}a^{-q}da$ with $q=3.5$ and the ratio of gas-to-dust mass of 100 as previously done in Section \ref{sec:adesp} but for composite grains of mass density $\rho=3 \g \cm^{-3}$.

Due to rotational disruption, which redistribute all grains of size $a=[a_{\rm disr}-a_{\rm disr,max}]$ to smaller sizes ($a<a_{\rm disr}$), the dust opacity is modified to
\bea
\kappa_{\rm abs,sca}(\lambda) = \frac{\int_{a_{\rm min}}^{a_{\rm disr}} X(a)a^{-q^{\prime}}da+\int_{a_{\rm disr,max}}^{a_{\max}} X(a)a^{-q}da}{\int_{a_{\rm min}}^{a_{\max}}Y(a)a^{-q}da},~~~~\label{eq:kappa1}
\ena
where $q^{\prime}$ indicates that new size distribution is assigned for the resulting grains and can be determined from the conservation of dust mass:
\bea
\int_{a_{\rm min}}^{a_{\rm max}} V(a) c a^{-q}da&=&\int_{a_{\rm min}}^{a_{\rm disr}} V(a) c a^{-q^{\prime}}da\nonumber\\
&&+ \int_{a_{\rm disr,max}}^{a_{\rm max}} V(a) c a^{-q}da
\label{eq:kappa2},
\ena
where $V(a) = 4 \pi \rho a^{3}/3$ is the grain volume. Here we assume $a_{\rm min}=0.01\mum$ and $a_{\rm max}=10\mum$.

\begin{figure*}
\centering
\includegraphics[width=0.45\textwidth]{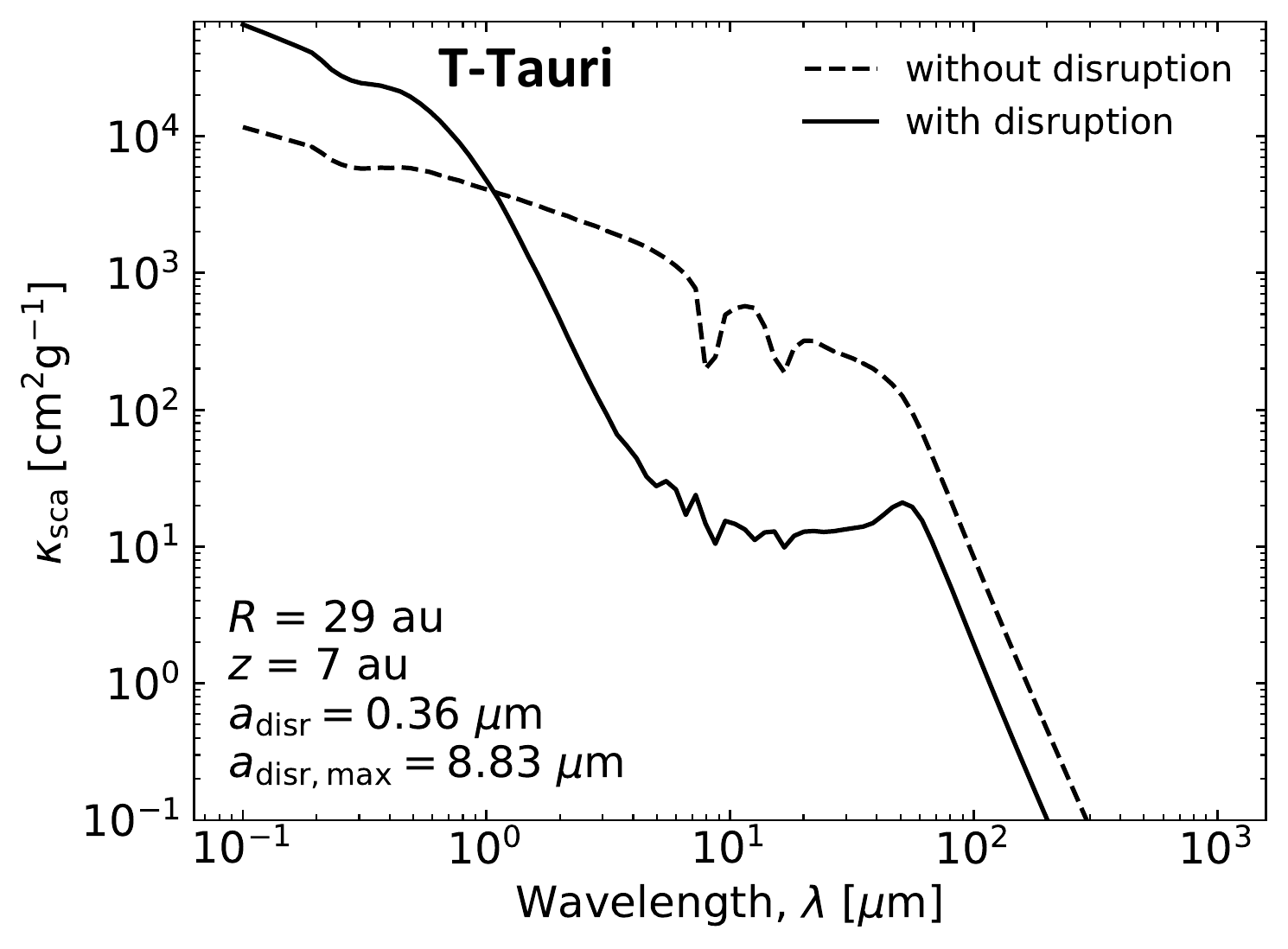}
\includegraphics[width=0.45\textwidth]{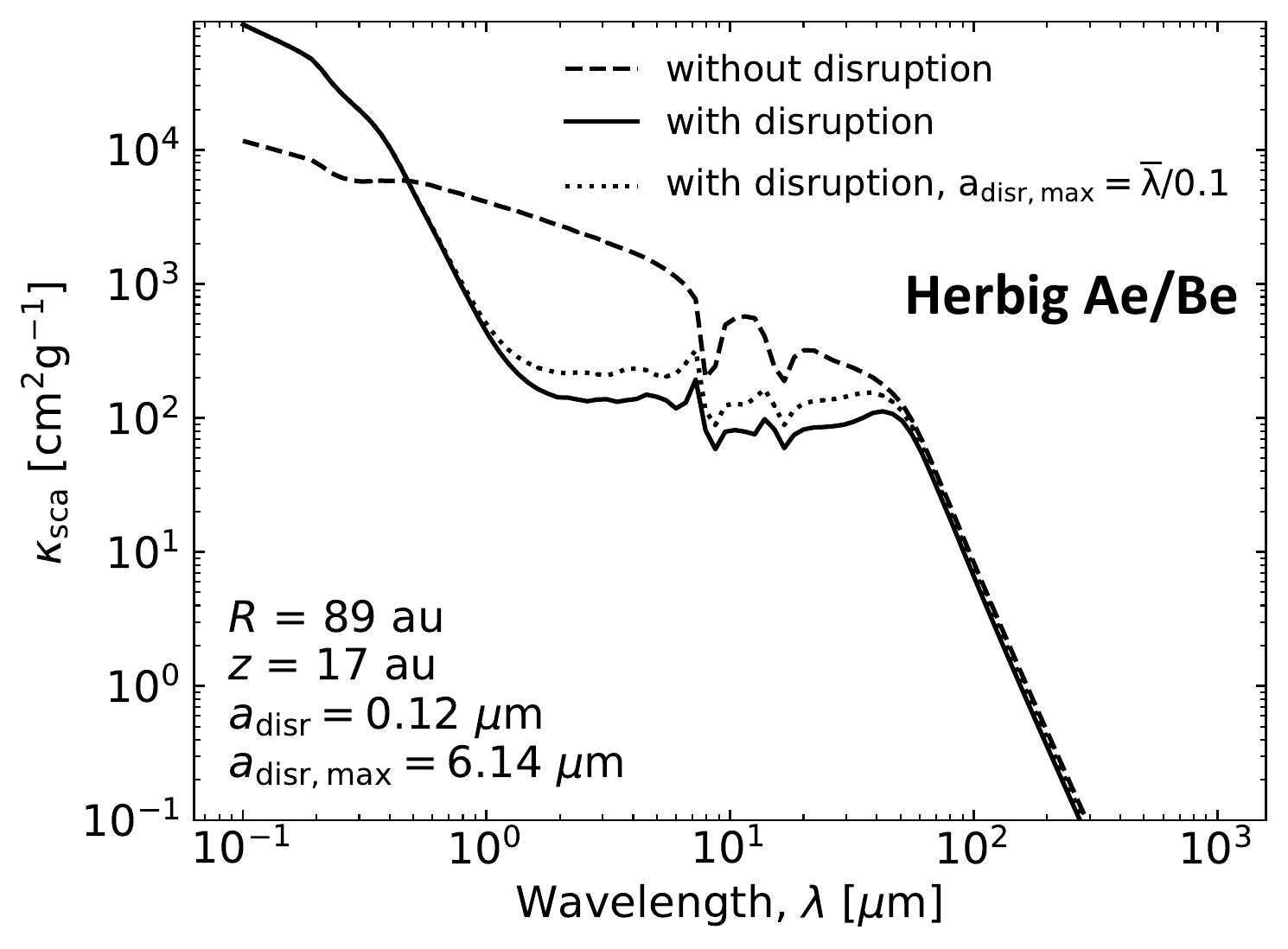}

\caption{Scattering opacity $\kappa_{\rm sca}$ for the different maximum grain sizes at selected positions in the surface layer of the disk as a function of the wavelength $\lambda$ with (solid lines) and without (dashed lines) the presence of disruption. The dotted line shows the results when considering $a_{\rm disr,max}=\overline{\lambda}/0.1$. Left and right panels show the results for T-Tauri disks and Herbig Ae/Be disks, respectively.}
\label{fig:ksca}
\end{figure*}

From the numerical results in Section \ref{sec:result}, we study how the dust opacity is modified by RATD in each cell, using the absorption/scattering cross-section for amorphous silicate grains as previously done in Section \ref{sec:Td-U}. Table \ref{tab:q} shows the obtained values of $q'$ at several cells in the disk, along with the locations of the cells and the disruption sizes calculated for those cells. In Figure \ref{fig:ksca} we show the scattering opacity $\kappa_{\rm sca}$ at a position in the disk surface for two types of PPDs, with $\Sigma_1 = 1 \g\cm^{-2}$. The lower and upper boundary for disruption sizes of grains by RATD in the numerical calculations without taking into account $a_{\rm disr,max} = \overline{\lambda}/0.1$ are shown for comparison in the case of Herbig disks. One can see that $\kappa_{\rm sca}$ decreases substantially at optical to mid-infrared ($\lambda \sim 0.5-50\mum)$ due to the disruption of dust grains. This arises from the fact that dust grains scatter efficiently photons with the wavelength comparable to the grain size, such that when those grains are removed by RATD, the scattering opacity is decreased accordingly. Note that in the case of Herbig disks, the upper boundary of disruption size at the chosen point exceeds $\overline{\lambda}/0.1 = 5.33\mum$. As a result, there is a considerable difference between the results with and without considering $a_{\rm disr,max}=\overline{\lambda}/0.1$.

\begin{table}
\begin{center}
\caption{Modified slopes of grain size distribution by RATD}\label{tab:q}
\begin{tabular}{lllllll} 
\hline\hline
{\it Disk type} & $R$ & $z$ & $a_{\rm disr}$ & $a_{\rm disr,max}$ & $q$ & $q'$ \\
 & $(\AU)$ & $(\AU)$ & $(\mum)$ & $(\mum)$ & & \\[1mm]

\hline
\multirow{4}{4em}{T-Tauri} & 5.31	& 0.50 & ND$^{a}$  &  ND & 3.5 & 3.5\\[1mm]
 & 5.31	& 1.05 & 0.6339  &  1.2218 & 3.5 & 3.53\\[1mm]
 & 5.31	& 2.01 & 0.0496  &  10.000 & 3.5 & 3.75\\[1mm]
 & 19.94	& 4.98 & 0.3488  &  9.8517 & 3.5 & 3.66\\[1mm]
\hline
\multirow{4}{4em}{Herbig\\Ae/Be} & 10.33	& 1.02 & ND  &  ND & 3.5 & 3.5\\[1mm]
 & 10.33	& 4.00 & 0.0270  &  10.000 & 3.5 & 3.79\\[1mm]
 & 52.05	& 9.85 & 0.1609  &  2.4039 & 3.5 & 3.63\\[1mm]
 & 102.09	& 20.20 & 0.1271  &  5.4929 & 3.5 & 3.68\\[1mm]
\hline
\multicolumn{7}{l}{$^a$~No disruption}\cr\\
\hline\hline
\end{tabular}
\end{center}
\end{table}

\begin{figure*}
\centering
\includegraphics[width=0.45\textwidth]{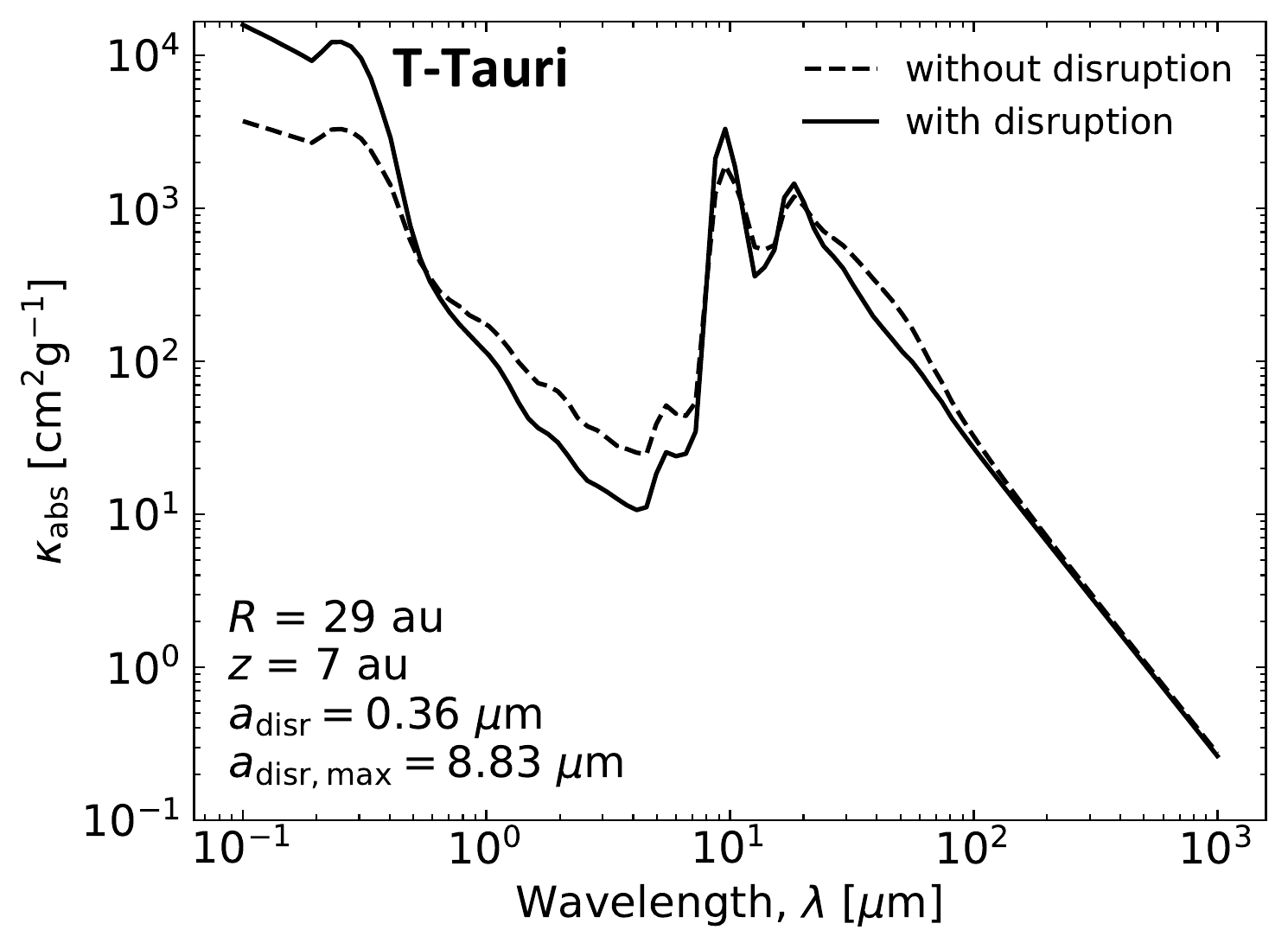}
\includegraphics[width=0.45\textwidth]{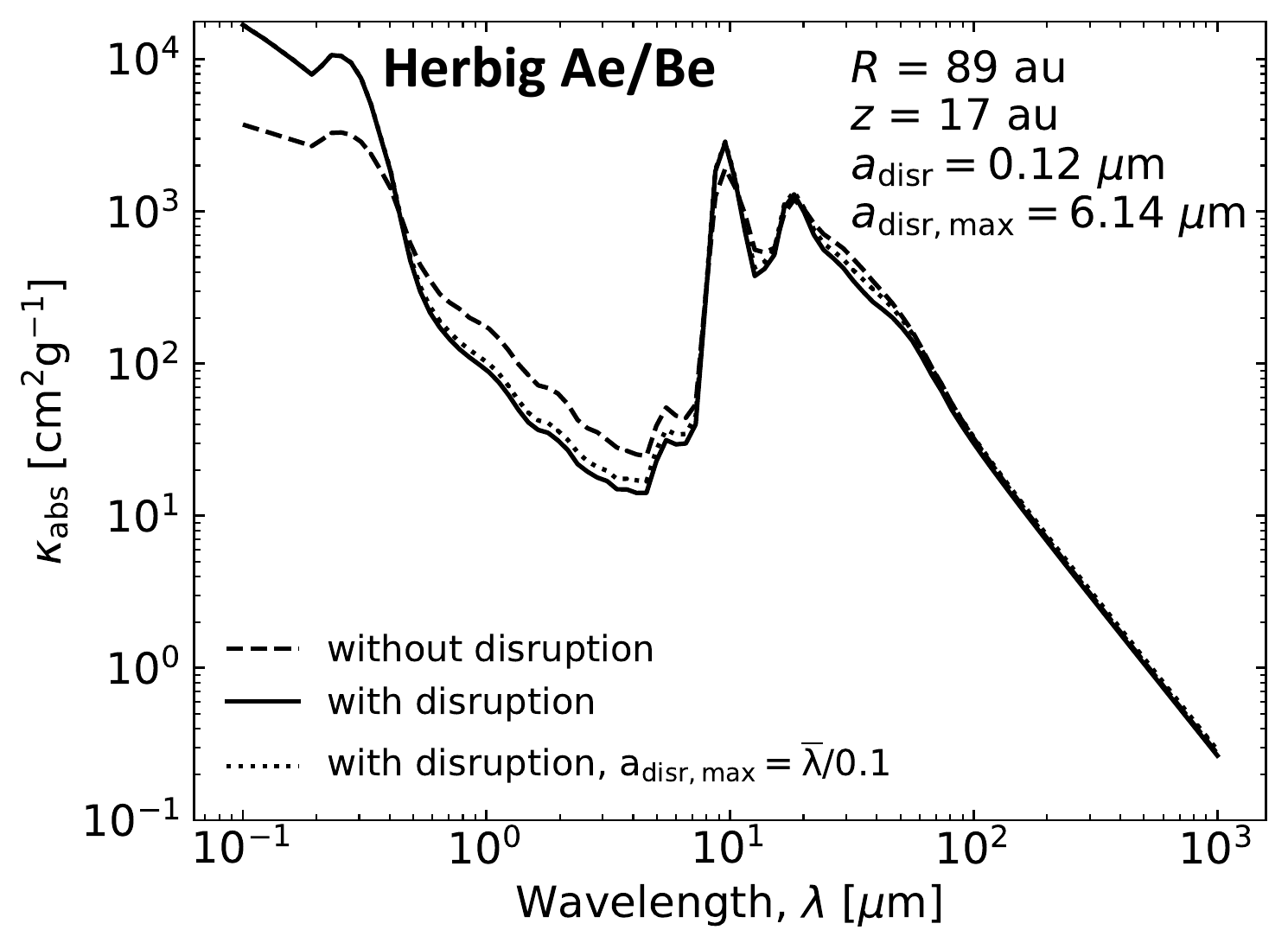}
\caption{Same as Figure \ref{fig:ksca}, but for the absorption opacity $\kappa_{\rm abs}$. Rotational disruption reduces $\kappa_{\rm abs}$ less effectively than $k_{\rm sca}$.}
\label{fig:kabs}
\end{figure*}

Figures \ref{fig:kabs} shows the absorption opacity $\kappa_{\rm abs}$ at a chosen disk radius and height in the surface layer of the disks around T-Tauri and Herbig Ae/Be stars for the cases with and without disruption. Though the difference is less pronounced than in the case of $\kappa_{\rm abs}$, there is a considerable decrease of $\kappa_{\rm abs}$ in the wavelengths between $\sim 0.4\mum$ and $\sim 8\mum$, which corresponds to optical and near/mid-IR region. This can be expected from the destruction of dust grains in the surface layer, where most of the mid-IR emission of the disks come from.

\begin{figure*}
\centering
\includegraphics[width=0.45\textwidth]{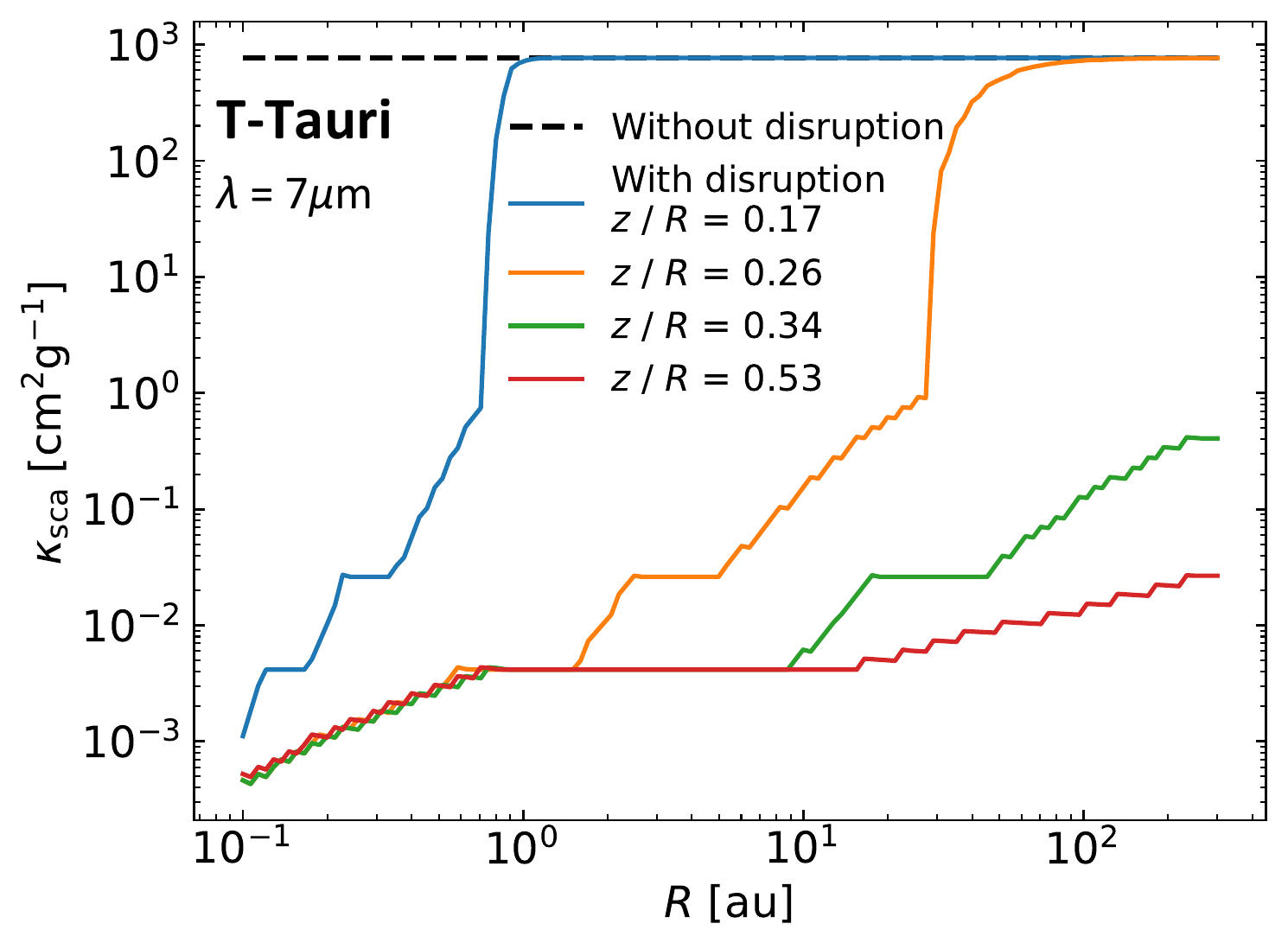}
\includegraphics[width=0.45\textwidth]{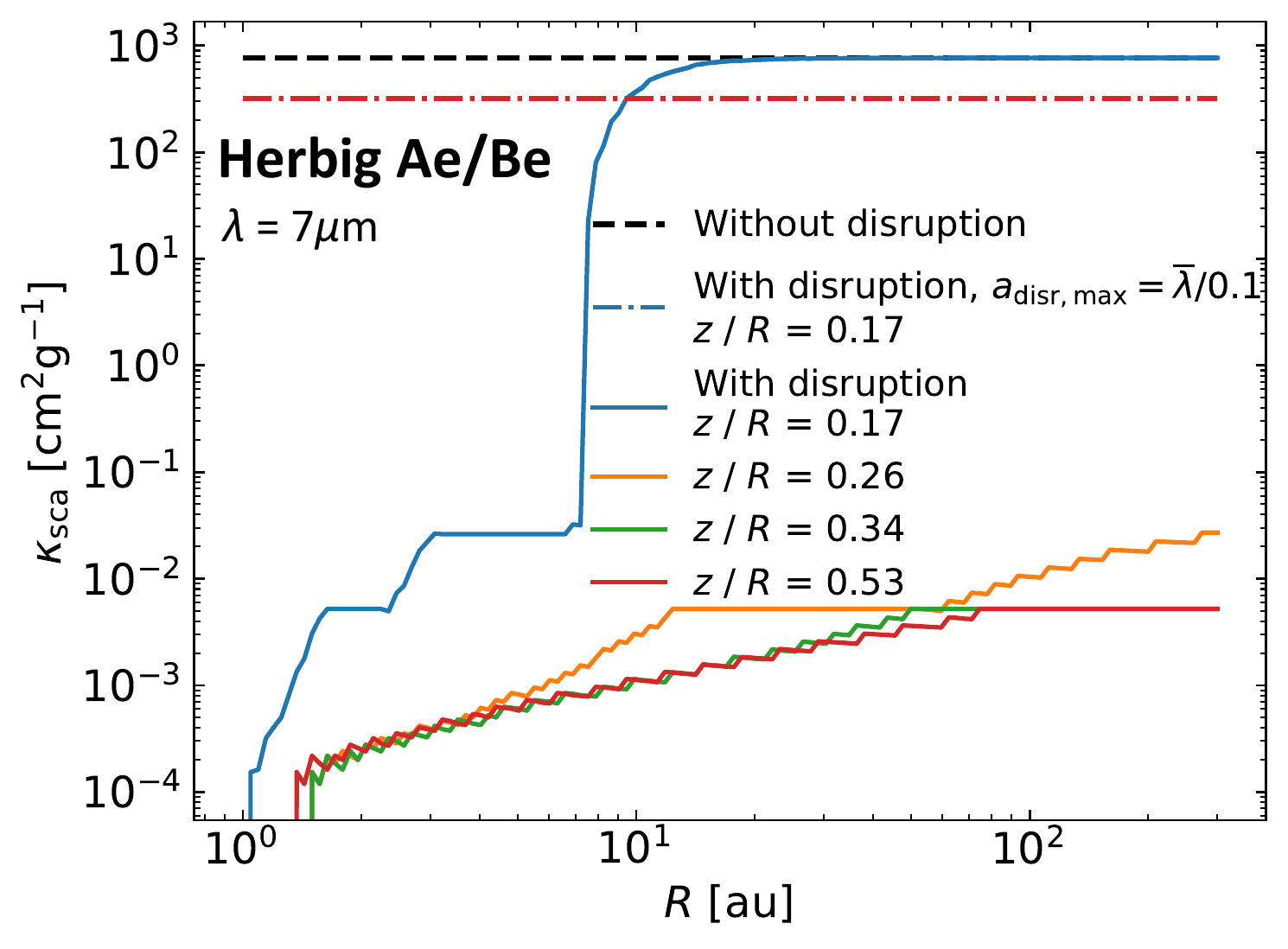}
\includegraphics[width=0.45\textwidth]{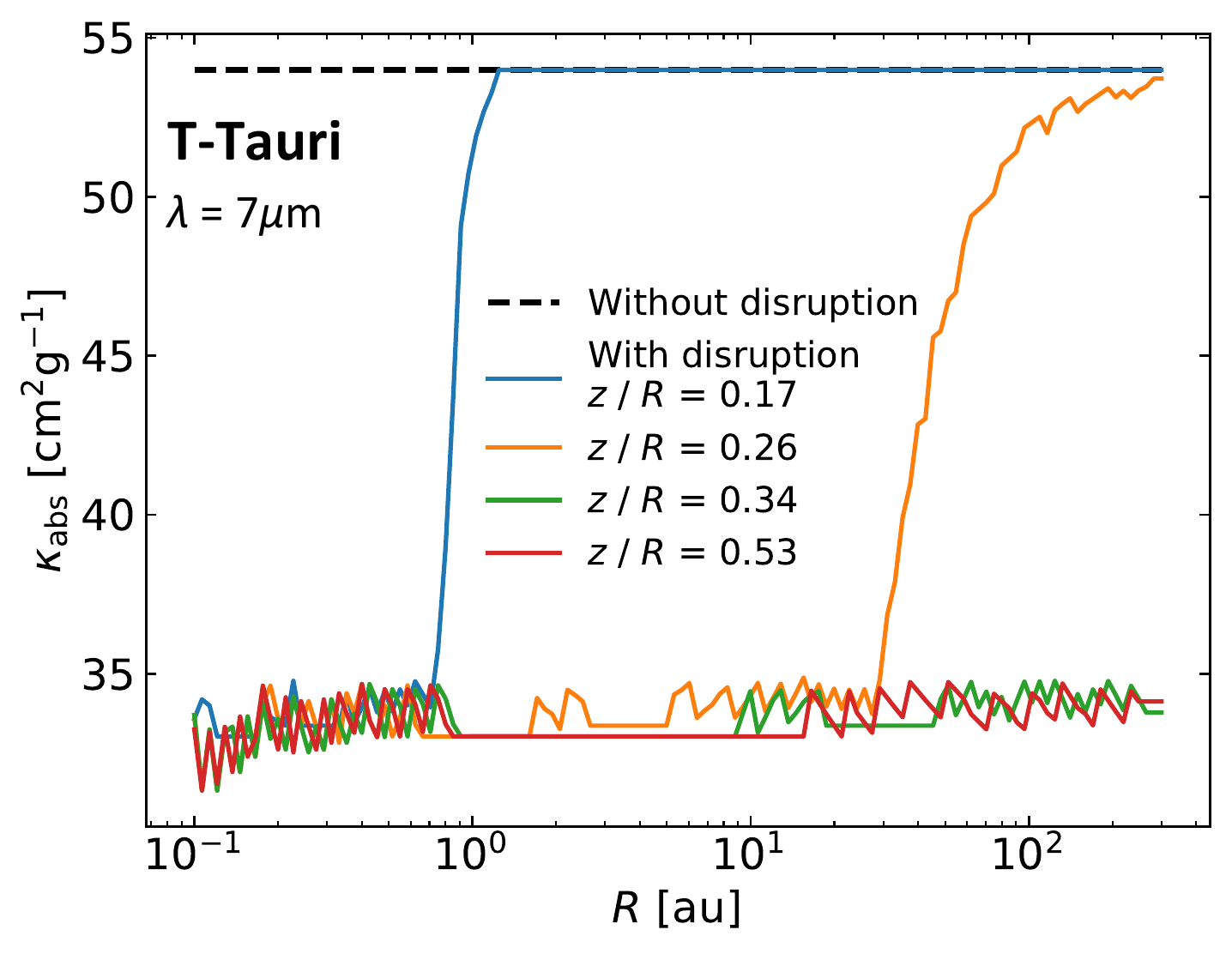}
\includegraphics[width=0.45\textwidth]{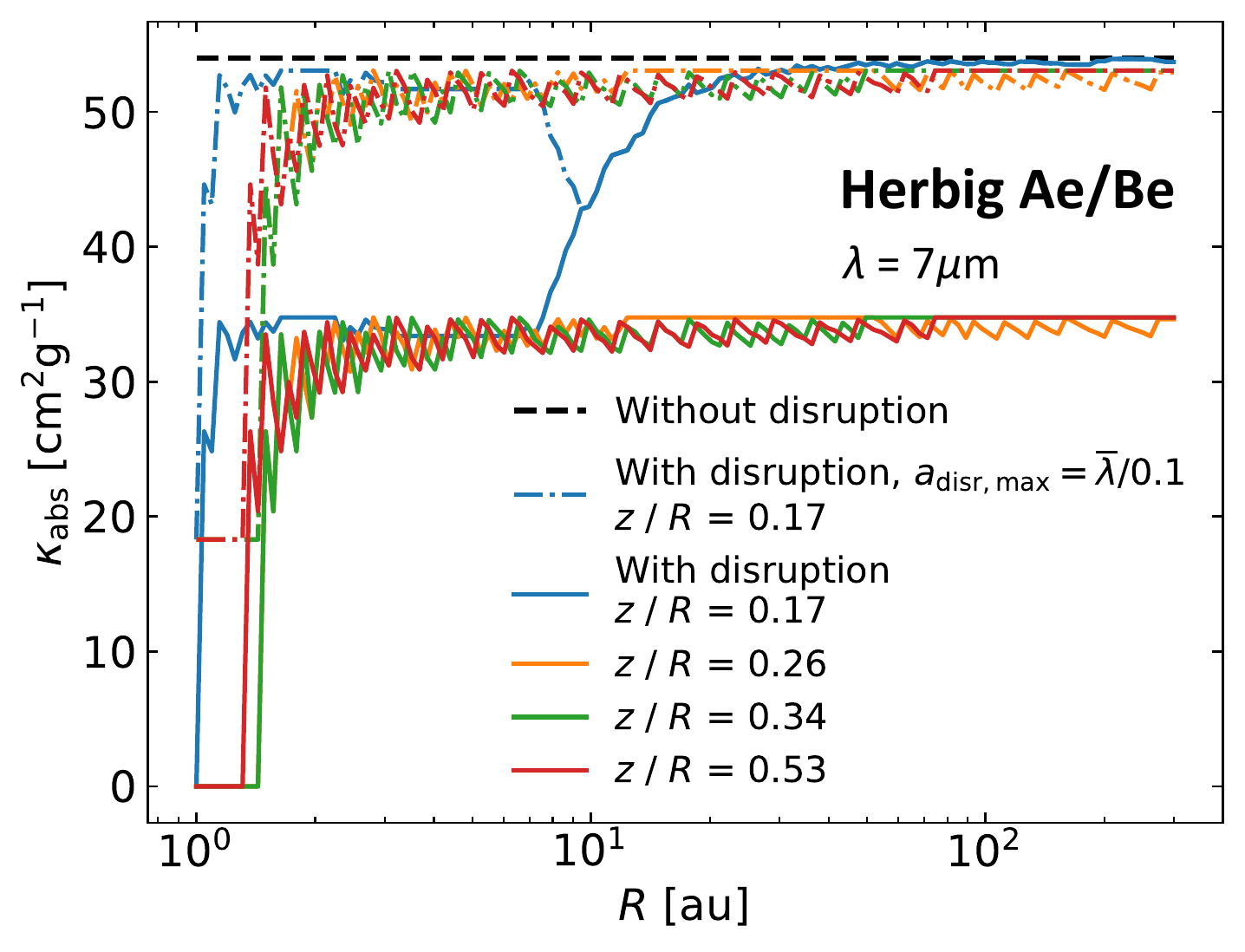}
\caption{Scattering opacity $\kappa_{\rm sca}$ (upper panels) and absorption opacity $\kappa_{\rm abs}$ (lower panels) as a function of the disk radius $R$ for different values of $z/R$ with (solid lines) and without (dashed black line) the presence of rotational disruption. Dotted lines show the results for $a_{\rm disr,max}=\overline{\lambda}/0.1$. Left and right panels show the results for T-Tauri and Herbig Ae/Be disks, respectively.}
\label{fig:opacvsR}
\end{figure*}

In Figure \ref{fig:opacvsR}, we keep $z/R$ constant and plot the opacities $\kappa_{\rm abs}$ and $\kappa_{\rm sca}$ versus the disk radius for the T-Tauri disks at $\lambda=7\mum$. Doing so, we are able to see how $\kappa_{\rm abs}$ and $\kappa_{\rm sca}$ vary with $R$. For higher value of $z/R$, the inner region experiences a significant decrease in opacity due to very strong effect of rotational disruption, while for lower values, the difference is not pronounced due to the absence of RATD in the interior. The disruption of large grains will consequently decrease the intensity of scattered light in NIR-MIR wavelengths from the surface layer of the disks. Many observations of PPDs have detected scattered light by dust grains on the surface layer, e.g., VLT/SPHERE NIR observation of the RY Lup disk around a T-Tauri star by \citet{2018A&A...614A..88L}, Gemini Planet Imager observation of Herbig Ae/Be stars HD 150193, HD 163296, and HD 169142 by \citet{2017ApJ...838...20M}. \citet{2016A&A...596A..70S} performed the scattered light mapping on the observational data from VLT/SPHERE and VLT/NACO of the disk around HD 100546 star and concluded that scattering opacity in the surface layer is mostly due to large, aggregate dust grains. Therefore, we expect that disruption of these grains by RATD which reduces $\kappa_{\rm sca}$ will have an important effect on NIR-MIR observations.

\section{Discussion}\label{sec:discussion}

\subsection{RATD as a top-down mechanism to form PAHs/nanoparticles in the disk surface layers}
The widespread presence of PAHs/nanoparticles in the surface layer of PPDs around Herbig Ae/Be stars, even in the inner gap (e.g., \citealt{2019A&A...623A.135B}) is difficult to explain because PAHs/nanoparticles in the surface layer are expected to be destroyed by EUV photons and the X-ray component of the star's radiation spectrum (\citealt{2010A&A...511A...6S}). To explain the observations of PAHs in the disk, \cite{2010A&A...511A...6S} suggested that PAHs are transported from the disk interior to the surface via turbulent mixing in the vertical direction, which requires the presence of PAHs in the disk interior. This scenario is difficult to reconcile with the fact that PAHs are expected to be depleted due to condensation into the ice mantle of dust grains in cold dense clouds (\citealt{1999Sci...283.1135B}; \citealt{2014A&A...562A..22C}; \citealt{2015ApJ...799...14C}). In this paper, we propose a top-down mechanism in which PAHs/nanoparticles are produced from the disruption of dust grains when they are being transported from the disk interior to the surface layers.

Our detailed modeling shows that large grains can be disrupted into nanoparticles in a vast area of the surface layer by the RATD mechanism. Moreover, PAHs/nanoparticles, which are frozen in the icy grain mantles, can desorb from the ice mantle via the ro-thermal desorption mechanism (\citealt{HoangTung}). Thus, if grains are small enough ($a\lesssim 1\mu$m) to be mixed with the gas by turbulence, they can escape grain settling and coagulation in the disk mid-plane and are transported to the surface layer (\citealt{1995Icar..114..237D}; \citealt{2004A&A...421.1075D}; \citealt{2006A&A...457..343F}; \citealt{2009A&A...496..597F}). When exposed to stellar radiation, large grains within the range size [$a_{\rm disr}$, $a_{\rm disr,max}$] could be disrupted by RATD into PAHs/nanoparticles (see Figure \ref{fig:adisrvsz}). Furthermore, our results suggest an increase in the relative abundance of nanoparticles with decreasing the distance to the star. Therefore, in the framework of rotational disruption and desorption, PAHs and nanoparticles would be continuously replenished in the surface layer, with the support of turbulent transport. This can resolve the longstanding puzzle about the ubiquitous presence of PAHs/nanoparticles in the hot surface layer of PPDs.

Finally, spectroscopic and polarimetric measurements of the debris disk HD 32297 from \citet{2019A&A...630A..85B} found an unusual blue NIR spectrum which corresponds to the presence of small grains of size much smaller than the bow-out limit due to radiation pressure \citep{1999ApJ...527..918W}. The authors explained this feature by a combination of several physical processes, including steady-state collisional evolution and the chain reaction of dust grain collisions triggered by the initial breakup of a large planetesimal near the central star, which \citet{2007A&A...461..537G} named the ``avalanche'' process. However, this process needs a two-belt structure for the breakups of massive planetesimals and the development and propagation of avalanche chain reaction \citep{2018A&A...609A..98T}.

\subsection{Implications for the detection of gas-phase water and complex molecules in PPDs}

Our numerical results for desorption sizes presented in Section \ref{subsec:adesp} indicate that rotational desorption is very efficient in the disk surface. The ice mantles of icy grains of size in the range $a_{\rm desp}<a<a_{\rm desp,max}$ are disrupted into small fragments. These tiny fragments then can be transiently heated to evaporation temperatures, inducing transient release of molecules, as theoretically demonstrated in \cite{Hoang:2019td}. Therefore, water ice and complex molecules frozen-in the ice mantle can be released into the gas phase even at low temperatures than their sublimation threshold. In particular, we predict that gas-phase water and COMs could be detected at locations well beyond the snowline of $R>3-100$ au (see Figures \ref{fig:a_desp_TTauri} and \ref{fig:a_disr_TTauri}).

To date, several complex molecules were detected in PPDs, including CH$_{3}$OH and (HCOOH) from TW Hydrae disk by \citep{2016ApJ...823L..10W} and \cite{Favre:2018kf}, and CH$_3$CN by \cite{Oberg2015}. Recently, ALMA observations toward the DG Tau disk by \citet{2019A&A...623L...6P} showed the lack of H$_2$CO emission in the warm inner disk, while the emission peak is located beyond the CO iceline ($\sim 40 \AU$). Interestingly, these molecules are detected far beyond the snowline at a distance above $10 \AU$. This cannot be explained by thermal sublimation, but is achieved via ro-thermal desorption as well as rotational desorption as shown in Figures \ref{fig:a_desp_TTauri} and \ref{fig:a_disr_TTauri}.

We note that non-thermal desorption mechanisms such as cosmic-ray-induced desorption and UV photodesorption are not ruled out in PPDs (see e.g., \citealt{Oberg2015}). Nevertheless, \citet{2010ApJ...722.1607W} carried out physical and chemical modeling of the PPD around the T-Tauri stars and found that cosmic-ray-induced desorption has little effect on the gas-phase abundances in the disk. 

\subsection{Implications for formation of comets and asteroids}
The water snowline characterized by the water sublimation temperature of $T_{d}\sim150\K$ divides the inner region of rocky planets from the outer region of gas giant planets.  Nevertheless, the precise location of the snowline is a longstanding problem in planetary science (\citealt{2006ApJ...640.1115L}; \citealt{Min:2011id}). In this paper, we studied the rotational desorption of ice mantles from the grain surface in PPDs.

Using the standard model of PPDs, we show that ice mantles from micron-sized grains ($a<100 \mum$) are disrupted and identify the location of snowline in the presence of rotational desorption. We find that the water snowline is extended outward. Figures \ref{fig:a_desp_TTauri} and \ref{fig:a_desp_Herbig} shows that rotational desorption takes place in the region beyond the snowline and thus can destroy the ice mantles of grains in this region. Indeed, observational detection of water ice from edge-on disks is reported in \citet{2007ApJ...667..303T} using water ice absorption features \citep{2008PASJ...60..557I}. Future telescopes such as SPHEREx, GMT, JWST, SUBARU, and WFIRST would be useful to test snowline. The rotational desorption of ice can have important implications for formation of planets, comets, and asteroids. Indeed, the total mass of ice rotationally disrupted is $M_{\rm ice,disr} = \int_{R_{\rm snowline}}^{R_{\rm out}} 2\pi r \Delta \Sigma_{\rm ice,disr}(r) dr \sim 10^{23}-10^{24} \g$ for T-Tauri disks and $\sim 10^{25} \g$ for Herbig Ae/Be disks. As a result, the number of comets formed in the solar nebula would be significantly lower than previous estimates. On the other hands, the number of asteroids would be higher accordingly.

\subsection{Rotational disruption and the problem of carbon deficit of the Earth and Rocky bodies}
Various studies indicate that the C/Si abundance ratio in the Earth and rocky bodies in the inner solar system is several orders of magnitudes lower than the primordial interstellar ratio (see e.g., \citealt{2010ApJ...710L..21L}; \citealt{2015PNAS..112.8965B}; \citealt{2017ApJ...845...13A}). Here we suggest that rotational disruption can explain the carbon deficit if there exist two separate populations of carbonaceous and silicate grains in PPDs. Note that the two separate populations originate naturally from the rotational disruption of large aggregates by RATD if these original grains are a carbon-silicate mixture as expected from grain coagulation in dense regions.

Indeed, our present studies imply that carbonaceous nanoparticles can be reproduced by the disruption of large carbonaceous grains which are being transported from the disk interior to the surface layer (see Figures \ref{fig:a_disr_TTauri} and \ref{fig:a_disr_Herbig}). When being exposed to harsh UV radiation, PAHs/carbonaceous nanoparticles are destroyed, converting solid carbon into the gas phase. The disruption of silicate grains by RATD is expected to be less efficient because of the effect of grain alignment. Indeed, silicate grains, which are paramagnetic material, are aligned by RATs along the ambient magnetic fields, so-called {\it B-RAT} alignment (see \citealt{Andersson2015} and \citealt{LAH15} for reviews). On the other hand, carbonaceous grains are diamagnetic material, such that they are unlikely aligned with the magnetic field but with the radiation direction (\citealt{2007MNRAS.378..910L}). If the stellar radiation makes an angle $\psi$ with the ambient magnetic field, the rotation rate driven by RATs for silicate grains is smaller than that of carbon grains by a factor $\cos\psi$ (\citealt{2009ApJ...695.1457H}; \citealt{2014MNRAS.438..680H}) if the Larmor precession rate ($\tau_{\rm Lar}^{-1}$) exceeds the precession rate around the radiation ($\tau_{k}^{-1}$ ;\citealt{2016ApJ...831..159H}). The B-RAT alignment can occur closer to the star if silicate grains contain iron inclusions (\citealt{2016ApJ...831..159H}). The magnetic field geometry in PPDs is uncertain, but for an hourglass-shaped magnetic field, one expects that the stellar radiation makes a considerable angle with the magnetic field in the surface layers, i.e., $\cos\psi< 1$. As a result, RATD is, in general, more efficient for carbonaceous grains than silicate grains, which enhances the conversion of large carbons into nanoparticles and the carbon depletion. A detailed study will be presented elsewhere.

\section{Summary}\label{sec:summary}
We have studied the disruption of dust and of ice mantles in PPDs using the newly discovered rotational disruption mechanism based on centrifugal force within rapidly spinning dust grains induced by radiative torques. The principal results are summarized as follow:
\begin{enumerate}
    
\item 
We performed two-dimensional modeling of rotational disruption of composite grains by radiative torques, namely the RATD effect, for the PPDs around T-Tauri and Herbig Ae/Be stars. We find that the RATD effect is efficient in disrupting large grains into nanoparticles. The disruption time is found much shorter than the grain-grain collision time in the warm and hot surface layers.

\item
We also modeled rotational disruption and desorption of ice mantles. We found that the ice mantles can be disrupted even in the warm intermediate layer with grain temperatures $T_{d}\sim 30-150\K$. We also found that the water and COMs can be desorbed rapidly as a result of ro-thermal desorption from tiny fragments produced by RATD. 

\item We quantified the effect of RATD on the dust opacity and found that rotational disruption of grains of sizes $a\sim 0.1-10\mum$ reduces the scattering opacity significantly in optical- MIR wavelengths. As a result, observations of scattered light from the surface layers of PPDs must take into account the effect of rotational disruption.

\item Our results suggest that rotational disruption of ice mantles and fluffy grains could reproduce PAHs/nanoparticles, which are usually observed in the hot surface layers of PPDs. The subsequent destruction of PAHs by extreme UV photons would induce the deficit of carbons in the PPDs, resolving the problem of C deficit in terrestrial planets.

\item We showed that the water snowline defined by rotational desorption of ice mantles is more extended than the classical snowline predicted by water evaporation temperature. The amount of water ice destroyed by rotational desorption is significant up to a distance of $100 \AU$, which decreases the number of comets and increases the number of asteroids.

\item Our detailed modeling shows that rotational desorption and ro-thermal desorption of ice mantles are still active beyond the snowline. Thus, future modeling of chemistry in disks should take into account the effect of rotational disruption and desorption of dust and ice.

\item We suggest that rotational disruption by RATs may resolve the carbon deficit problem in the Earth and rocky bodies in the inner solar system because carbonaceous grains may be disrupted more efficiently than silicates due to their higher rotation rate than silicate grains.
\end{enumerate}

\acknowledgments
We are grateful to the anonymous referee for helpful comments that improved our calculations and the presentation of the manuscript. N.D.T thanks Le Ngoc Tram, R. R. Rafikov and C. P. Dullemond for helpful discussions during the early stage of this work. This research was supported by the National Research Foundation of Korea (NRF) grants funded by the Korea government (MSIT) through the Basic Science Research Program (2017R1D1A1B03035359) and Mid-career Research Program (2019R1A2C1087045).

\end{document}